\documentclass[a4paper,11pt]{article}
\pdfoutput=1 

\usepackage{jcappub} 

\usepackage[T1]{fontenc} 
\usepackage[utf8]{inputenc}

\usepackage[english]{babel}
\usepackage{amsmath,amssymb,amsbsy,amstext,amsthm,booktabs,simplewick,exscale,relsize,slashed,graphicx,amsfonts,upgreek, xcolor}
\usepackage{multirow,color,dcolumn,bm,enumerate}
\usepackage{hyperref}
\usepackage{dsfont}
\usepackage{ragged2e} 
\DeclareUnicodeCharacter{2212}{-}
\usepackage{threeparttable}
\usepackage{makecell}

\usepackage[normalem]{ulem}

\title{Probing Primordial Black Hole Formation from Domain Wall Isocurvature Perturbations: Constraints and Implications}

\author[a,b,1]{Bo-Qiang Lu,\note{Corresponding author.}}
\author[c,d]{Cheng-Wei Chiang,}
\author[e,f,g]{and Tianjun Li}

\affiliation[a]{School of Science, Huzhou University, Huzhou, Zhejiang 313000, P. R. China}
\affiliation[b]{Zhejiang Key Laboratory for Industrial Solid Waste Thermal Hydrolysis Technology and Intelligent Equipment}
\affiliation[c]{Department of Physics, National Taiwan University, Taipei 10617, Taiwan}
\affiliation[d]{Physics Division, National Center for Theoretical Sciences, Taipei 10617, Taiwan}
\affiliation[e]{School of Physics, Henan Normal University, Xinxiang 453007, P. R. China}
\affiliation[f]{CAS Key Laboratory of Theoretical Physics, Institute of Theoretical Physics, Chinese Academy of Sciences, Beijing 100190, P. R. China}
\affiliation[g]{School of Physical Sciences, University of Chinese Academy of Sciences, No.~19A Yuquan Road, Beijing 100049, P. R. China}

\emailAdd{bqlu@zjhu.edu.cn}
\emailAdd{chengwei@phys.ntu.edu.tw}
\emailAdd{tli@itp.ac.cn}

\newcommand{\bvw}{\overline{v}_w}
\newcommand{\bL}{\overline{L}}
\newcommand{\brw}{\overline{\rho}_w}
\newcommand{\sgw}{\sigma_{w}}
\newcommand{\bsx}{\boldsymbol{x}}

\newcommand{\brho}{\bar{\rho}}

\abstract{Domain walls are topological defects produced by the spontaneous symmetry-breaking of discrete 
symmetry during cosmological phase transitions. Domain walls can significantly contribute to the energy density in the late-evolution stage. We propose that the density perturbations from the fluctuations in the number density of the domain walls could collapse to form primordial black holes. 
This mechanism becomes effective when the domain wall energy density ratio to that of the radiation reaches about 0.1 in the radiation-dominated Universe.
We find that models with $Z_2$ symmetry are excluded for interpreting pulsar timing array observations on the nano-Hz gravitational wave background since this model's domain wall number density fluctuations could lead to an overabundance of the primordial black holes. 
Moreover, the models, which generate approximately $N\sim 10$ domain walls from the spontaneous breaking of a discrete  $Z_N$  symmetry, are also subject to stringent constraints due to the overproduction of primordial black holes.
}

\begin{document}
\maketitle
\flushbottom

\section{Introduction}

It is widely believed that new physics is required to explain various phenomena beyond the Standard Model (SM), including the 
neutrino mass and mixing~\cite{King:2014nza}, dark matter~\cite{Bertone:2016nfn}, 
strong CP problem~\cite{Marsh:2015xka}, and matter-antimatter asymmetry~\cite{Dine:2003ax}, etc. 
As the cosmic temperature drops with the expansion of the Universe, the spontaneous breakdown of discrete symmetries during a phase transition in new physics models can generate sheet-like topological defects, the domain walls~\cite{Zeldovich:1974uw,Kibble:1976sj}.
The Universe after the phase transition is then randomly divided by them into distinct domains, which represent the clumps of the degenerate vacua. The domain walls tend to stretch in size and the neighboring domains with the same vacua merge to stabilize the network.
This evolution can be described by the percolation theory~\cite{Stauffer:1978kr}, whose fundamental results are based on the existence of a percolation threshold $p_c=0.311$ for a simple cubic lattice.
The domain wall's energy density scales as $\rho_{w}\propto t^{-1}$ in the scaling regime, which is much slower than the decay of radiation and dust, and eventually dominates the cosmological evolution.
In the domain wall era, the repulsive force from domain walls would lead to an inflation phase of the Universe with the scale factor going as $a\propto t^2$. Such an expansion history is not compatible with the observations since it could strongly alter the Big Bang nucleosynthesis (BBN), reduce the time for the galaxy formation, and lead to unacceptably large anisotropies of the cosmic microwave background (CMB)~\cite{Planck:2015fie}. Nevertheless, the research on the nature of dark energy in the context of 
domain wall has been continuously conducted by recent works~\cite{Friedland:2002qs,Conversi:2004pi,Mulki:2022tfw}.

Discrete symmetries exist widely in new physics models. A $Z_2$ symmetry can emerge from the spontaneous breakdown of a gauge group in the grand unified theory (GUT)~\cite{Preskill:1992ck,Dunsky:2021tih}. The $Z_2$ symmetry is also commonly imposed on the two-Higgs-doublet models for the spontaneous CP violation to suppress the Higgs-mediated flavor-changing neutral currents (FCNCs) at tree level~\cite{Lee:1973iz,Preskill:1991kd,Gunion:2002zf}.
The global $U(1)$ symmetry in the axion models for solving the strong CP problem is explicitly broken to the $Z_{\mathcal{N}}$ symmetry by the QCD instanton effect~\cite{Sikivie:1982qv,Preskill:1991kd}. Furthermore, to provide a dark matter candidate, a $Z_2$ symmetry is often imposed to make the dark matter particle stable~\cite{Chao:2017vrq,Huang:2017kzu,Grzadkowski:2018nbc,Chiang:2019oms,Chiang:2020yym,King:2023ayw,Pham:2024vso}.

A common solution to the domain wall problem is to introduce a bias potential $V_{\rm bias}$ to break the 
degenerate vacua~\cite{Dine:1993yw,Dvali:1994wv}. The domain wall network collapses as its surface energy is comparable with the bias potential, $\sigma H\simeq V_{\rm bias}$.
The bias potential can be generated by the instanton effects if the global symmetry is anomalous under the framework of the theory, with such examples as the QCD and quantum gravity~\cite{Preskill:1991kd,Dine:1993yw,Dvali:1994wv,Abel:1995wk,Larsson:1996sp,Lu:2023mcz}.
It is notable that, based on the successful and precise prediction in Ref.~\cite{Chiang:2020aui}, Ref.~\cite{Lu:2023mcz} suggests that the recent pulsar timing array (PTA) observations, including the NANOGrav 15-year dataset (NG15) and IPTA-DR2 dataset, strongly support the mechanism of the QCD instanton-induced domain wall annihilations.

Primordial black holes (PBHs) have attracted much attention since it was first suggested by Zel'dovich more than 50 years ago~\cite{Zeldovich:1967lct}.
Contrary to the black holes resulting from the death and collapse of stars in their final evolution stage, PBHs are formed by cosmological processes in the early Universe.  
In 1971, Hawking proposed the first mechanism of PBH production~\cite{Hawking:1971ei,Carr:1974nx}: 
the fluctuations of a radiation-dominated Universe can produce some overdense regions.  Once the density of the region $\delta=\delta\rho/\bar{\rho}$ exceeds a critical value $\delta_c\simeq w$, the overdense region can stop expanding and re-collapse to form PBHs when the region enters the horizon.
Cosmic inflation~\cite{Ivanov:1994pa,Randall:1995dj,Garcia-Bellido:1996mdl} is commonly considered a primary source of density perturbation. The large curvature perturbations generated by the quantum fluctuations of the scalar fields are dominated by the density perturbations once they exit the horizon~\cite{Garcia-Bellido:1996mdl}. For other PBH production mechanisms, we refer to Refs.~\cite{Sasaki:2018dmp,Carr:2020gox} for a recent review.

In contrast to PBH formation from the collapse of closed domain wall~\cite{Ipser:1983db,Rubin:2000dq,Ferrer:2018uiu,Liu:2019lul,Gouttenoire:2023gbn,Ge:2023rrq}, we show in this work that the density perturbation due to the number density fluctuations of domain walls in the late stage could lead to the formation of PBHs. 
The collapse of a closed domain wall needs to satisfy the Schwarzschild criterion, i.e., the closed wall has a Schwarzschild radius near collapse~\cite{Ferrer:2018uiu}. Our scenario is based on the critical collapse of horizon-size perturbations during the horizon crossing, as outlined above. We find that if the annihilations of the domain wall are the source of the nano-Hz stochastic gravitational wave background (SGWB) observed recently by PTAs, a significant PBH relic abundance could be produced via the number density fluctuations of the walls before their annihilations.

This work is arranged as follows. In section~\ref{sec:dwreview}, we review the evolution of the domain wall and emphasize the chopping effect on the domain wall size.  In section~\ref{sec:dwflucation}, we calculate the variance of density contrast and the power spectrum of the density perturbation from domain wall number density fluctuations.  In section~\ref{sec:PBHcc}, we review the critical collapse mechanism.  In section~\ref{sec:PBHabundance}, we calculate the PBH relic abundance from domain wall number density fluctuations.  In section~\ref{sec:EOS}, we define an effective equation of state for the mixture gas
containing domain walls and radiation.  A fit formula for the $\delta_c-\omega_{\rm eff}$ relation from the simulation is also provided.  We derive the gravitational wave background from the domain wall annihilations in section~\ref{sec:GWfromDW}, and perform the Bayesian statistics analysis of the IPTA-DR2 and NG15 datasets in section~\ref{sec:PTAdata}.  In section~\ref{sec:PBHconstraint}, we show the constraints on the domain wall interpretations for the PTAs using the PBH abundance from the domain wall number density fluctuations. 
Finally, in section~\ref{sec:conclusion}, we summarize our conclusions and discuss the uncertainty in the PBH formations. 
In appendix~\ref{app:toy}, appendix~\ref{app:surfaceenergy}, and appendix~\ref{app:VOS}, we provide respectively the details of calculations for the energy-momentum tensor of the domain wall in a $Z_2$ symmetry model, the surface energy tensor in the observer frame, and the velocity-dependent one-scale (VOS) model.  In appendix~\ref{app:btf}, we provide details for the complementary error function.
In appendix~\ref{app:params}, we present further details of our numerical strategy for the PTA dataset.
In appendix~\ref{app:models}, we provide detailed models that can generate \( N \) domain walls after spontaneous symmetry breaking.

\section{Domain wall}\label{sec:dwreview}

In this section, we first calculate the statistical state of the domain walls. To make sense of the domain wall evolution, we review the VOS model with the consideration of particle friction. The conclusions from this model are generally compatible with numerical simulations.

\subsection{Domain wall equation of state}\label{subsec:dweos}

The energy-momentum tensor of a plane domain wall lying in the $yz$ plane at $x=0$ (neglecting the thickness of the wall) can be estimated as~\cite{Vilenkin:1984ib,Gelmini:1988sf}
\begin{equation}\label{eq:tensorx}
    T_{\mu\nu}(x)=\sigma_w\delta (x){\rm diag}(1,0,-1,-1),
\end{equation}
where $\sigma_w$ is the surface tension of the wall. In appendix~\ref{app:toy}, we provide a concrete example for the calculation of the energy-momentum tensor in a toy model with a $Z_2$ symmetry.  Consider that within a volume $V\equiv L^3$, there are $N$ $yz$-plane walls moving along the $x$ direction with an average velocity $\overline{v}_w$.  The average energy-momentum tensor within a length $L$ can be estimated as
\begin{eqnarray}
    \begin{aligned}
    \overline{T}_{\mu\nu}\simeq \frac{1}{L}\int_{-L/2}^{L/2} dx T_{\mu\nu}(x,\overline{v}_w)
    =\frac{NS_{\mu\nu}^x(\overline{v}_w)}{L}=\frac{S_{\mu\nu}^x(\overline{v}_w)}{\overline{L}},
    \end{aligned}   
\end{eqnarray}
where $\overline{L}=L/N$ is the average wall separation, in which there exists on average one domain wall with the surface energy-momentum tensor
\begin{equation}
    S_{\mu\nu}^x(\overline{v}_w)=\int_{\overline{L}} dx T_{\mu\nu}(x,\overline{v}_w),
\end{equation}
with the superscript $x$ denoting the axis of wall motion. The energy-momentum tensor~\eqref{eq:tensorx} is in the rest frame of the wall. We can obtain the energy-momentum tensor in the observer frame by performing a Lorentz transformation.
In general, the domain wall can move along an arbitrary direction in the three-dimensional space and, with some algebra, we obtain the surface energy-momentum tensor in the observer frame as
\begin{equation}\label{eq:Smunu}
    S_{\mu\nu}(\overline{v}_w)=\frac{\sigma_w}{3}{\rm diag}(3\gamma^2,\overline{v}_w^2\gamma^2-2,
    \overline{v}_w^2\gamma^2-2,\overline{v}_w^2\gamma^2-2),
\end{equation}
where $\gamma=1/\sqrt{1-\bvw^2}$ is the Lorentz factor. See appendix~\ref{app:surfaceenergy} for the details of calculations.

The average energy-momentum tensor is determined by $\overline{T}_{\mu\nu}=S_{\mu\nu}/\overline{L}$.
Note that since the wall can extend along any direction, we interpret $\overline{L}$ as the correlation length of the wall.
The energy density and pressure of the wall are given by the temporal and spatial components of the energy-momentum tensor
\begin{equation}
    \rho_w=\overline{T}_{00}=\frac{\gamma^2\sigma_w}{\overline{L}},~~{\rm and}~~p_{w}=\overline{T}_{ii}=\frac{\sigma_w}{3\overline{L}}(\bvw^2\gamma^2-2).
\end{equation}
We then obtain the equation of state of the wall
\begin{eqnarray}\label{eq:dweos}
    \omega=\frac{p_w}{\rho_w}=\frac{-2+\bvw^2 \gamma^2}{3\gamma^2}.
\end{eqnarray}
For a relativistic wall with velocity $\bvw\sim 1$ and $\gamma\gg 1$, we have $\omega\simeq 1/3$, and the wall behaves like radiation. For a static wall, $\bvw\simeq 0$ and $\omega\simeq -2/3$. The equation of state is negative, $\omega<0$, for $\bvw\lesssim 0.8$.
In this case, the wall contributes negative pressure to the energy-momentum tensor and can lead to the inflation of the Universe if it dominates the energy density of the Universe.

\subsection{The evolution of a domain wall}\label{sec:VOS}

The correlation length \( \bar{L} \) (or curvature radius) and the root-mean-square velocity \( \bar{v}_w \) are two key physical quantities for describing the domain walls. 
We first adopt the so-called VOS model~\cite{Martins:2016book} (see appendix~\ref{app:VOS} for more details) to describe the evolution of \( \bar{L} \) and \( \bar{v}_w \) with cosmic expansion, which is given by
\begin{equation}\label{eq:VOS}
    \begin{aligned}
    \frac{d \bL}{d t}&=H \bL+\frac{\bL}{\ell_d} \bvw^2+c_w \bvw, \\
    \frac{d \bvw}{d t}&=\left(1-\bvw^2\right)\left(\frac{k_w}{\bL}-\frac{\bvw}{\ell_d}\right),
    \end{aligned}
\end{equation}
where the momentum parameter \( k_w = 0.66 \pm 0.04 \) and the chopping parameter \( c_w = 0.81 \pm 0.04 \) are derived from simulations of domain wall evolution in a radiation-dominated Universe~\cite{Martins:2016ois}.  
The momentum parameter \( k_w \) characterizes the curvature acceleration of the wall, while the chopping parameter \( c_w \) represents the interaction and merger between walls. The damping length is defined as 
\begin{equation}\label{eq:ell}
    \frac{1}{\ell_d}=3 H+\frac{1}{\ell_f} .
\end{equation}
The first term on the right-hand side of Eq.~\eqref{eq:ell} describes the Hubble damping effect due to the cosmic expansion and 
the second term denotes the particle friction, which is related to the pressure of the interaction by $\ell_f=\bvw \sigma_w/\Delta P$, where the pressure between the wall and particles $\Delta P$ is model-dependent.
We assume a pressure induced by friction from the bias potential, expressed as \(\Delta P = V_{\rm bias} \equiv V_{\rm false} - V_{\rm true}\). The vacuum pressure arising from the bias potential can drive the annihilation of the domain wall once it becomes comparable to the pressure exerted by the wall's surface tension. This condition determines the annihilation time of the domain wall as follows
\begin{equation}
    t_{\rm ann}=\frac{C_d\mathcal{A}\sgw}{V_{\rm bias}},
\end{equation}
where $\mathcal{A}$ is an $\mathcal{O}(1)$ parameter that depends on the number of domains.

\begin{figure}[t!]
    \centering
    \includegraphics[width=0.90\textwidth]{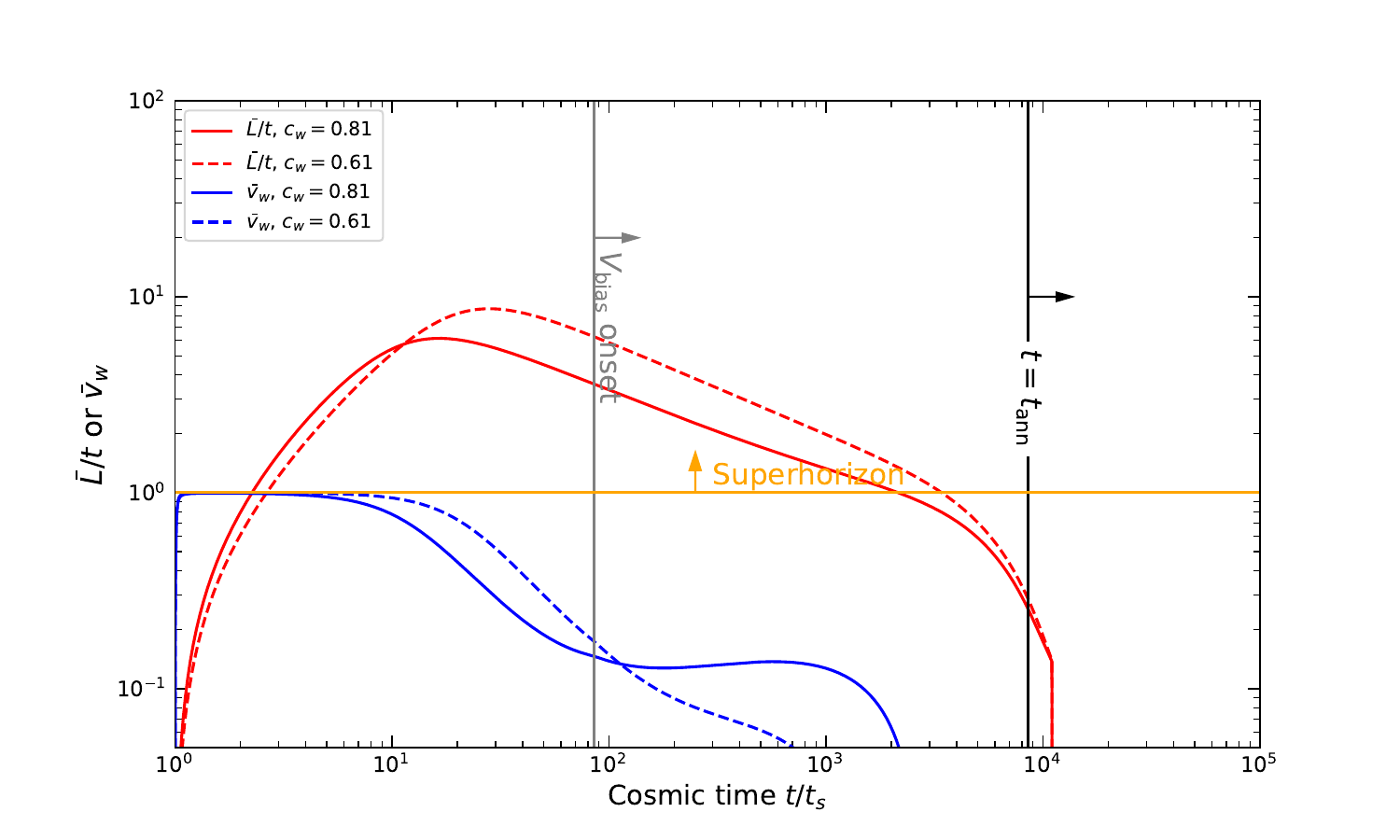}\\
    \includegraphics[width=0.90\textwidth]{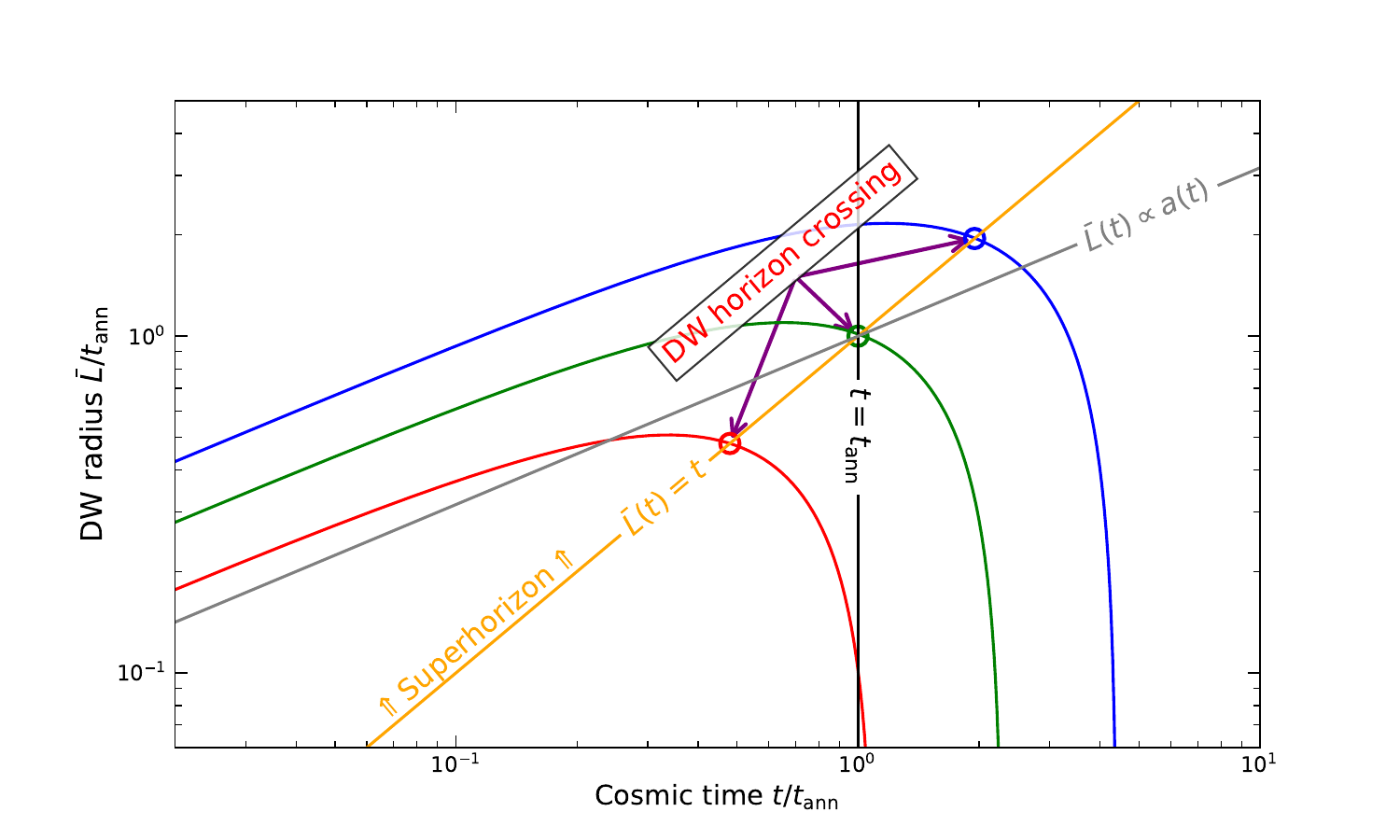}
    \caption{ Upper panel: The evolution of \( \bar{L} \) (red curves) and \( \bar{v}_w \) (blue curves) as a function of cosmic time \( t/t_s \) in the VOS model is shown. The solid curves and the dot-dashed curves represent the results with a chopping parameter $c_w=0.81$ and $0.61$, respectively.
    Lower panel: The evolution of the closed domain wall, governed by Eq.~\eqref{eq:SDW1}, is illustrated for different initial radii, represented by red, green, and blue curves.}
     \label{fig:VOS}
\end{figure}

In the upper panel of Fig.~\ref{fig:VOS}, we illustrate the evolution of the domain wall formed at time \( t_s \) with a symmetry-breaking scale of \( v = \sigma_w^{1/3} = 10^5 \)~GeV. The formation time of the domain wall is determined by \( t_s = M_{\rm Pl} / (3.32 g_*^{1/2} T_s^2) \), where \( g_*(T) \simeq 106.75 \) represents the effective degrees of freedom of the plasma, \( M_{\rm Pl} = 1.2 \times 10^{19} \)~GeV is the Planck mass, and the domain wall formation temperature is assumed to be \( T_s = v \). The initial conditions \( \bar{L}(t_s) = 0.01 t_s \) and \( \bar{v}_w \to 0 \) are adopted to solve Eqs.~\eqref{eq:VOS}. We further assume a vacuum pressure \( V_{\rm bias} = (10~\rm GeV)^4 \), with the onset of the bias potential occurring at \( t_{\rm on} = 0.01 t_{\rm ann} \).

The red curves illustrate the evolution of \( \bar{L}(t) \), while the blue curves denote that of \( \bar{v}_w \). To estimate the impact of the chopping effect, the chopping parameter \( c_w \) is set to 0.81 and 0.61 for the solid and dashed curves, respectively.
We summarize the key observational results from Fig.~\ref{fig:VOS} as follows:
\begin{itemize}
    \item Initially, domain walls are produced as a radiation-like fluid, exhibiting \( \bar{v}_w = 1 \) and \( \omega = 1/3 \) (see Eq.~\eqref{eq:dweos}). This dynamical feature is consistent with the velocity evolution equation in the VOS model (Eq.~\eqref{eq:VOS}): when \( \bar{L} \to 0 \), the curvature acceleration \( k_w / \bar{L} \to \infty \), which would otherwise lead to an unphysical infinite velocity. To enforce a finite initial velocity, we impose \( \bar{v}_w = 1 \) for the domain walls so that $d\bar{v}_w/dt=0$. However, due to limitations in numerical resolution, simulations in Ref.~\cite{Martins:2016ois} fail to explicitly capture this initial relativistic behavior.
    \item Driven by curvature acceleration induced by wall tension, the domain walls rapidly expand to a superhorizon scale with \( \bar{L} \sim 6t \). To clarify this evolution, we fix \( \bar{v}_w = 1 \) in the first equation of the VOS model (Eq.~\eqref{eq:VOS}) before \( \bar{L}/t \) attains its maximum value. Prior to the activation of the bias potential, the damping length satisfies \( 1/\ell_d = 3H \); combined with the radiation-dominated Universe condition \( H = 1/(2t) \), the characteristic scale of the domain walls initially grows as \( \bar{L} \propto t^2 \). As \( \bar{L} \) increases, the curvature acceleration gradually weakens, while the relativistic speed of the domain walls can be sustained until the curvature acceleration is overtaken by Hubble damping. This transition occurs when the characteristic scale reaches
    \begin{equation}
        \bar{L}=\frac{6k_w}{\bar{v}_w}t\sim 6t~,
    \end{equation}
    which is determined by the balance condition \( k_w/\bar{L} = \bar{v}_w/\ell_d \). Notably, owing to the wall’s curvature acceleration, domain walls can retain extreme relativistic speeds ($\bar{v}_w \simeq 1$) even after their characteristic scale enters the superhorizon regime (i.e., $\bar{L} \gtrsim t$).
    \item Once the domain walls reach the superhorizon scale of \( \bar{L} \sim 6t \), the condition \( k_w/\bar{L} \lesssim \bar{v}_w/\ell_d \) is satisfied, meaning that the local deceleration due to friction becomes more dominant than curvature acceleration. Subsequently, the correlation length of the domain walls decreases to the horizon scale under the combined influence of Hubble friction, chopping effects, and friction from the bias potential. As cosmic time approaches \( t_{\rm ann} \), \( \bar{L}(t) \) experiences a substantial decrease, re-entering the horizon prior to the domain walls’ annihilation at \( t_{\rm ann} \).
    \item Furthermore, we note that at the early stages, chopping between domain walls can cause an increase in \( \bar{L} \), which is a signature of domain wall mergers. Following the attainment of a maximum value, chopping effectively reduces the domain wall scale.
\end{itemize}

It is important to emphasize that the curvature acceleration of domain walls is a global mechanical effect induced by wall tension—analogous to the stretching of an elastic membrane—whereby each segment of the wall undergoes synchronous stretching under the tension, independent of local information propagation. Consequently, the curvature radius, or equivalently the characteristic scale of the domain wall, can extend to a superhorizon regime (\(\bar{L} \sim 6t\)) without violating causality constraints. In contrast, the propagation speed of damping effects, e.g., the Hubble friction, cannot exceed the speed of light and thus is subject to causality limitations. As a result, the characteristic length associated with damping effects must not be larger than the Hubble scale. For the Hubble friction term in Eq.~\eqref{eq:VOS}, the damping length is \(\ell_d = 1/(3H) = 2t/3\), which is explicitly subhorizon.

In the initial phase, the curvature acceleration dominates the expansion of domain walls, driving them to superhorizon scales while sustaining extremely relativistic speeds (\(\bar{v}_w \simeq 1\)). As the characteristic scale of the domain walls expands to \(\bar{L} \sim 6t\), the efficiency of curvature acceleration diminishes significantly. Both friction and chopping are local effects; once curvature acceleration weakens, these local processes take over the evolution of the superhorizon-sized domain walls, ultimately leading to their re-entry into the horizon.

The primary role of the chopping effect is not to maintain the scaling regime, but to regulate the transition into it by controlling when superhorizon-sized walls re-enter the horizon. We therefore argue that its key significance lies in setting the timing of this horizon re-entry. As shown below, this timing directly determines the amplitude of the resultant density perturbation, and consequently the abundance of PBHs. In contrast, consistent with simulations~\cite{Martins:2016ois,Dankovsky:2024zvs}, chopping is not an essential mechanism for sustaining the scaling regime itself.
While the chopping effect is expected to be time-dependent, the analysis in Ref.~\cite{Martins:2016ois} employs a constant chopping parameter for simplicity.
The chopping effect between walls can result in the efficient production of closed configurations, as confirmed by the simulations~\cite{Lazanu:2015fua}.
The generation of closed domain walls could, in turn, suppress the chopping effects between walls. Once closed domain walls form at a superhorizon scale, the onset of the scaling regime for the walls could be hindered by the reduced chopping effect. 
It has been suggested that superhorizon-sized closed domain walls might never attain the scaling regime prior to the onset of appreciable pressure from the bias potential~\cite{Gouttenoire:2023gbn}.
To explore this, we adopt the following formula to describe the evolution of a spherical domain wall by neglecting the chopping effect~\cite{Deng:2020mds}
\begin{equation}\label{eq:SDW1}
\ddot{\chi}+\left(4-3 a^2\dot{\chi}^2\right) H \dot{\chi}+\frac{2}{a^2 \chi}\left(1-a^2 \dot{\chi}^2\right) 
=-\left(\frac{V_{\rm bias}}{\sgw}-6 \pi G \sgw\right) \frac{\left(1-a^2 \dot{\chi}^2\right)^{3 / 2}}{a},
\end{equation}
where the dot above a variable denotes a time derivative \( d/dt \), and \( \chi(t) \) is related to the curvature radius of the spherical wall by \( \bar{L}(t) = a \chi(t) \), with the scale factor \( a = (t / t_i)^{1/2} \) (where \( t_i \) denotes the initial time). Using \( \rho_w = \mathcal{A} \sigma_w / t \) and \( \rho_{\rm tot} = 3H^2 / (8\pi G) = 3 / (32\pi G t^2) \), the energy fraction of the domain wall is given by
\begin{equation}
    f_w(t)=\frac{\rho_w}{\rho_{\rm rad}}\simeq \frac{\rho_w}{\rho_{\rm tot}}=\frac{32\pi\mathcal{A}G\sgw t}{3}.
\end{equation}
Then, we obtain \( G \sigma_w t = 3 f_w(t) / (32\pi \mathcal{A}) \). Multiplying both sides of Eq.~\eqref{eq:SDW1} by \( t_{\rm ann} \), we derive the dimensionless form of Eq.~\eqref{eq:SDW1} as
\begin{equation}\label{eq:SDW2}
\ddot{\tilde{\chi}}+\left(4-3 a^2 \dot{\tilde{\chi}}^2\right) \tilde{H} \dot{\tilde{\chi}}+\frac{2}{a^2 \tilde{\chi}}\left(1-a^2 \dot{\tilde{\chi}}^2\right)
=-\left(C_d\mathcal{A}-\frac{9 f_w(t_{\rm ann})}{16 \mathcal{A}}\right) \frac{\left(1-a^2 \dot{\tilde{\chi}}^2\right)^{3 / 2}}{a},
\end{equation}
where the dot now denotes the derivative \( d/d\tilde{t} \) with \( \tilde{t} \equiv t / t_{\rm ann} \), \( \tilde{\chi}(\tilde{t}) \equiv \chi(\tilde{t}) / t_{\rm ann} \), and \( \tilde{H}(\tilde{t}) = 1 / (2\tilde{t}) \). We assume that the closed domain wall is generated at the initial time \( t_i \), occurring after the formation of the domain wall, i.e., \( t_i > t_s \). We then solve the equation of motion, Eq.~\eqref{eq:SDW2}, for the spherical domain wall with the initial conditions \( \bar{L}(t_i) \equiv a(t_i) \tilde{\chi}(t_i) \gg t_s \), \( \tilde{\chi}^{\prime}(t_i) = 0 \), and \( a(t_i) = 1 \).

In the lower panel of Fig.~\ref{fig:VOS}, we plot the trajectories of the curvature radius for different initial radii, represented by red, green, and blue lines.  We observe that well before the horizon crossing of the domain wall, the wall's radius scales as \( \bar{L}(t) \propto a(t) \), indicating a superhorizon size above the horizon line \( \bar{L}(t) = t \). As time progresses, the pressure from the bias potential becomes significant, causing the radius trajectory to cross the horizon line at a time \( t_e \). If \( t_e \lesssim t_{\rm ann} \), the wall network enters the horizon before significant annihilation occurs.
For the case where \( t_e \gtrsim t_{\rm ann} \), as denoted by the blue trajectory, the annihilation occurs at the horizon crossing time \( t_e \), rather than at \( t_{\rm ann} \), leading to the phenomenon of later annihilation~\cite{Gouttenoire:2023gbn}. Thus, we confirm that for closed domain walls with negligible chopping effects, the trajectory of the curvature radius remains at a superhorizon scale and does not enter the scaling regime (\( \bar{L}(t) \simeq t \)) until shortly before annihilation starts.

\section{Density fluctuations from domain walls}\label{sec:dwflucation}

After the production of domain walls via the Kibble mechanism~\cite{Kibble:1976sj}, the size of the walls rapidly grows to a superhorizon scale through the merger and expansion of the walls. Subsequently, the walls re-enter the horizon due to the combined effects of chopping, Hubble friction, and the pressure from the bias potential. 
During the horizon crossing of the domain walls, the network ultimately becomes a few-body system composed of individual domain walls. Consider a domain wall network consisting of \( N \) identical walls, each with mass \( M_w \), randomly distributed within the Hubble volume \( V_H \equiv 4\pi r_H^3 / 3 \), where \( r_H = 1/H \). We employ the Poisson distribution to describe the probability of finding \( n \) walls within a volume of \( V \equiv 4\pi r^3 / 3 \)
\begin{equation}\label{eq:PD}
    \mathbb{P}_n(V)
    =
    \left(\frac{V}{V_w}\right)^{n}\frac{e^{-V/V_w}}{n!},
\end{equation}
where \( V_w = 4\pi \bar{L}^3 / 3 \) is the volume of a single wall, and \( V = n V_w \). The applicability of the Poisson distribution to the domain wall network has been validated through simulations in Refs.~\cite{Takahashi:2020tqv,Kitajima:2022jzz}. 
The spatially averaged energy density of the domain wall within the volume \( V \) is given by \( \bar{\rho}_w(r) = nM_w / V \). The ensemble average of \( \bar{\rho}_w(r) \) is then expressed as
\begin{equation}\label{eq:agbrho}
   \langle \bar{\rho}_w(r) \rangle
   =
   \sum_{n=0}^{\infty}\bar{\rho}_w(r)\mathbb{P}_n(V)
   =
   \frac{M_w}{V_w}.
\end{equation}
Therefore, the quantity \( \langle \bar{\rho}_w(r) \rangle \equiv \langle \bar{\rho}_w \rangle \) is independent of the position \( r \). The density perturbation arising from Poisson fluctuations is defined as 
\begin{equation}
    \delta\bar{\rho}_w(r) 
    = 
    \bar{\rho}_w(r) - \langle \bar{\rho}_w \rangle.
\end{equation}
The variance of the energy density is then given by
\begin{equation}\label{eq:drho2}
    \langle (\delta \bar{\rho}_w(r))^2 \rangle
    =\langle \bar{\rho}_w^2(r) \rangle-\langle \bar{\rho}_w \rangle^2
    =\sum_{n=0}^{\infty}\bar{\rho}_w^2(r)\mathbb{P}_n(V)-\langle \bar{\rho}_w \rangle^2
    =\frac{M_w^2}{V_wV}.
\end{equation}
We define the relative energy density fluctuation, or the density contrast, as \( \delta = \delta\bar{\rho} / \rho_{\rm tot} \). For a two-component Universe consisting of radiation and domain walls, the total density perturbation is given by \( \delta\bar{\rho} = \delta\bar{\rho}_w + \delta\bar{\rho}_r \), where \( \delta\bar{\rho}_w \) and \( \delta\bar{\rho}_r \) are the perturbations induced by the domain wall and radiation, respectively. In this work, we neglect the radiation density perturbation, and the mean square of the density contrast is then given by
\begin{equation}
    \langle \delta^2(r) \rangle
    =
    \frac{\langle (\delta \bar{\rho}_{w}(r))^2 \rangle}{\rho_{\rm tot}^2}
    =
    \frac{\langle \bar{\rho}_w\rangle^2V_w}{\rho_{\rm tot}^2V},
\end{equation}
where we have used Eqs.~\eqref{eq:agbrho} and~\eqref{eq:drho2} for the second equality. If we consider the fluctuation within the Hubble volume, i.e., \( V \to V_H \), then we obtain
\begin{equation}\label{eq:vvh}
    \left. \langle \delta^2 \rangle \right|_{V \to V_H} 
    = 
    \frac{f_w^2}{N},
\end{equation}
where we have used the relations \( f_w(t) = \langle \bar{\rho}_w \rangle / \rho_{\rm rad} \simeq \langle \bar{\rho}_w \rangle / \rho_{\rm tot} \) and \( V_H = N V_w \).
Our rigorous result in Eq.~\eqref{eq:vvh} confirms the density fluctuation $\delta\rho/\rho \sim G\sigma_w t$ estimated by Vilenkin (see Eq.~(13.5.13) in Ref.~\cite{Vilenkin:2000jqa}) for a $Z_2$ network ($N\sim 1$), and thereby generalizes it to the case of an arbitrary domain wall number $N$.

Let us now calculate the power spectrum of the density perturbation arising from Poisson fluctuations. Denote the domain wall energy density at a location \( \bsx \) within the volume \( V \) as \( \rho_w(\bsx) \). The spatially averaged quantities can then be expressed as
\begin{equation}
\bar{\rho}_{w}(r)
=\frac{1}{V} \int_{V} \mathrm{~d}^3 \bsx\rho_{w}(\bsx)~~{\rm and}~~
\overline{\delta\rho}_{w}(r)
=\frac{1}{V} \int_{V} \mathrm{~d}^3 \bsx\delta\rho_{w}(\bsx).
\end{equation}
Using the definition $\delta\rho_w(\bsx)=\rho_w(\bsx)-\langle \bar{\rho}_w\rangle$, we find that $\overline{\delta\rho}_{w}(r)\equiv \delta\bar{\rho}_w(r)=\bar{\rho}_{w}(r)-\langle \bar{\rho}_w\rangle$ and 
$\langle (\overline{\delta\rho}_{w}(r))^2\rangle\equiv\langle (\delta\bar{\rho}_{w}(r))^2\rangle=\langle \bar{\rho}_w^2(r)\rangle-\langle \bar{\rho}_w\rangle^2$. Together with Eq.~\eqref{eq:drho2}, we obtain
\begin{equation}\label{eq:spatialcor}
    \left \langle \int_{V} \mathrm{~d}^3 \bsx\int_{V}\mathrm{~d}^3 \bsx^{\prime}\delta\rho_{w}(\bsx)\delta\rho_{w}(\bsx^{\prime})\right \rangle =\frac{M_w^2V}{V_w}.
\end{equation}

The randomness of the spatial distribution implies that fluctuations at two different locations are independent~\cite{Papanikolaou:2020qtd}. Furthermore, the spatial distribution is assumed to be homogeneous and isotropic. With these properties, we can express the density correlator as 
\begin{equation}
    \langle \delta\rho_{w}(\bsx) \delta\rho_{w}(\bsx^{\prime}) \rangle = g \, \delta(\bsx - \bsx^{\prime}),
\end{equation}
where \( \delta(\bsx - \bsx^{\prime}) \) is the Dirac delta function, and \( g \) may depend on \( |\bsx - \bsx^{\prime}| \). Integrating over the volume \( V \), we obtain
\begin{equation}
    \int_{V} \mathrm{d}^3 \bsx \int_{V} \mathrm{d}^3 \bsx^{\prime} \, \langle \delta\rho_{w}(\bsx) \delta\rho_{w}(\bsx^{\prime}) \rangle = V g.
\end{equation}
Using Eq.~\eqref{eq:spatialcor}, we find \( g = M_w^2 / V_w = \langle \bar{\rho}_w \rangle^2 V_w \). Substituting this into the correlator, we obtain the density contrast correlator as
\begin{equation}\label{eq:dr1}
    \left \langle \frac{\delta\rho_{w}(\bsx)}{\rho_{\rm tot}} \frac{\delta\rho_{w}(\bsx^{\prime})}{\rho_{\rm tot}} \right \rangle = f_w^2 V_w \, \delta(\bsx - \bsx^{\prime}).
\end{equation}

Expanding the density perturbation in the Fourier space, we have 
\begin{equation}\label{eq:Fexpand}
    \delta_{w}(t) \equiv \frac{\delta \rho_{w}(\boldsymbol{z})}{\bar{\rho}_{w}} = \int \frac{\mathrm{d}^3 \boldsymbol{k}}{(2 \pi)^{3/2}} \, \delta_{w,\boldsymbol{k}}(t) \, e^{i \boldsymbol{k} \cdot \boldsymbol{z}},
\end{equation}
where \( \boldsymbol{k} \) and \( \boldsymbol{z} \) denote the wave number and coordinates in the comoving space, respectively. Using Eq.~\eqref{eq:dr1}, the correlator in the comoving coordinates is given by
\begin{equation}\label{eq:dr2}
    \left \langle \frac{\delta\rho_{w}(\boldsymbol{z})}{\bar{\rho}_{w}} \frac{\delta\rho_{w}^*(\boldsymbol{z}^{\prime})}{\bar{\rho}_{w}} \right \rangle = \frac{4\pi}{3} \left(\frac{\bar{L}}{a}\right)^3 \delta(\boldsymbol{z} - \boldsymbol{z}^{\prime}).
\end{equation}
The reduced power spectrum \( \mathcal{P}_{\delta_w}(k) \) is defined in terms of 
\begin{equation}
    \left\langle \delta_{w,\boldsymbol{k}} \delta_{w,\boldsymbol{k}^{\prime}}^* \right\rangle \equiv \frac{2\pi^2}{k^3} \mathcal{P}_{\delta_w}(k) \delta\left(\boldsymbol{k} - \boldsymbol{k}^{\prime}\right).
\end{equation}
With Eqs.~\eqref{eq:Fexpand} and~\eqref{eq:dr2}, we obtain
\begin{equation}
    \mathcal{P}_{\delta_w}(k) = \frac{2}{3\pi} \left(\frac{k}{k_e}\right)^3 \simeq \left(\frac{k}{\mathcal{H}(t_e)}\right)^3 \quad \text{for} \quad k < k_e,
\end{equation}
where \( k_e = a / \bar{L}(t_e) \sim \mathcal{H}(t_e) \), and \( t_e \) denotes the horizon crossing time of the wall. The second approximate equality considers a horizon-sized domain wall during the horizon crossing. The power spectrum for the density contrast is then given by
\begin{equation}
    \mathcal{P}_{\delta}(k) = f_w^2(t) \mathcal{P}_{\delta_w}(k) \simeq f_w^2(t) \left(\frac{k}{\mathcal{H}(t_e)}\right)^3 \quad \text{for} \quad k \lesssim \mathcal{H}(t_e).
\end{equation}
As discussed above, the horizon crossing of the domain wall occurs when the pressure from the bias potential becomes comparable to the surface energy of the wall. Therefore, it is expected that the annihilation of the domain wall takes place near the time of the horizon crossing, i.e., \( t_e \simeq t_{\rm ann} \). We also consider the scenario in which PBHs are formed at the horizon crossing time of the wall. Thus, we have \( t_f = t_e \simeq t_{\rm ann} \). For simplicity, in the following analysis, we adopt the domain wall annihilation time \( t_{\rm ann} \) for evaluating the power spectrum.

In the inflationary theory for PBH production, the variance of the density fluctuations is generally related to the power spectrum by 
\begin{equation}\label{eq:varpw}
    \sigma^2 = \int_{0}^{\infty} W^2(kR) \, \mathcal{P}_{\delta}(k) \, \frac{dk}{k},
\end{equation}
where \( W(kR) \) is a volume-normalized Fourier transform of the window function smoothed over a comoving scale \( R \), often assumed to be Gaussian. Instead of using Eq.~\eqref{eq:varpw}, which depends on the assumption of \( W(kR) \), we adopt Eq.~\eqref{eq:vvh} as the variance of the density perturbation from the domain wall network during its horizon crossing
\begin{equation}\label{eq:variance}
    \sigma^2 = \left. \langle \delta^2 \rangle \right|_{V \to V_H} = \frac{f_w^2}{N}.
\end{equation}
Note that if the primordial black hole is formed during the horizon crossing of the domain wall, i.e., \( k \simeq \mathcal{H}(t_e) \), the power spectrum is \( \mathcal{P}_{\delta}(k) \simeq f_w^2(t_e) \). We observe that the variances given by Eqs.~\eqref{eq:varpw} and \eqref{eq:variance} are of the same order, \( \sim f_w^2(t_e) \), except for a statistical factor \( 1/N \). The case of the horizon crossing of the domain wall and the superhorizon mode \( k \) occurring at the same time \( t_e=t_f\simeq t_{\rm ann} \) provides the primary scenario for the production of PBHs, as will be considered below.

Then the probability density distribution function of the density contrast, $P(\delta)$, is featured by a zero 
ensemble averaged value, $\langle \delta \rangle=0$, and a variance $\sigma$ given by Eq.~\eqref{eq:variance}.
In this work, we consider a Gaussian density distribution function
\begin{equation}\label{eq:gaussP}
    P(\delta)=\frac{1}{\sqrt{2\pi}\sigma}\exp\left(-\frac{\delta^2}{2\sigma^2}\right).
\end{equation}
Simulations~\cite{Press:1989yh,Hiramatsu:2012sc,Lazanu:2015fua} indicate that the domain walls tend to merge with the neighboring ones in the evolution, resulting in a wall network with a small number of objects. For a model with a $Z_2$ symmetry, there eventually exists $N = 1$ domain wall that separates two domains~\cite{Press:1989yh}. For models with $\mathcal{N}\sim 10$ domains, one expects that $N\sim \mathcal{N}$ domain walls exist in the late stage of the network evolution.
Therefore, once the domain walls contribute significantly to the total energy density, with \( f_w \sim 0.1 \), near the time of domain wall annihilation, the energy density perturbation arising from the number density fluctuations of the domain walls can become remarkable.

\section{Isocurvature perturbation from domain wall network}\label{sec:isoperturb}

In this section, we consider cosmological perturbations in the background spacetime, described by the flat Friedmann–Robertson–Walker (FRW) Universe. The FRW metric, including the first-order scalar perturbations in the conformal-Newtonian gauge, is given by
\begin{equation}
d s^2 = a(\eta)^2 \left[ -(1 + 2 \Phi) d \eta^2 + (1 - 2 \Psi) \delta_{i j} d x^i d x^j \right],
\end{equation}
where \( \Phi \) represents the Newtonian potential and \( \Psi \) denotes the intrinsic curvature perturbation, both of which are gauge-invariant quantities.
The background evolution is governed by the Friedmann equation
\begin{equation}
    \mathcal{H}^2 = \frac{8\pi G}{3} a^2 \bar{\rho},
\end{equation}
and the energy continuity equation
\begin{equation}\label{eq:cont}
    \bar{\rho}^{\prime} = -3 \mathcal{H} \bar{\rho} (1 + \omega),
\end{equation}
where \( \omega = \bar{p} / \bar{\rho} \) is the equation of state parameter, and the prime denotes the derivative with respect to the conformal time.
From these equations, we can derive several useful background relations. Here, we list two of them
\begin{eqnarray}
\label{eq:bkg3}
\frac{w^{\prime}}{1+w} & =&3 \mathcal{H}\left(w-c_s^2\right) \\
\label{eq:bkg4}
\bar{p}^{\prime}=w \bar{\rho}^{\prime} +w^{\prime} \bar{\rho}&=&-3 \mathcal{H}(1+w) c_s^2 \bar{\rho},
\end{eqnarray}
where the sound speed $c_s^2=\bar{p}^{\prime}/\brho^{\prime}$.
In Fourier space, the Einstein equations for scalar perturbations are given by~\cite{Suonio:2014cp}
\begin{eqnarray}
\label{eq:Ein1}
&&\left(\frac{k}{\mathcal{H}}\right)^2 \Psi = -\frac{3}{2} \left[\delta^N + 3(1+w) \frac{\mathcal{H}}{k} v^N\right], \\
\label{eq:Ein2}
&&\left(\frac{k}{\mathcal{H}}\right)^2 (\Psi - \Phi) = 3 w \Pi, \\
\label{eq:Ein3}
&&\mathcal{H}^{-1} \Psi^{\prime} + \Phi = \frac{3}{2} (1+w) \frac{\mathcal{H}}{k} v^N, \\
\label{eq:Ein4}
&&\mathcal{H}^{-2} \Psi^{\prime \prime} + \mathcal{H}^{-1} \left(\Phi^{\prime} + 2 \Psi^{\prime}\right) + \left(1 + \frac{2 \mathcal{H}^{\prime}}{\mathcal{H}^2}\right) \Phi - \frac{1}{3} \left(\frac{k}{\mathcal{H}}\right)^2 (\Phi - \Psi) = \frac{3}{2} \frac{\delta p^N}{\bar{\rho}},
\end{eqnarray}
where the subscript \( N \) denotes the conformal-Newtonian gauge. Here, the relative energy density perturbation is defined as \( \delta = \delta\rho / \bar{\rho} \) with \( \delta\rho = \rho - \bar{\rho} \), \( v \) is the fluid velocity, and \( \Pi \) represents the anisotropic stress of the fluid.
The energy-momentum continuity equation \( T^{\mu}_{\nu;\mu} = 0 \) yields the density perturbations in the conformal-Newtonian gauge
\begin{eqnarray}
\label{eq:dNvN1}
\left(\delta^N\right)^{\prime} & =& (1+w) \left(-k v^N + 3 \Psi^{\prime}\right) + 3 \mathcal{H} \left(w \delta^N - \frac{\delta p^N}{\bar{\rho}}\right), \\
\label{eq:dNvN2}
\left(v^N\right)^{\prime} & =& -\mathcal{H} (1-3w) v^N - \frac{w^{\prime}}{1+w} v^N + \frac{k \delta p^N}{\bar{\rho} + \bar{p}} - \frac{2}{3} \frac{w}{1+w} k \Pi + k \Phi.
\end{eqnarray}

Let us consider a two-component universe with the total fluid energy density and pressure given by \( \bar{\rho} = \sum \bar{\rho}_i \) and \( \bar{p} = \sum \bar{p}_i \), respectively. The component fluids are labeled by \( i \) and \( j \), while quantities without a subscript denote variables of the total fluid. We assume that energy transfer between the fluid components in the background universe is negligible. In this case, each component fluid separately satisfies the energy continuity equation~\eqref{eq:cont}, the background relations~\eqref{eq:bkg3}-\eqref{eq:bkg4}, and the energy-momentum continuity equations~\eqref{eq:dNvN1}-\eqref{eq:dNvN2}.
The quantities for the total and component fluids are related by
\begin{equation}
    \omega = \frac{\sum \bar{\rho}_i \omega_i}{\bar{\rho}},
    ~~c_s^2 = \sum \frac{\bar{\rho}_i + \bar{p}_i}{\bar{\rho} + \bar{p}} c_i^2,
    ~~\delta = \frac{\sum \bar{\rho}_i \delta_i}{\bar{\rho}},
    ~~{\rm and}~~
    v = \sum \frac{1 + w_i}{1 + w} \frac{\bar{\rho}_i}{\bar{\rho}} v_i.
\end{equation}
We also define the total entropy perturbation as
\begin{equation}\label{eq:totalS}
\mathcal{S} \equiv \mathcal{H} \left( \frac{\delta p}{\bar{p}^{\prime}} - \frac{\delta \rho}{\bar{\rho}^{\prime}} \right)
= \frac{1}{3(1 + w)} \left( \frac{\delta \rho}{\bar{\rho}} - \frac{1}{c_s^2} \frac{\delta p}{\bar{\rho}} \right).
\end{equation}
Here, we have used Eq.~\eqref{eq:cont} and \( \bar{p}^{\prime} = c_s^2 \bar{\rho} \) for the second equality. This definition is also valid for the component fluids, as there is no energy transfer between them.
The relative entropy perturbation, or isocurvature perturbation, is defined as
\begin{equation}
    S_{ij} \equiv \mathcal{H} \left( \frac{\delta \rho_i}{\bar{\rho}_i^{\prime}} - \frac{\delta \rho_j}{\bar{\rho}_j^{\prime}} \right)
    = \frac{\delta_i}{1 + \omega_i} - \frac{\delta_j}{1 + \omega_j}.
\end{equation}
In the following, we omit the subscript and use \( S \) to denote the isocurvature perturbation.

Using Eqs.~\eqref{eq:bkg3},~\eqref{eq:dNvN1}, and~\eqref{eq:totalS}, we obtain
\begin{equation}\label{eq:Sijp1}
    S^{\prime} = -k v_{\rm rel} - 9 \mathcal{H} \left(c_i^2 \mathcal{S}_i - c_j^2 \mathcal{S}_j\right),
\end{equation}
where the relative velocity \( v_{\rm rel} \equiv v_i^N - v_j^N = v_i - v_j \) is a gauge-independent quantity. 
We further assume the relation \( p_i = \omega_i \bar{\rho}_i = c_i^2 \bar{\rho}_i \) for both fluid components. Consequently, the internal entropy perturbation \( \mathcal{S}_i \) in Eq.~\eqref{eq:Sijp1} vanishes. 
Using Eqs.~\eqref{eq:dNvN2} and~\eqref{eq:Sijp1}, we derive the second derivative of \( S \)
\begin{equation}\label{eq:Spp1}
\begin{aligned}
S^{\prime \prime} =  -k v_{\rm rel}^{\prime} 
= &~ k \mathcal{H} \left[ \left(1 - 3 w_i\right) v_i^N - \left(1 - 3 w_j\right) v_j^N \right] + k \left( \frac{w_i^{\prime}}{1 + w_i} v_i^N - \frac{w_j^{\prime}}{1 + w_j} v_j^N \right) \\
& - k^2 \left( \frac{c_i^2 \delta_i^N}{1 + w_i} - \frac{c_j^2 \delta_j^N}{1 + w_j} \right) + \frac{2}{3} k^2 \left( \frac{w_i}{1 + w_i} \Pi_i - \frac{w_j}{1 + w_j} \Pi_j \right).
\end{aligned}
\end{equation}
In the above derivation, the momentum transfer between components has been neglected.
Defining the energy fraction of component \( i \) as $f=\bar{\rho}_i/\bar{\rho}_j$, we can express the component perturbations in terms of the total and relative perturbations through the following transformations
\begin{eqnarray}\label{eq:matrix1}
\left\{\begin{matrix}
  \delta_i=a_{11}\delta+a_{12}S, \\
  \delta_j=a_{21}\delta+a_{22}S,
\end{matrix}\right.\quad\quad{\rm and}\quad\quad
\left\{\begin{matrix}
  v_i=b_{11}v+b_{12}v_{\rm rel}, \\
  v_j=b_{21}v+b_{22}v_{\rm rel},
\end{matrix}\right.
\end{eqnarray}
where
\begin{eqnarray}\label{eq:matrix2}
\left\{\begin{matrix}
    a_{11}=\frac{(1+\omega_i)(1+f)}{\lambda},~\\
    a_{12}=\frac{(1+\omega_i)(1+\omega_j)}{\lambda},
\end{matrix}\right.\quad\quad{\rm and}\quad\quad
\left\{\begin{matrix}
    a_{21}=-\frac{(1+\omega_j)(1+f)}{\lambda},\\
    a_{22}=\frac{(1+\omega_i)(1+\omega_j)f}{\lambda},
\end{matrix}\right.
\end{eqnarray}
and
\begin{eqnarray}\label{eq:matrix3}
\left\{\begin{matrix}
    b_{11}=\frac{(1+\omega)(1+f)}{\lambda},\\
    b_{12}=\frac{1+\omega_j}{\lambda},~~~~~~~~
\end{matrix}\right.\quad\quad{\rm and}\quad\quad
\left\{\begin{matrix}
    b_{21}=\frac{(1+\omega)(1+f)}{\lambda},\\
    b_{22}=\frac{(1+\omega_i)f}{\lambda},~~~~
\end{matrix}\right.
\end{eqnarray}
with $\lambda=1+f+f\omega_i+\omega_j$.
We have used the relation \( c_i^2 = \omega_i \), which implies \( \omega_i^{\prime} / (1 + \omega_i) = 0 \). Neglecting the anisotropic stress and replacing the component quantities with the total and relative ones in Eq.~\eqref{eq:Spp1}, we obtain
\begin{equation}\label{eq:KSE}
    \frac{S^{\prime \prime}}{\mathcal{H}^{2}} + \frac{(1 - 3\mu_2) S^{\prime}}{\mathcal{H}} = -\left(\frac{k}{\mathcal{H}}\right)^2 \left(\mu_1 \delta^C + \mu_2 S\right),
\end{equation}
where
\begin{equation}\label{eq:dC}
    \delta^C = \delta^N + 3 \left(\frac{\mathcal{H}}{k}\right) (1 + \omega) v^N,
\end{equation}
with \( C \) representing the comoving gauge, and the parameters
\begin{equation}
    \mu_1 = \frac{(\omega_i - \omega_j)(1 + f)}{\lambda} \quad {\rm and} \quad \mu_2 = \frac{\omega_i (1 + \omega_j) - \omega_j (1 + \omega_i) f}{\lambda}.
\end{equation}
If the components are cold matter and radiation, the parameters reduce to \( \mu_1 = -1 / (1 + w) \) and \( \mu_2 = (1 - 3c_s^2) / 3 \). In this case, we reproduce the so-called Kodama-Sasaki equation~\cite{Kodama:1986ud}.

The Bardeen equation determines the evolution of $\delta^C$ in the comoving gauge
\begin{equation}
\begin{aligned}
    &
    \mathcal{H}^{-2} \delta^{C\prime \prime}+\left(1-6 w+3 c_s^2\right) \mathcal{H}^{-1} \delta^{C\prime}-\frac{3}{2}\left(1+8 w-6 c_s^2-3 w^2\right)\delta^C
    \\
    &=-c_s^2 \left(\frac{k}{\mathcal{H}}\right)^2 \left[\delta^C-3(1+w) \mathcal{S}\right].
\end{aligned}
\end{equation}
With Eqs.~\eqref{eq:matrix1}-\eqref{eq:matrix3} and considering the conditions $f\ll 1$ and $\delta^C\ll S$ for the isocurvature perturbation at the initial time, the Bardeen equation can be written as 
\begin{equation}\label{eq:dC2}
    \mathcal{H}^{-2} \delta^{C\prime \prime}-\frac{3\delta^{C\prime}}{4\mathcal{H}}-\frac{17}{4}\delta^C
    =
    -\frac{1}{12} \left(\frac{k}{\mathcal{H}}\right)^2 \left(\delta^C-\frac{14}{3}fS\right),
\end{equation}  
where we have assumed $\omega_w=-2/3$ for the domain wall. Together with Eq.~\eqref{eq:KSE}, we can obtain the solution for the evolution of $\delta^C$ and $S$. Let us consider the isocurvature modes, i.e., $\delta^C\ll S$ and the condition $f\to 0$. Then Eqs.~\eqref{eq:dC2} and \eqref{eq:KSE} can be reduced to 
\begin{eqnarray}\label{eq:dC3}
    \mathcal{H}^{-2} \delta^{C\prime \prime}-\frac{3\delta^{C\prime}}{4\mathcal{H}}-\frac{17}{4}\delta^C & =&\frac{17}{48}\left(\frac{k}{\mathcal{H}}\right)^2 f S,
    \\
    \label{eq:S}
    \mathcal{H}^{-2} S^{\prime \prime}+3\mathcal{H}^{-1} S^{\prime} & =&0.
\end{eqnarray}
Requiring \( S(0) \) to be finite at the initial time \( \tau = 0 \), the solution to Eq.~\eqref{eq:S} is a constant, i.e., \( S = S(0) \). With this result, we can determine \( \delta^C \) by solving Eq.~\eqref{eq:dC3}, yielding
\begin{equation}\label{eq:dctau}
    \delta^C(\tau) = \frac{7}{90} k^2 \tau^2 f_w S(0).
\end{equation}
Using the Einstein equations~\eqref{eq:Ein1}-\eqref{eq:Ein4}, we derive various relations among perturbation quantities, summarized as follows
\begin{eqnarray}\label{eq:p1}
    \Phi &=& -\frac{3}{2} \left(\frac{\mathcal{H}}{k}\right)^2 \delta^C, \\
    \label{eq:v1}
    v^N &=& \frac{2}{3(1+w)} \left(\frac{k}{\mathcal{H}}\right) \left(\mathcal{H}^{-1} \Phi^{\prime} + \Phi\right), \\
    \label{eq:dN1}
    \delta^N &=& \delta^C - 3 \left(\frac{\mathcal{H}}{k}\right) (1+w) v^N,
\end{eqnarray}
where we have assumed a radiation-dominated universe, in which the anisotropic stress is negligible and \( \Phi = \Psi \). Together with Eq.~\eqref{eq:dctau}, we find
\begin{equation}
    \delta^N \equiv \delta \simeq \frac{1}{3}f_w S(0) \quad {\rm for} \quad k \to 0.
\end{equation}

For the superhorizon mode of the isocurvature perturbation \( k \ll \mathcal{H}(t) \), we can neglect the term on the right-hand side of Eq.~\eqref{eq:KSE}. Considering a radiation-dominated universe and requiring \( S(0) \) to be finite with \( S^{\prime}(0) = 0 \), we again find the solution \( S = S(0) \) for the superhorizon mode.

We assume that the density perturbation is primarily induced by the Poisson fluctuation of the domain wall network and neglect the radiation density fluctuation. Consequently, \( S \simeq \delta_w / (1 + \omega_w) = 3 \delta / f_w \). 
Based on the discussions in Section~\ref{sec:VOS}, we find that the domain walls are initially generated as a radiation gas, and therefore, the isocurvature perturbations are not produced at the initial time. The size (correlation length) of the domain wall rapidly grows to a superhorizon scale, and the closed configurations are formed before the domain wall is slowed down. The correlation length of the wall can be significantly reduced when the bias potential plays an important role in the domain wall evolution. As a result, isocurvature perturbations are produced only during the horizon crossing of the domain wall, which occurs near the annihilation time of the domain wall, as the bias potential has already become comparable to the surface tension of the wall. This implies that the superhorizon mode of the density perturbation arising from the isocurvature perturbation of the wall is given by
\begin{equation}\label{eq:dfS}
    \delta = f_w(t_{\rm ann}) \delta_w \simeq \frac{1}{3}f_w(t_{\rm ann}) S(0),
\end{equation}
where the relation \( t_e=t_f\simeq t_{\rm ann} \) has been adopted.
With \( \Phi = \Psi \) in the radiation-dominated universe, Eq.~\eqref{eq:Ein2} provides the Newtonian potential during the horizon crossing of mode \( k \)
\begin{equation}\label{eq:NPtf}
    \Phi(t_f) \simeq -\frac{3}{2} \left(\frac{k}{\mathcal{H}(t_{\rm ann})}\right)^2 \delta^N(t_{\rm ann}).
\end{equation}
Furthermore, from Eq.~\eqref{eq:dC}, we also have \( \delta^C \sim \delta^N \) when \( k \sim \mathcal{H}(t_{\rm ann}) \). Therefore, we expect that the gauge dependence of the density contrast becomes weak during the horizon crossing of mode \( k \).


\section{Critical collapse}\label{sec:PBHcc}

Let us assume that the perturbed region is spherically symmetric and has an initial radius $R(t)$ larger than the Hubble scale. 
The locally overdense region can then constitute a spatially closed Universe with a metric given by~\cite{Carr:1975qj,Yokoyama:1995ex}
\begin{equation}
    d s^2=-d t^{\prime 2}+R^2\left(t^{\prime}\right)\left[\frac{d r^2}{1-\kappa r^2}
    +r^2\left(d \theta^2+\sin ^2 \theta d \varphi^2\right)\right], 
\end{equation}
where $\kappa > 0$ denotes the perturbed total energy per unit mass. Therefore, the density of the perturbed region is 
\begin{equation}
    \rho=\frac{3}{8 \pi G} \left( H^2+\frac{\kappa}{R^2} \right).
\end{equation}
The initial expansion rates of the background and the perturbed region can be set to be the same by choosing the 
coordinates~\cite{Harrison:1969fb,Yokoyama:1995ex} so that the initial density contrast is given by
\begin{equation}
    \delta_i =\frac{\rho_i-\bar{\rho}_i}{\bar{\rho}_i}=\frac{3 \kappa}{8 \pi G \bar{\rho}_i R_i^2}
    =\frac{\kappa}{H_i^2 R_i^2} ,
\end{equation}
where the subscript $i$ is used to denote the initial time $t_i$ and the initial background density $\bar{\rho}_i=3H_i^{2}/(8\pi G)$.
Integrating the continuity equation
\begin{equation}
    \frac{d \rho}{d t}+3 H(\rho+p)=0,
\end{equation}
we obtain the evolution of the energy density 
\begin{equation}\label{eq:rho}
    \rho=\rho_i\left(\frac{R}{R_i}\right)^{-3(1+\omega)},
\end{equation}
with a constant equation of state $\omega\equiv p/\rho$.
At a critical time $t_c$, the overdense regions stop expanding, i.e. $H(t_c)=0$, with a critical density $\rho_c=\kappa/R_c^2$.
The ratio between the energy density at the initial time and the critical time is then given by 
\begin{equation}\label{eq:fracrho}
    \frac{\rho_{c}}{\rho_i}=\frac{\kappa}{R_i^2 H_i^2+\kappa} \frac{R_i^2}{R_{c}^2}
    =\frac{\delta_i}{1+\delta_i} \frac{R_i^2}{R_{c}^2}.
\end{equation}
Together with Eqs.~\eqref{eq:rho} and~\eqref{eq:fracrho}, we obtain the critical scale factor for which the overdense region stops expanding
\begin{equation}\label{eq:Rc}
    \frac{R_c}{R_i}=\left(\frac{1+\delta_i}{\delta_i}\right)^{\frac{1}{1+3 \omega}} .
\end{equation}
The scale factor evolves as $R(t)\propto t^{2/3(1+\omega)}$, therefore the critical time is given by 
\begin{equation}\label{eq:tc}
    \frac{t_c}{t_i}=\left( \frac{1+\delta_i}{\delta_i} \right)^{\frac{3(1+\omega)}{2(1+3\omega)}}.
\end{equation}

The overdense region could contrast further against the pressure gradient if the self-gravitation $\Phi$ of the mass within the critical radius $R_c$ is larger than the internal energy $U$ of the overdense region, which gives~\cite{Carr:1974nx}
\begin{equation}
    \Phi+U\sim -G\rho^2R_c^5+\omega \rho R_c^3\gtrsim 0.
\end{equation}
For a radiation-dominated Universe, $\rho\sim t^2/G$, then one obtains the lower limit on the scale of the overdense region
\begin{equation}\label{eq:Rc2}
    R_c\gtrsim R_J\simeq c_s t_c,
\end{equation}
where $R_J$ is the Jeans scale and $c_s\simeq \sqrt{\omega}$. 
With Eqs.~\eqref{eq:Rc},~\eqref{eq:tc}, and~\eqref{eq:Rc2}, one finds the critical density criterion 
for PBH formation in the radiation-dominated Universe:
\begin{equation}
    \delta_i\gtrsim \delta_c\simeq c_s^2\simeq\frac{1}{3}.
\end{equation}
A more precise value is $\delta_c=0.453$ for $\omega=1/3$ from the numerical calculation (see section~\ref{sec:EOS}).
If the overdense region stops expanding at a radius larger than the Hubble horizon, i.e., $R_c\gtrsim t_c$, the PBHs formed in 
this region would be separated from our Universe and, thus, make no contributions to the observations.
We therefore have the limit on the size of the overdense region, $c_st_c\lesssim R_c\lesssim t_c$, which corresponds to the condition
\begin{equation}
    \frac{1}{3}\lesssim \delta_i\lesssim 1.
\end{equation}

The Hawking and Carr arguments for PBH production can be succinctly represented by the Newton potential criterion
\begin{equation}\label{eq:NPcriterion}
    |\Phi | = \frac{3}{2} \left(\frac{k}{\mathcal{H}}\right)^2 \delta \gtrsim 1,
\end{equation}
where we have omitted the subscript \( N \) in Eq.~\eqref{eq:NPtf}. During the horizon crossing of mode \( k \) with \( k \simeq \mathcal{H} \), the Newton potential criterion equivalently requires \( \delta \gtrsim \delta_c \simeq 0.5 \).
Using the Poisson distribution in Eq.~\eqref{eq:PD} with the variance in Eq.~\eqref{eq:variance}, the ensemble-averaged density fluctuation from the isocurvature perturbation is given by
\begin{equation}\label{eq:aved}
    \bar{\delta}(t_{\rm ann}) = \int_0^{\infty} \delta P(\delta) d\delta = \frac{f_w(t_{\rm ann})}{\sqrt{2\pi N}},
\end{equation}
where \( t_e = t_f \simeq t_{\rm ann} \) has been used.
Comparing this with Eq.~\eqref{eq:dfS}, we observe that the mean initial isocurvature perturbation of the superhorizon mode is
\begin{equation}
    \bar{S}(0) = \frac{3}{\sqrt{2\pi N}}.
\end{equation}
For a network containing \( N \lesssim 10 \) domain walls, the isocurvature perturbation from the domain wall is \( S(0) \sim \mathcal{O}(1) \). 
From Eq.~\eqref{eq:aved}, we observe that the density perturbation could exceed the critical value and form a PBH if the energy density fraction of the domain wall \( f_w(t_e) \sim \mathcal{O}(0.1) \) during the horizon crossing of the domain wall.

\section{PBH formation}\label{sec:PBHabundance}

We first calculate the PBH relic abundance from the critical collapse. We then review the power spectrum of the density perturbation from the inflation and summarize the constraints of current CMB observations on the inflationary scenarios. Finally, we calculate the variance of the density contrast from the number density fluctuations of the domain walls.

\subsection{PBH relic abundance}

Let us define a parameter $\beta$ to represent the fraction of total energy consisting of PBHs
at their formation time $t_f$:
\begin{equation}\label{eq:bt0}
    \beta(t_f)=\frac{\rho_{\mathrm{PBH}}(t_f)}{\rho_{\rm tot}(t_f)}=\Omega_{\mathrm{PBH}}\left(\frac{H_0}{H_f}\right)^2
    \left(\frac{a_0}{a_f}\right)^{3},
\end{equation}
where $\Omega_{\mathrm{PBH}} \equiv \rho_{\mathrm{PBH}}\left(t_0\right) / \rho_{\text {crit}}$ (with $\rho_{\rm crit}=3H_0^2/(8\pi G)$) is the density of the PBH relic abundance at present and $0$ denotes the present time.
In the second equality of Eq.~\eqref{eq:bt0} we have used $\rho_{\rm PBH}(t_f)=\rho_{\rm PBH}(t_0)(a_0/a_f)^3$.
It is commonly related to the dark matter relic abundance by $\Omega_{\mathrm{PBH}}=f_{\mathrm{PBH}}\Omega_{\mathrm{CDM}}$, with the 
fraction of dark matter $f_{\rm PBH}$ in the form of PBHs.

It is expected that the PBHs formed from the Hubble horizon (re-)entry of the large density perturbations could have masses 
at the same order as the associated horizon mass $M_H\simeq 2\times 10^{15}(t/1s)M_{\odot}$.
To further estimate the PBH production, let's take the naive relation between the PBH mass $M$ and the horizon mass from 
critical phenomena~\cite{Sasaki:2018dmp,Carr:2020gox}
\begin{equation}\label{eq:Mpbh}
    M=\gamma_p M_H,
\end{equation}
where the numerical factor $\gamma_p\simeq 0.2$. During the radiation domination, the expansion rate of the Universe 
is $H=1/(2t)$, and the energy density of a flat Universe is 
\begin{equation}\label{eq:rhorad}
    \rho_{\rm rad}=\frac{3H^2}{8\pi G}=\frac{\pi^2}{30}g_*(T)T^4,
\end{equation}
where $g_*(T)$ is the total effective degrees of freedom, and the Newtonian gravitational constant $G$ is related to the 
Planck mass $M_{\rm pl}$ by $G=1/M_{\rm pl}^2$ with $M_{\rm pl}=1.22\times 10^{19}$~GeV. The horizon mass is then given by
\begin{equation}
    M_{H}=\frac{4\pi}{3}\rho_{\rm rad}r_H^{3}=\frac{1}{2GH}=\frac{3\sqrt{5}M_{\rm pl}^3}{4\pi^{3/2}g_{*}^{1/2}(T)T^2}.
\end{equation}
Taking advantage of the Press-Schechter formalism, the PBH energy fraction can be expressed as~\cite{Niemeyer:1997mt}
\begin{equation}\label{eq:PSF}
    \beta(M)=2\int_{\delta_c}^{\infty}\frac{M}{M_H}P(\delta)d \delta ,
\end{equation}
where we consider a Gaussian probability distribution~\eqref{eq:gaussP} for the density contrast.
Note that we use an upper limit of infinity for the integration and confirm that it is not different from the integration result with an upper limit of $\delta=1$ from the requirement of avoiding separate Universes~\cite{Carr:1974nx,Yokoyama:1995ex}.
A recent work~\cite{Kopp:2010sh} argued that this upper limit was a result of the geometry of the selected slicing rather than as a consequence of avoiding separate Universes, which instead led to a divergence of density fluctuations on the superhorizon scale.
With the PBH mass given by Eq.~\eqref{eq:Mpbh}, the fraction of PBHs at the formation time is 
\begin{equation}
    \beta(M)=\gamma_p{\rm erfc}\left(\frac{\delta_c}{\sqrt{2}\sigma}\right),
\end{equation}
where ${\rm erfc}(x)$ is the complementary error function (see appendix~\ref{app:btf} for more details).
The critical density contrast $\delta_c$ varies with the equation of state $\omega$. 
For radiation phase $\delta_c=0.453$. We determine $\delta_c$ with the $\delta_c-\omega$ relation revealed by the 
simulations in Ref.~\cite{Musco:2012au} (see section~\ref{sec:EOS}).

A more precise PBH mass formation was obtained from the numerical computations on the collapse of a relativistic fluid, 
which indicated a simple scaling law for the PBH mass distribution function near the critical 
threshold~\cite{Choptuik:1992jv,Evans:1994pj,Koike:1995jm}
\begin{equation}\label{eq:Mpbh2}
    M=CM_H(\delta-\delta_c)^{b},
\end{equation}
where the constant parameter $C=3.3$ and $b=0.36$ for the formation of PBH during radiation domination~\cite{Byrnes:2018clq}.
Note that these parameters may be affected by the phase of cosmic matter. Then the PBH energy fraction is given by
\begin{equation}\label{eq:bt}
    \beta(M)=2\int_{\delta_c}^{\infty}C(\delta-\delta_c)^{b}P(\delta)d \delta.
\end{equation}
In this work, we use the PBH mass~\eqref{eq:Mpbh} by default, unless otherwise stated.

Once the energy fraction in PBHs at their formation is obtained, the relic abundance of PBHs today can be determined by taking into account the cosmic expansion (and neglecting the evaporation and accretion of PBHs).
The radiation density red-shifts as $a^{-4}$ while the PBH density scales as $a^{-3}$.
Using the fact of entropy conservation in the comoving volume, i.e.,
\begin{equation}\label{eq:entcon}
    s_fa_f^3=s_0a_0^3,~~{\rm with}~~s(T)=\frac{2\pi^2}{45}g_{*,s}T^3,
\end{equation}
where the current Universe entropy density $s_0=2891.2~\rm cm^{-3}$. 
Together with Eqs.~\eqref{eq:bt0}-\eqref{eq:rhorad} we obtain the the current PBH relic density abundance
\begin{equation}\label{eq:omgpbh}
    \begin{aligned}
    \Omega_{\mathrm{PBH}}&=\frac{2\pi G\beta(M)s_0g_*(T)T}{H_0^2g_{*,s}(T)} 
    =(3\sqrt{5\pi})^{1/2}\frac{\beta(M)s_0}{H_0^2(M_{\rm pl}M_H)^{1/2}}\frac{g_{*}^{3/4}(T)}{g_{*,s}(T)}\\
    &\simeq \left(\frac{\beta(M)}{5.74 \times 10^{-9}}\right)\left(\frac{0.68}{h}\right)^{2} 
    \left(\frac{g_{*}(T)}{10.75}\right)^{3/4}\left(\frac{10.75}{g_{*,s}(T)}\right)\left(\frac{M_{\odot}}{M_H}\right)^{1/2},
    \end{aligned}
\end{equation}
where the Hubble expansion rate is $H_0=100 h~\mathrm{km~s}^{-1}~\mathrm{Mpc}^{-1}$ with $h=0.68$ from the Planck observation~\cite{Planck:2018vyg}.
With the relation $M=\gamma_p M_H$ and the approximation $g_{*}(T)\simeq g_{*,s}(T)$, we reproduce $\Omega_{\rm PBH}$ given in Ref.~\cite{Carr:2009jm}.
We observe that the PBH energy fraction at the formation should be of order $\beta(M)\sim 10^{-9}$ so that the PBH relic abundance today does not exceed the required dark matter relic abundance. Thus the variance of the density contrast should be small and the probability density distribution should be concentrated so that the probability for the production of large density contrasts is exponentially suppressed in the tail of the probability density distribution.

\subsection{Density fluctuation from inflation-induced curvature perturbation}

From the above, we know that one of the most important parameters to determine the PBH abundance is the variance $\sigma$, which is directly related to the physics of the production of density perturbations.

Let's first briefly review the PBH production in the inflation scenario.
For the density fluctuations from the curvature perturbations during inflation, the variance is in general estimated by Eq.~\eqref{eq:varpw}.
On comoving hypersurfaces, the density perturbation can be related to the curvature perturbation by~\cite{Green:2004wb}
\begin{equation}
    \mathcal{P}_{\delta}(k)=\frac{4(1+\omega)^2}{(5+3\omega)^2}\left(\frac{k}{aH}\right)^2\mathcal{P}_{\mathcal{R}_c}(k).
\end{equation}
The PBHs are formed at the horizon crossing $k=aH$. The power-law ansatz is commonly adopted for the primordial curvature power spectrum:
\begin{equation}
    \mathcal{P}_{\mathcal{R}_c}(k)=A_{\mathcal{R}_c} \left( \frac{k}{k_*}\right)^{n_s-1},
\end{equation}
where the amplitude $A_{\mathcal{R}_c}$ is a constant that depends on the inflation physics, and $k_*$ is the reference scale.
Current CMB and LSS observations on large scales ($k\lesssim {\rm Mpc^{-1}}$) have restricted the primordial curvature power spectrum at the order of around $10^{-9}$. To have a sizable PBH formation for LIGO observation, the primordial curvature power spectrum should be of order $\mathcal{P}_{\mathcal{R}_c}(k)\sim 10^{-2}-10^{-1}$ 
(corresponding to a variance in the range $10^{-4}\lesssim \sigma^2\lesssim 10^{-2}$ which gives $\beta\sim 10^{-9}-10^{-6}$, 
see Fig.~\ref{fig:bt} in appendix~\ref{app:btf}), which requires a blue-tilted spectral index $n_s\simeq 1.85$ 
at scale $k\sim 10^6~\rm Mpc^{-1}$ (corresponding to a PBH mass $M\sim 30 M_{\odot}$)~\cite{Sasaki:2018dmp,Byrnes:2018clq}.
On the other hand, the CMB observations have constrained the spectral index to be red-tilted, $n_s=0.968\pm0.006$ at the scale $k_*=0.05~\rm Mpc^{-1}$.
The relatively large blue-tilted spectral index at a small scale cannot be realized in the single-field slow-roll inflation but may be achieved in the running mass inflation or multi-scalar inflation models, in which the scale-dependent spectral index can be realized
(see Refs.~\cite{Stewart:1997wg,Drees:2011hb,Alabidi:2012ex,Silk:1986vc,Randall:1995dj,GarciaBellido:1996mdl,Clesse:2015wea} for example).

\subsection{Density fluctuation from isocurvature perturbation induced by domain wall network}

As demonstrated in section~\ref{sec:VOS}, the isocurvature perturbations from the domain wall network are produced during the horizon crossing of the domain wall. We consider the PBHs due to the density fluctuation arising from the isocurvature perturbation at the horizon crossing time of the domain wall. At this stage, the bias potential becomes significant, which could lead to the annihilation of the domain wall network. Therefore, we have \( t_e = t_f \simeq t_{\rm ann} \).
The energy density of the domain wall is related to the wall's scale by \( \rho_w = \sigma_w / \bar{L} \). At the horizon crossing time \( t_e \), \( \bar{L} \simeq t / \mathcal{A} \), and the domain wall energy density is then given by 
\begin{equation}\label{eq:rhodw}
    \rho_{w}=\frac{\mathcal{A}\sigma_w}{t}=2\mathcal{A}H\sigma_w,
\end{equation}
where $\mathcal{A}$ depends on the number of walls. For $N\le 6$ we take the values of $\mathcal{A}$ from the simulations~\cite{Kawasaki:2014sqa}, and for $N> 6$ we take $\mathcal{A}=0.4N$. Therefore, the energy fraction of the wall is 
\begin{equation}\label{eq:dwfrac}
    f_{w}=\frac{\rho_w}{\rho_{\rm rad}}=\frac{16\pi G\mathcal{A}\sigma_w}{3H}=\frac{32\pi G\mathcal{A}\sigma_w t}{3}.
\end{equation}
The variance of density contrast induced by the domain wall is then given by 
\begin{eqnarray}\label{eq:varMH}
    \sigma&=&\frac{f_w}{\sqrt{N}}=\frac{32\pi G\mathcal{A}\sigma_w t}{3\sqrt{N}}=\frac{32\pi G^2 M_H\mathcal{A}\sigma_w}{3\sqrt{N}}\nonumber\\
    &\simeq & 1.81\times10^{-3}\frac{\mathcal{A}}{\sqrt{N}}\frac{M_H}{M_{\odot}}\left(\frac{\sigma_w^{1/3}}{10^5~\rm GeV}\right)^3.
\end{eqnarray}
To make sure that the energy density of the domain wall does not dominate the Universe, we require $f_w\lesssim 1$, which 
gives the following constraint on the PBH horizon mass 
\begin{equation}
    M_H\lesssim 5.52\times 10^2 M_{\odot}\left(\frac{10^5~\rm GeV}{\sigma_w^{1/3}}\right)^3.
\end{equation}

\begin{figure}[t!]
    \centering
    \includegraphics[width=0.49\textwidth]{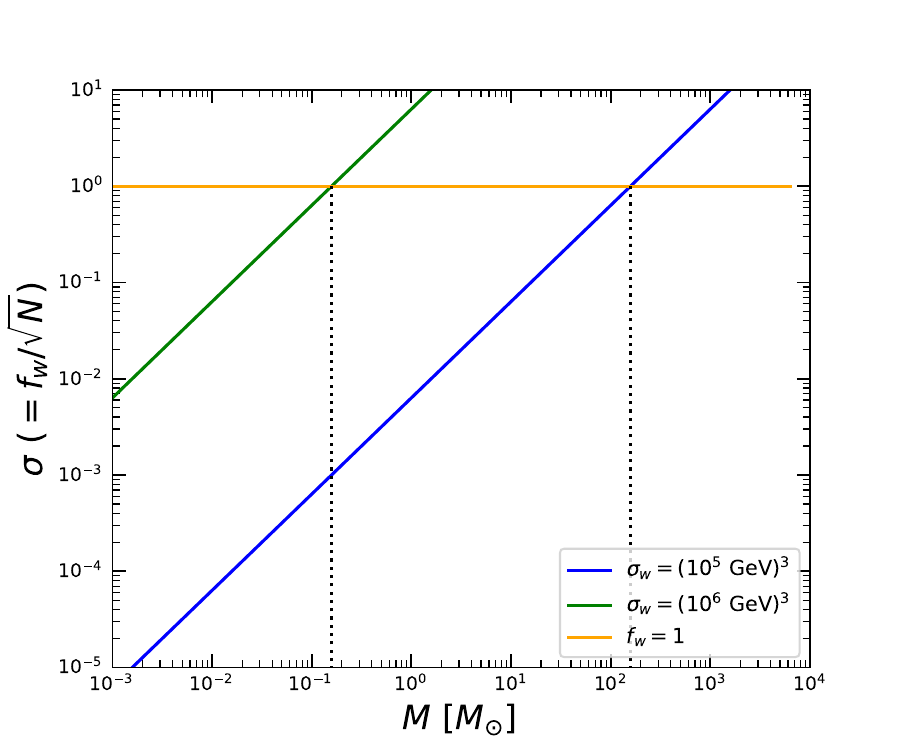}
    \includegraphics[width=0.49\textwidth]{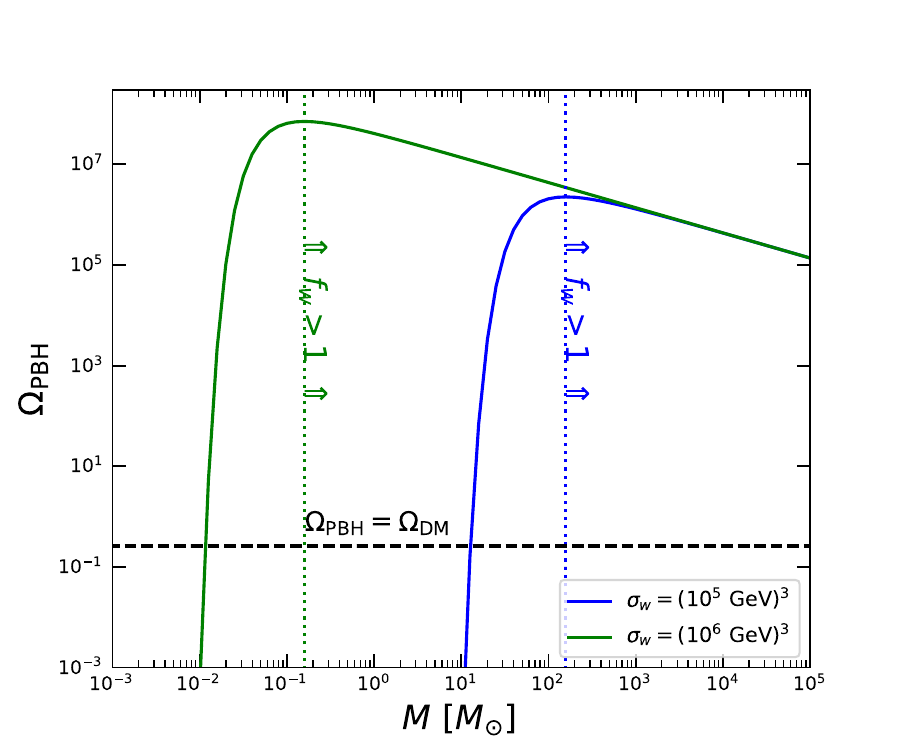}
    \caption{Left: The variance $\sigma$ as a function of the PBH mass $M$.
    Right: The PBH relic abundance $\Omega_{\rm PBH}$ as a function of the PBH mass $M$.  We take the wall number $N=1$.}
     \label{fig:sigma}
\end{figure}

In Fig.~\ref{fig:sigma}, we plot the variance~\eqref{eq:varMH} and PBH relic abundance as a function of the PBH mass with the relation $M=\gamma_p M_{H}$ and assuming the domain wall number $N=1$. The blue and green solid curves represent the results with the wall surface tension $\sigma_w^{1/3}=10^5$~GeV and $10^6$~GeV, respectively.
As shown in the figure, the PBH relic abundance from the domain wall number fluctuations becomes significant when the variance $\sigma (=f_w/\sqrt{N})\sim 10^{-1}$, which would correspond to a PBH mass of $M\sim 10^1~M_{\odot}$ and $10^{-2}~M_{\odot}$ for $\sigma_w^{1/3}=10^5$~GeV and $10^6$~GeV, respectively, as shown in the right panel of the figure.  The PBH abundance sharply increases with the domain wall energy fraction $f_w$ and reaches its maximum at $f_w=1$.
For $f_w\simeq 1$, the PBH energy fraction reaches its maximum $\beta(M)\simeq 2\times 10^{-1}$ (see the right plot of Fig.~\ref{fig:bt} in appendix~\ref{app:btf}).
Beyond the maximum point, $\Omega_{\rm PBH}$ scales as $M^{-1/2}$. However, we note that the results with $f_w\gtrsim 1$ are not reliable.
This is because the domain wall energy density dominates the radiation and the Universe enters into a solid-like phase with the 
equation of state $\omega\simeq -2/3$. In this phase, the Universe would undergo an inflation stage before the annihilation of the domain walls.
In this work, we assume a radiation-dominated Universe with $f_w\lesssim 1$ and, therefore, do not give a further discussion on this scenario.

\section{Effective equation of state}\label{sec:EOS}

\begin{figure}[t!]
    \centering
    \includegraphics[width=0.8\textwidth]{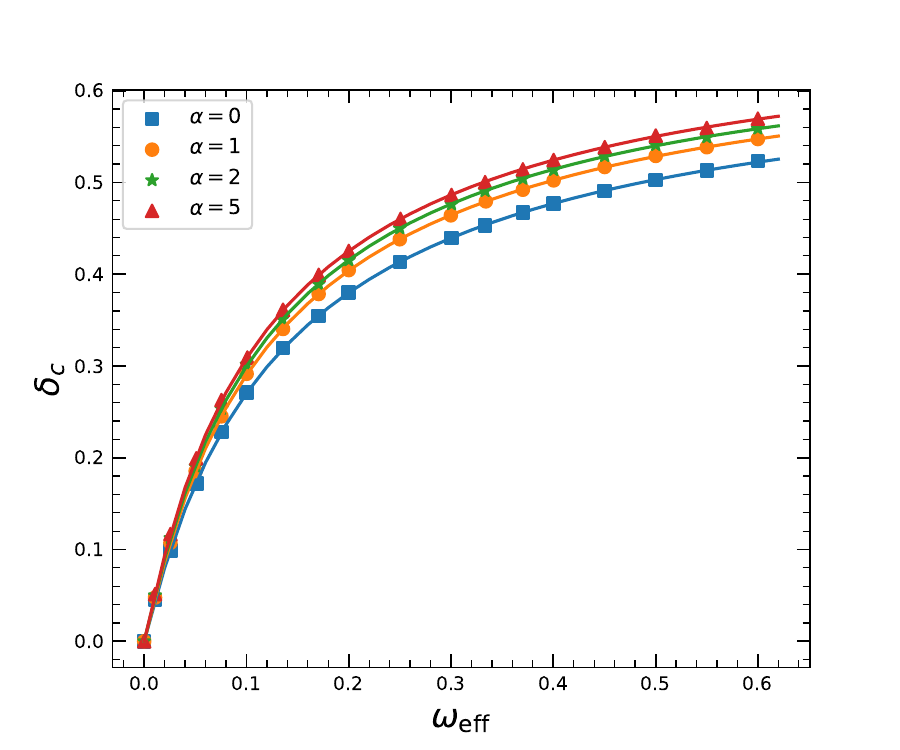}
    \caption{The solid curves represent the fit to the simulation results, denoted by the colored points, given in Ref.~\cite{Musco:2012au} with Eq.~\eqref{eq:fit}.}
     \label{fig:dcweff}
\end{figure}

The critical collapse scenario states that the horizon-size perturbations could stop expanding and collapse to form PBHs 
when they enter into the horizon with a density contrast exceeding a critical threshold $\delta_c$.
Therefore, the exact value of the critical density contrast $\delta_c$, which depends on the equation of state $\omega$, is nontrivial for the PBH formation. Effects of $\omega$ on PBH formation have been investigated in the context of first-order phase transition~\cite{Jedamzik:1999am,Jedamzik:2024wtq} and QCD phase transition~\cite{Byrnes:2018clq}.
The main force to resist the collapse of an overdense region comes from the gas pressure, which is reduced with the decrease of the equation of state during a phase transition. Therefore, the probability of PBH formation increases with the decrease of $\omega$. 
Eq.~\eqref{eq:dweos} shows that domain walls with $\bvw\simeq 0.4$ in the scaling regime have a negative equation 
of state, $\omega\simeq -2/3$. This is because domain walls have a negative surface pressure for $\bvw\lesssim 0.8$ 
(corresponding to $\bvw^2\gamma^2\simeq 2$), as shown by Eq.~\eqref{eq:Smunu}. We define an effective equation of state for 
the mixture gas of radiation and domain walls:
\begin{equation}\label{eq:weff}
    \omega_{\rm eff}=\frac{p_w+p_{\rm rad}}{\rho_w+\rho_{\rm rad}}=\frac{1-2f_w}{3(1+f_w)}.
\end{equation}
The cosmic equation of state can get smaller when the domain walls become a nontrivial component of the Universe.
The variation of the critical density contrast $\delta_c$ as a function of the equation of state has been
calculated in Ref.~\cite{Musco:2012au}. 
We find these simulation results can be reproduced by the following fit function 
\begin{equation}\label{eq:fit}
    \delta_c=\frac{\delta_0}{\left[ 1+(\omega_0/\omega_{\rm eff})^{s_1} \right]^{s_2}},
\end{equation}
where the fit parameters are provided in Tab.~\ref{tab:dcfit}. In Fig.~\ref{fig:dcweff}, we present the fit results of 
the simulations with different values of the perturbation profile parameter $\alpha$. We observe that all the simulation results are nicely reproduced with the fit formula. In this work, we adopt $\alpha=0$ as the default choice, which represents a Gaussian curvature profile.

\begin{table}[t!]\centering
	\large
	\begin{tabular}{|c|c|c|c|c|}  \hline
	Parameter  & $\delta_0$ & $\omega_0$ & $s_1$ & $s_2$ \\ \hline
    $\alpha=0$ & 0.65188 & 0.11922 & 0.94511 & 1.12954 \\ \hline
    $\alpha=1$ & 0.67491 & 0.10604 & 0.93996 & 1.16995 \\ \hline
    $\alpha=2$ & 0.67756 & 0.1109  & 0.97455 & 1.09519 \\ \hline
    $\alpha=5$ & 0.69254 & 0.10163 & 0.94824 & 1.15388 \\ \hline
	\end{tabular}\caption{Fit results of simulations in Ref.~\cite{Musco:2012au} with the fit formula~\eqref{eq:fit} 
    for different values of $\alpha$.}
    \label{tab:dcfit}
\end{table}

\begin{figure}[t!]
    \centering
    \includegraphics[width=0.49\textwidth]{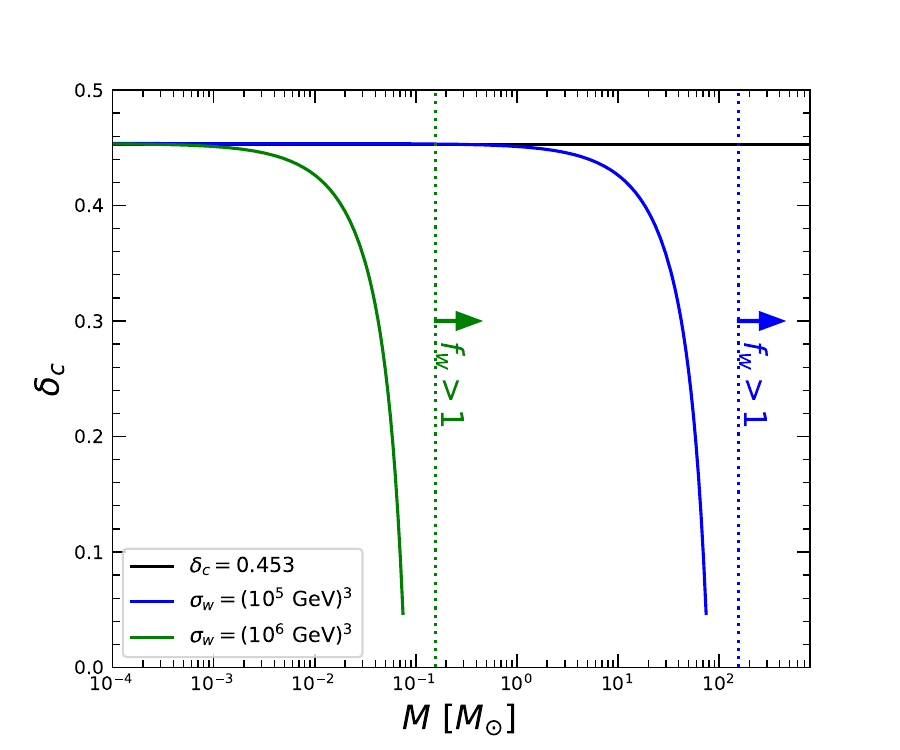}
    \includegraphics[width=0.49\textwidth]{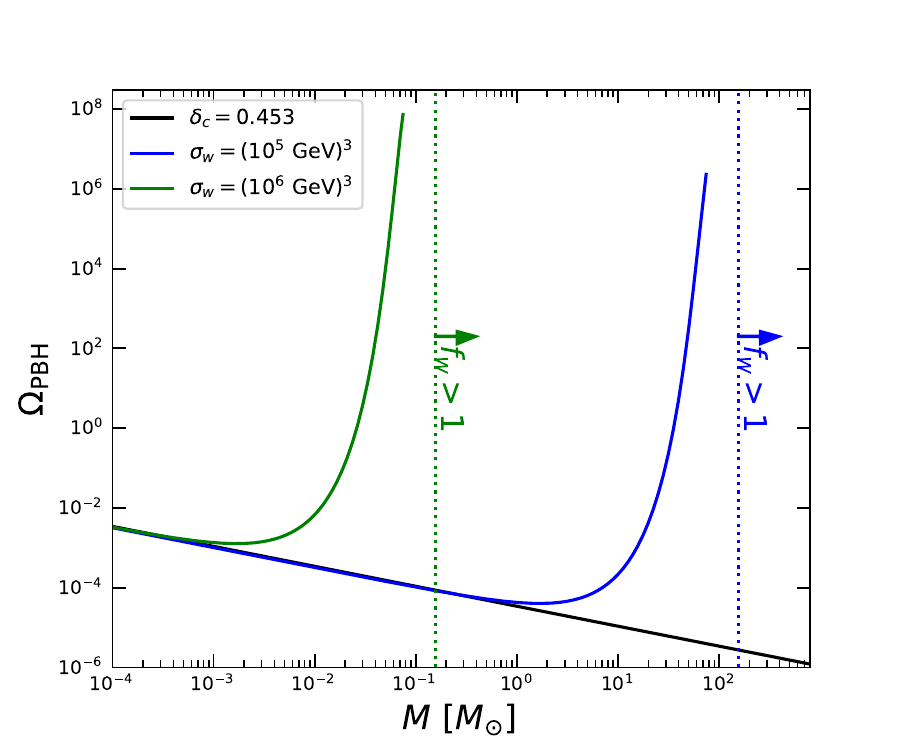}
    \caption{Left: The critical density contrast $\delta_c$ as a function of the PBH mass $M$. 
    Right: The PBH relic abundance $\Omega_{\rm PBH}$ as a function of the PBH mass $M$, assuming $N=1$ and a constant 
    variance $\sigma^2=0.004$. The black lines represent the results with a constant $\delta_c$ in the radiation phase.
    The effects on $\delta_c$ and $\Omega_{\rm PBH}$ from the domain walls are represented by the blue and green curves for 
    $\sigma_{w}^{1/3}=10^5$~GeV and $10^6$~GeV, respectively.}
     \label{fig:mdcomg}
\end{figure}

In the left plot of Fig.~\ref{fig:mdcomg}, we show the critical density contrast $\delta_c$ in Eq.~\eqref{eq:fit} as a function of the PBH mass.
We observe that $\delta_c$ sharply decreases when $M\sim 10^{1}~M_{\odot}$ (blue curve with $\sigma_w^{1/3}=10^5$~GeV) and $M\sim 10^{-2}~M_{\odot}$ (green curve with $\sigma_w^{1/3}=10^6$~GeV), corresponding to $f_w\sim 10^{-1}$.  This leads to an enormous improvement in the PBH formation probability, as shown in the right plot of Fig.~\ref{fig:mdcomg}, where we adopt a constant variance $\sigma^2=0.004$.
Therefore, once a fraction $f_w\sim 10^{-1}$ of the cosmic energy density is composed of the domain walls, the softened equation of state by the domain wall could also lead to a significant enhancement of the PBH relic abundance.

\section{Gravitational waves from domain wall annihilation}\label{sec:GWfromDW}

It is known that the domain wall's energy density scales as $\propto t^{-1}$ while the energy density of radiation scales 
as $t^{-2}$. As a result, the domain walls would dominate the evolution of the Universe if they existed for long enough.
A common solution to the problem is to introduce a bias potential which lifts the degenerate vacua by an energy 
$V_{\rm bias}=V_{\rm flase}-V_{\rm true}$ so that the domain walls can annihilate at some time $t_{\rm ann}$ before it over-close 
the Universe. The annihilation of the domain wall takes place when the bias energy becomes comparable to the surface energy of 
the domain wall, i.e., $V_{\rm bias}\simeq \rho_{w}$, which gives the annihilation temperature
\begin{equation}\label{eq:Tann}
    T_{\rm ann}\simeq 334.7~{\rm MeV}\left( \frac{1}{\mathcal{A}} \right)^{1/2}\left( \frac{10.75}{g_*(T_{\rm ann})} \right)^{1/4}
    \left( \frac{V_{\rm bias}^{1/4}}{0.1~{\rm GeV}} \right)^{2}\left( \frac{10^5~{\rm GeV}}{\sigma_w^{1/3}} \right)^{3/2}.
\end{equation}
This corresponds to a horizon mass 
\begin{equation}
    M_H(T_{\rm ann})\simeq \frac{\mathcal{A}\sigma_w}{GV_{\rm bias}}=1.34~M_{\odot}\left(\frac{\mathcal{A}}{1}\right)
    \left(\frac{0.1~{\rm GeV}}{V_{\rm bias}^{1/4}} \right)^4\left(\frac{\sigma_w^{1/3}}{10^5~{\rm GeV}} \right)^3.
\end{equation}

Simulations have indicated that the emissions of gravitational waves and radiation from domain walls in the scaling regime 
are negligible. Nearly all the domain wall energy is converted into gravitational waves when the domain walls are destroyed 
by the bias potential.  The peak frequency of gravitational waves is obtained at the annihilation time of the domain walls, given by $f_{\rm peak}(t_{\rm ann})\simeq H(t_{\rm ann})$, and is red-shifted by the cosmic expansion to the present as
\begin{equation}\label{eq:fpeak}
    f_{\rm peak}(t_0)\simeq 1.21 \times 10^{-8} \mathrm{~Hz}\left(\frac{g_{*}\left(T_{\mathrm{ann}}\right)}{10.75}\right)^{1/2}
    \left(\frac{g_{*s}\left(T_{\mathrm{ann}}\right)}{10.75}\right)^{-1/3} \left(\frac{T_{\mathrm{ann}}}{0.1~\rm{GeV}}\right).
\end{equation}
The peak gravitational wave amplitude produced at the annihilation time is determined as
\begin{equation}
    \Omega_{\mathrm{GW}}\left(f_{\mathrm{peak}}\left(t_{\mathrm{ann}}\right)\right)
    \simeq \frac{8 \pi \tilde{\epsilon}_{\mathrm{gw}} G^{2} \mathcal{A}^{2} \sigma_w^{2}}
    {3 H^2(t_{\rm ann})},   
\end{equation}
with $\tilde{\epsilon}_{\rm{gw}} \simeq 0.7 \pm 0.4$~\cite{Hiramatsu:2013qaa}.
The peak gravitational wave amplitude today is estimated as
\begin{equation}\label{eq:homg1}
    h^2\Omega_{\rm{GW}}^{\rm peak}\left(t_{0}\right)=2.4\times 10^{-10}\left(\frac{\tilde{\epsilon}_{\rm{gw}}}{0.7}\right)
    \left(\frac{\mathcal{A}}{1}\right)^2\left(\frac{g_{*,s}(T_{\rm ann})}{10.75}\right)^{-4/3}\left(\frac{\sigma_w^{1/3}}{10^5~{\rm GeV}}\right)^6
    \left(\frac{T_{\rm ann}}{0.1~{\rm GeV}}\right)^{-4}.
\end{equation}
Using Eq.~\eqref{eq:dwfrac} we have
\begin{equation}\label{eq:sgfw}
    \left(\frac{\sigma_w^{1/3}}{10^5~{\rm GeV}}\right)^3=39.6f_w(T)\left(\frac{1}{\mathcal{A}}\right)\left(\frac{g_{*}(T)}{10.75}\right)^{1/2} 
    \left(\frac{T}{0.1~{\rm GeV}}\right)^{2}.
\end{equation}
Combining Eqs.~\eqref{eq:homg1} and~\eqref{eq:sgfw}, we have 
\begin{equation}\label{eq:gwpeak}
    h^2\Omega_{\rm{GW}}^{\rm peak}\left(t_{0}\right)=3.8\times 10^{-9}\left(\frac{\tilde{\epsilon}_{\rm{gw}}}{0.7}\right)
    \left(\frac{g_{*,s}(T_{\rm ann})}{10.75}\right)^{-4/3}\left(\frac{g_{*}(T_{\rm ann})}{10.75}\right)
    \left(\frac{f_w(T_{\rm ann})}{0.1}\right)^2.
\end{equation}

The causality requires the gravitational wave spectrum scale as $\propto (f/f_{\rm peak})^3$ for the 
frequency below $f_{\rm peak}$~\cite{Caprini:2009fx}. On the other hand, for $f\gtrsim f_{\rm peak}$, 
numerical simulations~\cite{Hiramatsu:2013qaa} indicate a scaling behavior $\propto (f/f_{\rm peak})^{-1}$.
We adopt the following fit formula for the gravitational wave spectrum from the domain wall annihilation~\cite{Ferreira:2022zzo}
\begin{equation}\label{eq:gwspec}
    h^2\Omega_{\rm{GW}}=h^2\Omega_{\rm{GW}}^{\rm peak}\frac{(\gamma_c+c_1)^{c_2}}{(c_1x^{-\gamma_c/c_2}+\gamma_c x^{c_1/c_2})^{c_2}},
\end{equation}
where $x=f/f_{\rm peak}$, $\gamma_c=3$ from the requirement of causality, and $c_1,~c_2\simeq 1$ from the simulations.
However, $c_1$ and $c_2$ may depend on the physics model, as indicated by the simulations~\cite{Hiramatsu:2012sc}.

Recent simulations~\cite{Ferreira:2024eru,Notari:2025kqq} suggest that the gravitational wave emission from domain wall annihilation, driven by a bias potential, can be delayed by approximately an order of magnitude relative to the annihilation time itself. Consequently, the peak gravitational wave amplitude may be enhanced by one to two orders of magnitude. While such an intensified signal would lead to stronger constraints on the parameter space, we adopt the commonly used gravitational wave spectrum in Eq.~\eqref{eq:gwspec} for our subsequent data analysis to provide a conservative estimate.

\section{PTA data analysis}\label{sec:PTAdata}

The recent PTA observations from NANOGrav~\cite{NANOGrav:2023gor,NANOGrav:2023hvm}, CPTA~\cite{Xu:2023wog}, EPTA~\cite{EPTA:2023fyk}, 
and PPTA~\cite{Reardon:2023gzh} provide $\sim 3\sigma$ evidence for the existence of 
stochastic gravitational wave background (SGWB) at the nano-Hz frequency band. Various scenarios have been proposed to interpret 
the SGWB signal (see e.g.,~\cite{Madge:2023dxc,Bian:2023dnv,Wu:2023hsa,Ellis:2023oxs} for a summary).
The domain wall annihilation is a compelling interpretation for PTAs because the bias potential induced by the 
QCD phase transition can naturally lead to nano-Hz gravitational waves from the annihilation of domain walls~\cite{Chiang:2020aui,Lu:2023mcz}.
It has been noticed in Ref.~\cite{Lu:2023mcz} that the domain wall should annihilate before it dominates the Universe so that the gravitational wave signal could be loud enough for the PTA experiments.
As shown in this work, this would also lead to significant PBH productions due to the domain wall number density fluctuations and the equation 
of state reduction. Our PBH production mechanism is different from that by the collapse of closed domain walls, which depends heavily 
on the annihilation process of the walls~\cite{Ipser:1983db,Ferrer:2018uiu}.

For our purpose, we follow the standard Bayesian statistics to analyze the PTA data, including the IPTA-DR2 dataset~\cite{IPTA:datalink} 
and the recent NG15 dataset~\cite{NANOGrav:2023gor}. 
The timing-residual cross-power spectral density measured by the PTAs is given by 
\begin{equation}
    S_{a b}(f)=\Gamma_{a b} \frac{h_c^2(f)}{12 \pi^2 f^3},
\end{equation}
where the characteristic strain $h_c(f)$ is related to the gravitational wave background relic abundance via 
\begin{equation}
    \Omega _{\mathrm{GW}}(f)=\frac{2\pi^2 f^2 h_c^2(f)}{3 H_0^2}.
\end{equation}
We consider an isotropic and unpolarized GW background (GWB) overlap reduction function $\Gamma_{a b}$, 
which is described by the Hellings-Downs function.

We employ the package \texttt{PTArcade}~\cite{Mitridate:2023oar}, which wraps the code \texttt{enterprise}~\cite{Ellis:2020zenodo} and 
\texttt{ceffyl}~\cite{Lamb:2023jls}, to implement the GW signal model.
The Bayesian inference is adopted here to derive information on the parameters by fitting to the pulsars' timing residuals $\delta t$,
which is given by the sum of white noise and red noise, including the pulsar-intrinsic red noise and common 
red noise produced by a GWB (see appendix for more detail). 
Furthermore, the package \texttt{PTMCMCSampler}~\cite{Ellis:2017} is employed to perform the Markov chain Monte Carlo (MCMC) sampling.
In the fit to each dataset, we separately generate $2\times 10^6$ MC sample points. 
The parameter posterior distributions are derived using \texttt{GetDist}~\cite{Lewis:2019xzd}.
Following Refs.~\cite{Antoniadis:2022pcn,NANOGrav:2023gor} to avoid the pulsar-intrinsic excess noise at high frequencies, 
only the first 13 and 14 frequencies of the IPTA-DR2 and NG15 datasets are included in our analysis. 
The priors of the parameters for the signal model and noise are provided in table~\ref{tab:priors},  
together with the posteriors from the analysis presented in table~\ref{tab:posteriors} and Fig.~\ref{fig:conN1} in appendix~\ref{app:params}.
It has been previously found that the domain wall scenario can improve the fit to the NANOGrav data by a Bayesian 
factor $B\simeq 10$~\cite{NANOGrav:2023hvm}. The purpose of our fit in this work is not to repeat the model comparison but to 
estimate the parameter regions of our concerns.

\begin{figure}[t!]
    \centering
    \includegraphics[width=0.49\textwidth]{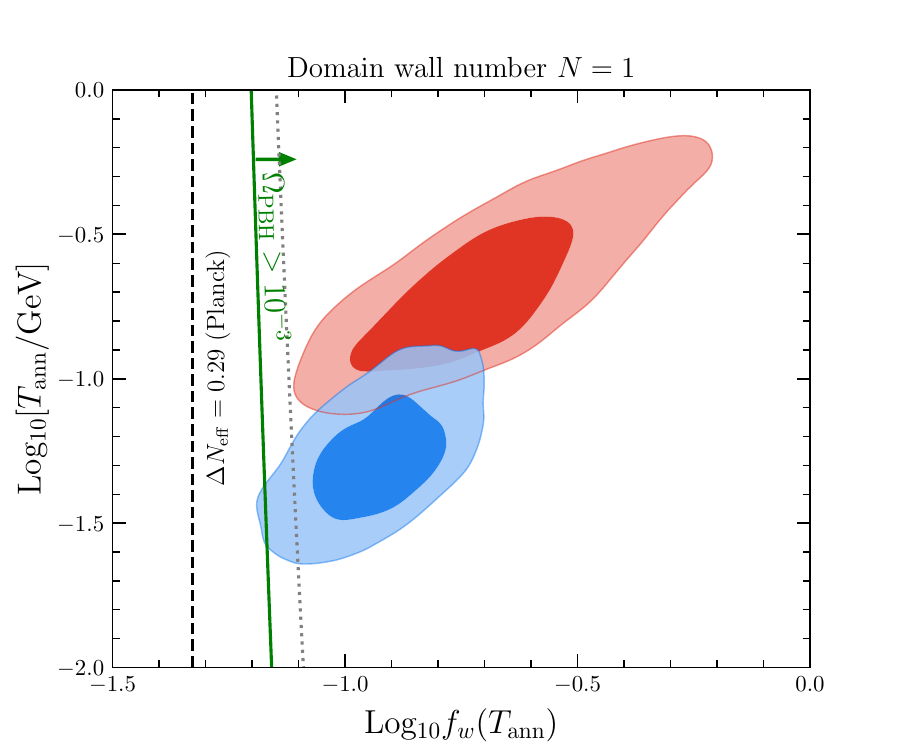}
    \includegraphics[width=0.49\textwidth]{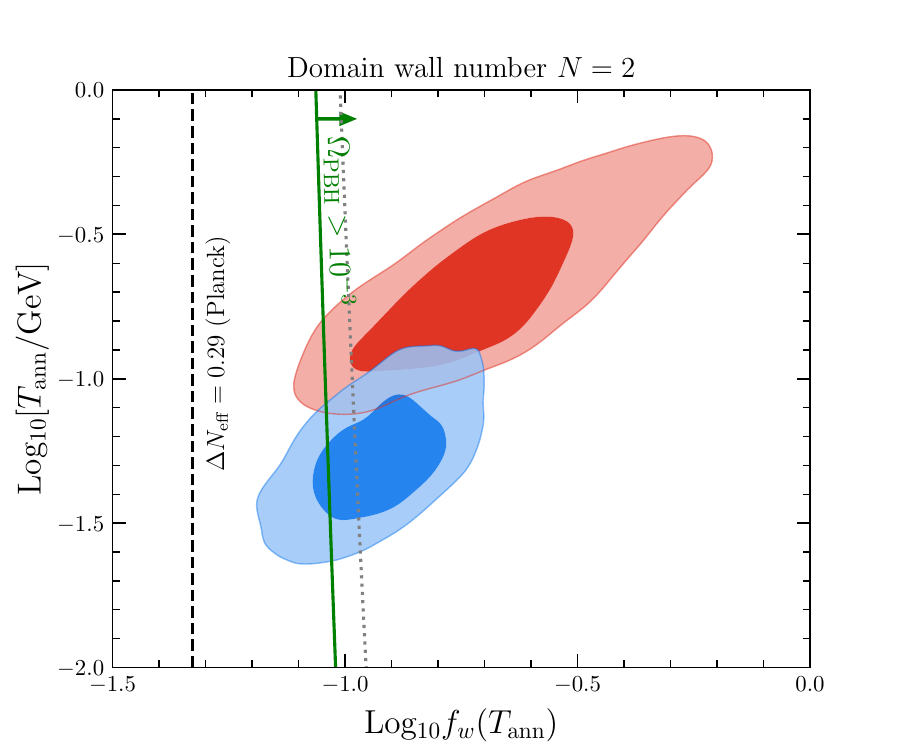}\\
    \includegraphics[width=0.49\textwidth]{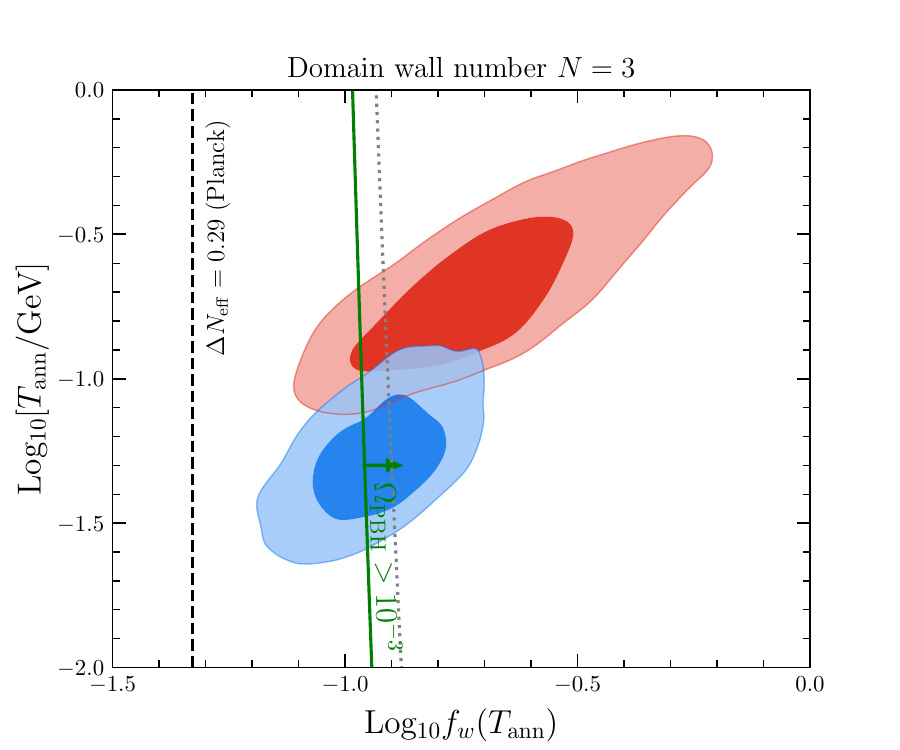}
    \includegraphics[width=0.49\textwidth]{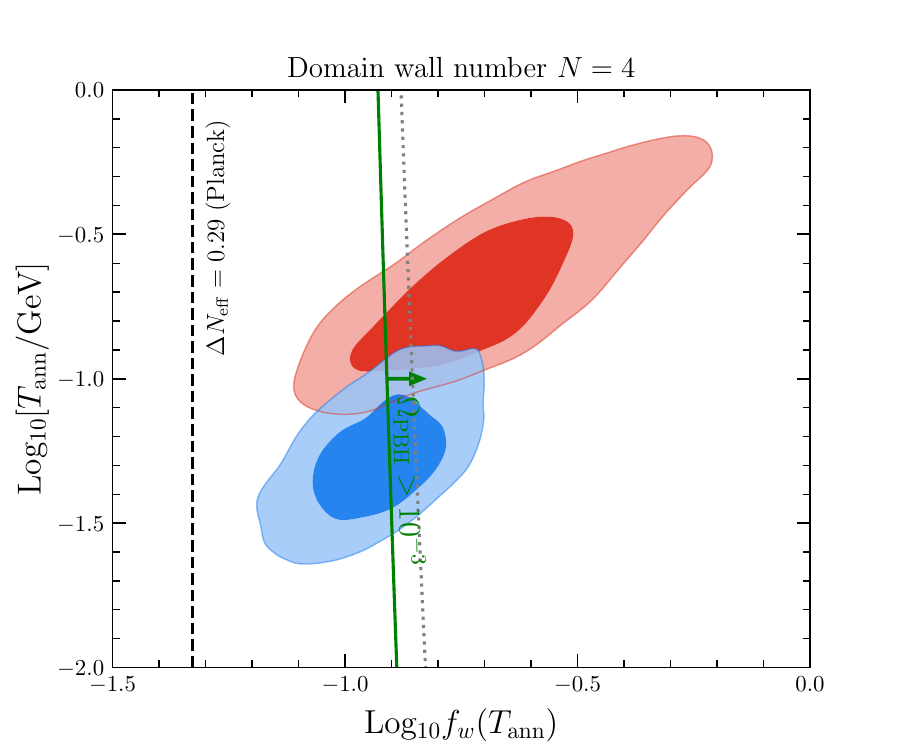} 
    \caption{The red (light red) and blue (light blue) regions represent the $1\sigma$ ($2\sigma$) regions of the 2D posterior distributions of the annihilation temperature $T_{\rm ann}$ and the domain wall density fraction $f_w(\rm ann)$ by fit to the NG15 and IPTA-DR2 datasets, respectively. The black dashed lines represent the $\Delta N_{\rm eff}$ constraints.  The solid green line and dotted gray line represent contours of $\Omega_{\rm PBH}=10^{-3}$ and $0.25$ from the domain wall fluctuations, respectively.}
    \label{fig:fTN1}
\end{figure}

\begin{figure}[t!]
    \centering
    \includegraphics[width=0.49\textwidth]{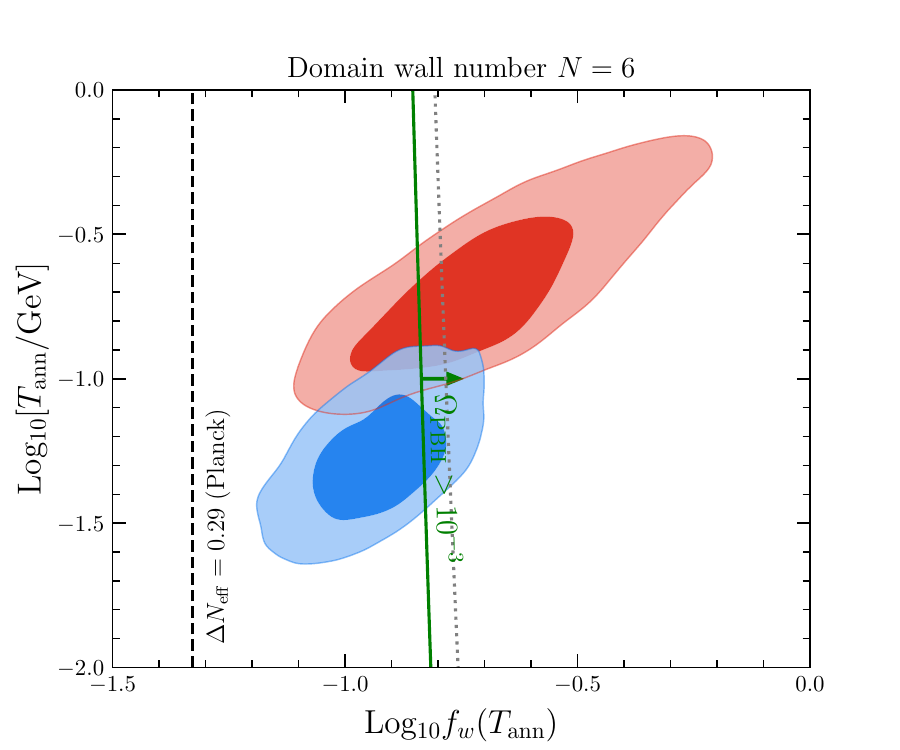}
    \includegraphics[width=0.49\textwidth]{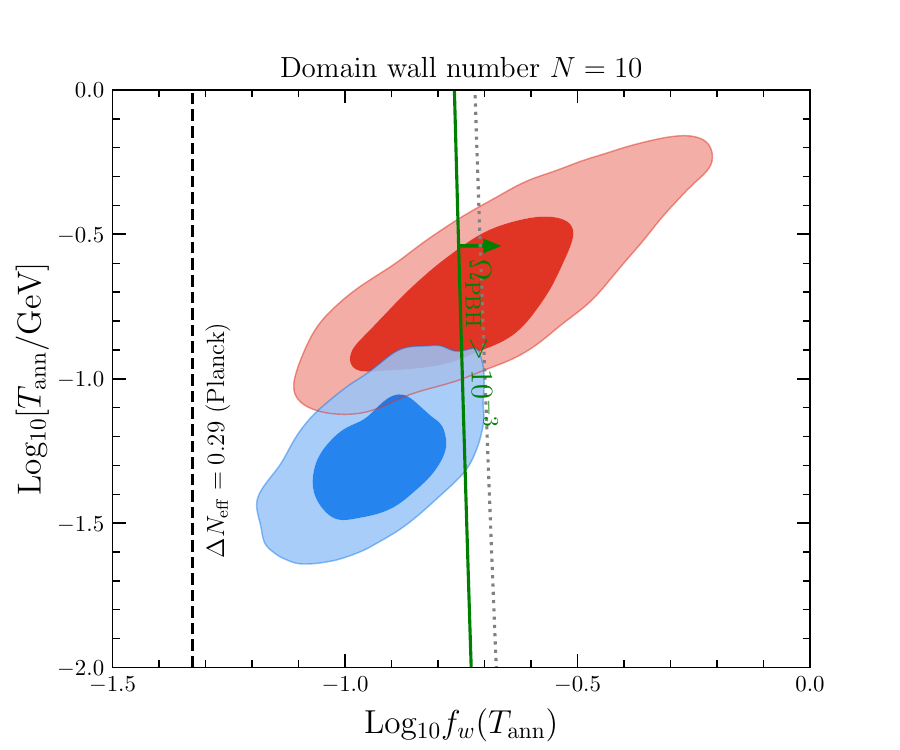}
    \caption{The same as Fig.~\ref{fig:fTN1}, with $N=6$ (left panel) and $N=10$ (right panel).}
    \label{fig:fTN2}
\end{figure}

In Fig.~\ref{fig:fTN1} and Fig.~\ref{fig:fTN2}, we show the interpretations of the domain wall scenario for the PTA observation, noting that our fit based on Eqs~\eqref{eq:fpeak} and~\eqref{eq:gwpeak} is $N$-independent.
The red and blue regions represent the 2D posterior distributions of $T_{\rm ann}$ and $f_w(T_{\rm ann})$ by fit to the NG15 and IPTA-DR2 datasets, respectively.
In addition, the fit to the observations indicates that a significant cosmic energy density consists of the domain walls.  The relativistic quanta released by the domain walls before their annihilation could leave an observable imprint on the standard cosmology. In order not to spoil the observations of BBN and CMB, 
we require the following conditions on the domain wall scenario, depending on the specific model:
\begin{itemize}
    \item Case {\it I}: Domain walls decay to standard model particles. In this case, the domain wall should annihilate before 
    the onset of BBN at $T_{\rm BBN}\simeq 2.7$~MeV.
    \item Case {\it II}: Domain walls mainly decay to dark radiation in a dark sector, which weakly contacts with or totally 
    decouples from the visible plasma. The dark radiation could alter the expansion rate of the Universe by contributing to 
    the effective number of neutrino species $\Delta N_{\rm eff}= \rho_{\rm DR}(t_{\rm rec})/\rho_{\nu}(t_{\rm rec})$, 
    where $t_{\rm rec}$ is the epoch of recombination and the energy density of a single neutrino species is given by 
    \begin{equation}
        \rho_{\nu}(t_{\rm rec})=2 \times \frac{7}{8} \times \frac{\pi^2}{30}\left(\frac{4}{11}\right)^{4/3}T_{\rm rec}^4,
    \end{equation}
    where the recombination temperature $T_{\rm rec}\simeq 0.3$~eV. With Eqs.~\eqref{eq:rho},~\eqref{eq:rhorad}, and~\eqref{eq:entcon},
    the effective number can be estimated as
    \begin{eqnarray}\label{eq:DNeff}
        \Delta N_{\rm eff}&=&\frac{\rho_{\rm DR}(t_{\rm ann})}{\rho_{\nu}(t_{\rm rec})}
        \left(\frac{g_{*,s}(T_{\rm rec})T_{\rm rec}^3}{g_{*,s}(T_{\rm ann})T_{\rm ann}^3}\right)^{4/3}\nonumber\\
        &\simeq & f_w(T_{\rm ann})\left(\frac{p_{\rm DR}}{0.16}\right)\left(\frac{g_{*}(T_{\rm ann})}{10.75}\right)
        \left(\frac{10.75}{g_{*,s}(T_{\rm ann})}\right)^{4/3},
    \end{eqnarray}
    where we have used $g_{*,s}(T_{\rm rec})\simeq 3.91$ and $\rho_{\rm DR}(T_{\rm ann})=p_{\rm DR}f_{w}\rho_{\rm rad}(T_{\rm ann})$, 
    with $0\leq p_{\rm DR}\leq 1$ being the energy fraction of domain walls that is transferred to dark radiation. We adopt $p_{\rm DR}=1$ for a conservative estimate.
\end{itemize}

For case {\it I}, Fig.~\ref{fig:fTN1} and Fig.~\ref{fig:fTN2} show that to interpret PTAs, the annihilation of domain walls should take place during the QCD scale before the onset of BBN, i.e., the annihilation temperature $T_{\rm ann}$ should fall in the range of $\sim 0.01-1$~GeV. 
Therefore, domain walls do not affect the BBN processes if the domain wall decays mainly into standard model particles.  For case {\it II}, $\Delta N_{\rm eff}$ observations can put useful constraints if the domain wall decays into dark radiation.  The current CMB from Planck and BAO observations provide an upper bound on the number of extra neutrino species $\Delta N_{\rm eff}\leq 0.29$ at 95\% 
confidence level (CL)~\cite{Planck:2018vyg} (the black dashed lines). 
The upcoming CMB experiments including Simons Observatory~\cite{SimonsObservatory:2018koc} and the CMB-S4 experiment~\cite{CMB-S4:2022ght}
could further tighten this upper bound to 0.11 and 0.06, respectively.
Indeed, numerical estimates have indicated that at most about $(10-20)\%$ of the wall energy $\rho_w$ would radiate as relativistic 
quanta during its evolution in the scaling regime. Therefore, constraints from extra neutrino species by Eq.~\eqref{eq:DNeff} with 
$p_{\rm DR}=1$ are less conservative.  We expect the observations from future Simons Observatory and CMB-S4 experiments can probe 
part of the parameter space in case {\it II}.

\begin{figure}[t!]
    \centering
    \includegraphics[width=0.9\textwidth]{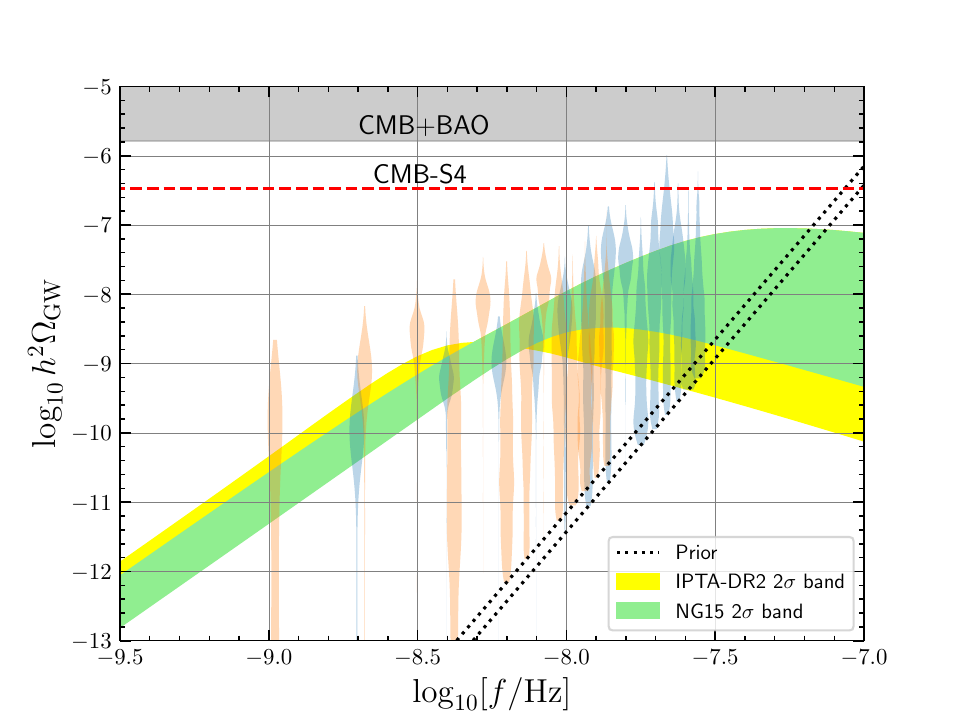}
    \caption{The gravitational wave spectra as a function of frequency $f$. 
    The light-orange and light-blue violins reproduce the posteriors of the free spectrum for IPTA-DR2~\cite{Antoniadis:2022pcn} 
    and NG15~\cite{NANOGrav:2023gor} with the lower prior given by the dotted lines, respectively. 
    The green region and yellow regions represent the 2$\sigma$ band for the interpretations of the NG15 and IPTA-DR2 datasets 
    within the domain wall scenarios, respectively.}
    \label{fig:DWfit}
\end{figure}

In Fig.~\ref{fig:DWfit}, we show the fit results for the IPTA-DR2 and NG15 gravitational spectra. We observe that the domain wall scenario can well explain the PTA observations while a large uncertainty remains at frequencies $f\gtrsim 10$~nHz.
Note that the domain walls annihilate to gravitational radiation before BBN can also make a contribution to the extra neutrino species and affect the expansion rate of the Universe. The bound on the relic abundance of gravitational radiation today is related to the effective number of neutrino species by~\cite{Caprini:2018mtu}
\begin{equation}
    h^2\Omega_{\mathrm{GW}}(t_0)\lesssim 5.6 \times 10^{-6} \Delta N_{\mathrm{eff}}.
\end{equation}
The gray region and red dashed line represent the upper bounds of the gravitational radiation relic abundance from Planck+BAO and future CMB-S4, respectively. The gravitational spectra at frequencies $\gtrsim 10^{-7.5}-10^{-7}$~Hz might have some overlaps with the CMB-S4 experiment, although the pulsar-intrinsic excess noise exists in this frequency band. Therefore, the upcoming CMB experiments will shed more light on the nano-Hz SGWB.

\begin{figure}[t!]
    \centering
    \includegraphics[width=0.49\textwidth]{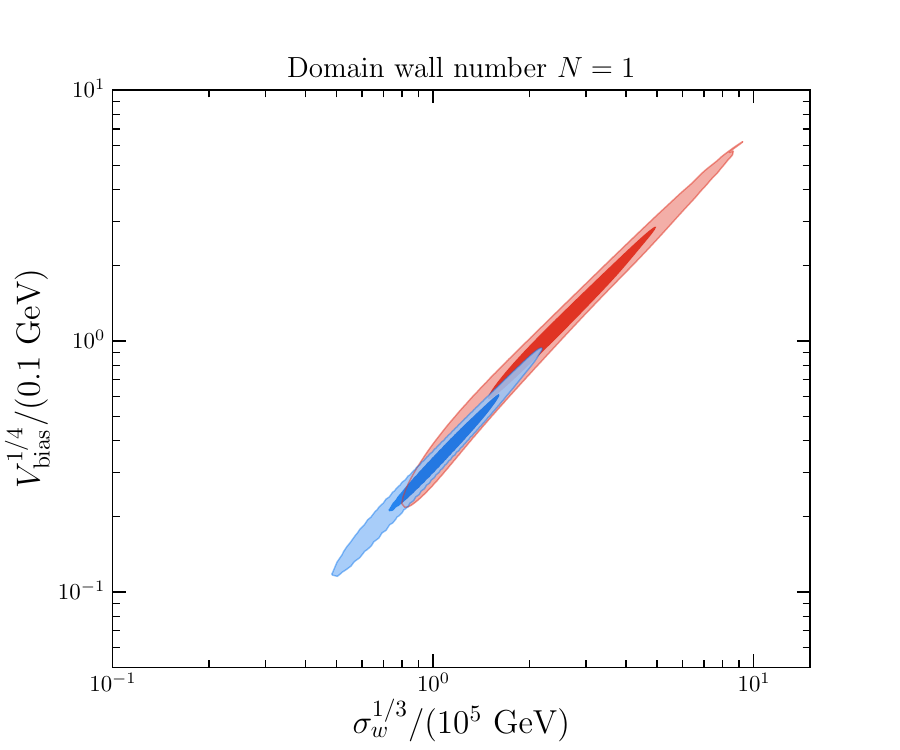}
    \includegraphics[width=0.49\textwidth]{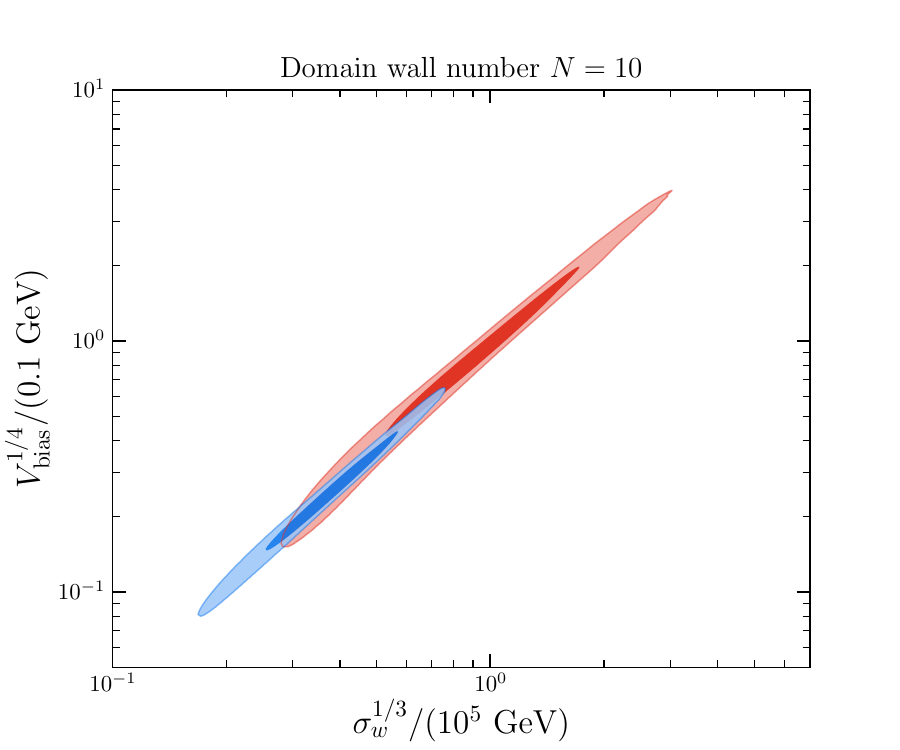}
    \caption{$1\sigma$ and $2\sigma$ contours of posterior distributions in the $\sigma_w^{1/3}-V_{\rm bias}^{1/4}$ plane, with $N=1$ and $N=10$ in the left and right panels, respectively.}
    \label{fig:sigV}
\end{figure}

Combining Eq.~\eqref{eq:Tann} and Eq.~\eqref{eq:sgfw}, we obtain the bias potential in terms of $f_w(T_{\rm ann})$ and $T_{\rm ann}$
\begin{eqnarray}\label{eq:Vbft}
    \frac{V_{\rm bias}^{1/4}}{0.1~{\rm GeV}}=1.37(f_w(T_{\rm ann}))^{1/4}\left( \frac{T_{\rm ann}}{0.1~{\rm GeV}} \right)
    \left( \frac{g_*(T_{\rm ann})}{10.75} \right)^{1/4}.
\end{eqnarray}
Together with Eq.~\eqref{eq:sgfw} we can therefore transfer the fit results in the $f_w-T_{\rm ann}$ space to the $\sigma_w-V_{\rm bias}$
plane. We show these results in Fig.~\ref{fig:sigV}, which indicates that the typical bias potential 
$V_{\rm bias}^{1/4}\sim \Lambda_{\rm QCD}\sim 0.1$~GeV and the surface tension $\sigma_w^{1/3}\sim 10^5$~GeV and they present a strong correlation.

\section{Constraints and implications from PBH formation}\label{sec:PBHconstraint}

Most of the domain wall energy $\rho_w$ is released at the annihilation temperature $T_{\rm ann}$, beyond which $\rho_w$ scales as 
$T^{\alpha}$ as indicated by the simulations. Let us parametrize the temperature-dependent domain wall energy density 
as follows~\cite{Kawasaki:2014sqa,Ferrer:2018uiu}
\begin{eqnarray}
    \rho_w(T)  = 
      \left\{\begin{matrix}
       \rho_w &~~{T\geq T_{\rm ann}},
       \\
       \rho_w\left ( \frac{T}{T_{\rm ann}} \right )^{\alpha} &~~{T<T_{\rm ann}},
       \end{matrix}\right.
\end{eqnarray}
where $\rho_w=\mathcal{A}\sigma_w/t$. A fit to the simulation of the domain wall evolution gives $\alpha\simeq 7$~\cite{Kawasaki:2014sqa}.
Then the evolution of the fraction of domain wall energy density is given by
\begin{eqnarray}\label{eq:fwevolu}
    f_w=
    \left\{\begin{matrix}
     1.81\times10^{-3}\frac{\mathcal{A}}{1}\frac{M_H(T)}{M_{\odot}}
    \left ( \frac{\sigma_w^{1/3}}{10^5~\rm GeV} \right )^3 & M_H\leq M_H(T_{\rm ann})\\
     1.81\times10^{-3}\frac{\mathcal{A}}{1}\frac{M_H(T)}{M_{\odot}}
    \left ( \frac{\sigma_w^{1/3}}{10^5~\rm GeV} \right )^3 
    \left( \frac{M_H(T_{\rm ann})\sqrt{g_*(T_{\rm ann})}}{M_H(T)\sqrt{g_*(T)}} \right)^{\alpha/2}  & M_H> M_H(T_{\rm ann}).
    \end{matrix}\right.
\end{eqnarray}
It is worth mentioning that, unlike the PBHs from the collapse of closed domain walls, our mechanism is not sensitive to the evolution of the network after the onset of domain wall annihilation, whose details remain to have a large uncertainty~\cite{Ferrer:2018uiu}.

\begin{figure}[t!]
    \centering
    \includegraphics[width=0.75\textwidth]{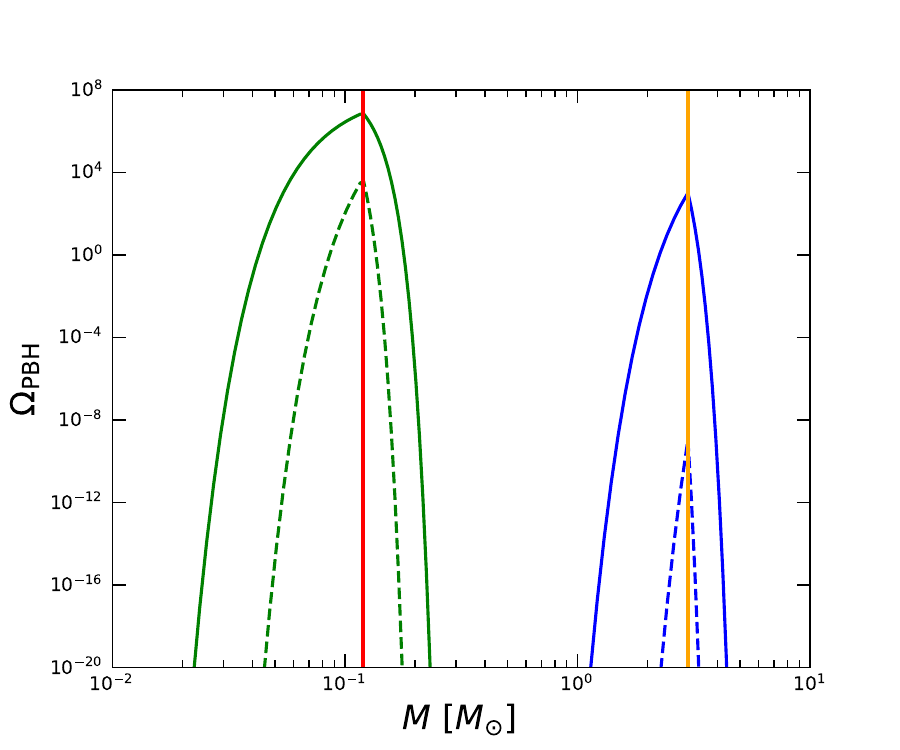}
    \caption{The PBH relic abundance produced from the number density fluctuations of domain walls with $(T_{\rm ann},~f_w(T_{\rm ann}))=(0.1~{\rm GeV},~0.1)$ (blue curves) and $(0.5~{\rm GeV},~0.2)$ (green curves).
    The critical density contrast $\delta_c=0.453$.
    The solid and dashed curves represent the results with $N=1$ and $N=4$, respectively. The red and orange vertical lines represent 
    the PBH mass at the temperature of domain wall annihilation for the two scenarios, respectively.}
     \label{fig:omegapbh}
\end{figure}

In Fig.~\ref{fig:omegapbh}, we show the PBH relic abundance $\Omega_{\rm PBH}(f_w,T)$ from the domain wall number density fluctuations, with the evolution of wall energy fraction according to Eq.~\eqref{eq:fwevolu}. The annihilation temperatures $T_{\rm ann}$ and the energy fractions $f_w$
at $T_{\rm ann}$ are taken to be $(0.1~{\rm GeV},~0.1)$ (blue curves) and $(0.5~{\rm GeV},~0.2)$ (green curves), which are two benchmarks for PTA observations.  The solid and dashed curves correspond to the results with the domain wall number $N=1$ and $N=4$, respectively.
The surface tension can be determined by Eq.~\eqref{eq:sgfw} at $T_{\rm ann}$ and the horizon mass at the annihilation temperature is $M_H(T_{\rm ann})=1/(2GH(T_{\rm ann}))$. For our choices, the PBH masses at annihilation are $M(0.1~{\rm GeV})\simeq 0.12 M_{\odot}$ (orange vertical 
line) and $M(0.5~{\rm GeV})\simeq 3.0 M_{\odot}$ (red vertical line). We observe that the PBH relic abundance $\Omega_{\rm PBH}$ 
has a narrow spike around $M(T_{\rm ann})$, below and above which $\Omega_{\rm PBH}$ sharply decreases with the PBH mass.
Therefore, PBHs are significantly produced at the moment close to the domain wall annihilation. This fact will be adopted to put useful constraints on the model parameter space of the domain wall scenario.

\begin{table}[t!]\centering
	\large
	\begin{tabular}{|c|c|c|c|c|}  \hline
	Parameter  & $N=1$ & $N=4$ & $N=6$ & $N=10$ \\ \hline
    $f_w(T_{\rm ann}=0.1~{\rm GeV})$  & 0.07  & 0.12  & 0.15  & 0.18 \\ \hline
	\end{tabular}\caption{Upper limits on the domain wall energy fraction $f_w$ from the PBH abundance, 
    with various values of the domain wall number $N$.  }
    \label{tab:fwupper}
\end{table}

From Eq.~\eqref{eq:omgpbh} and Fig.~\ref{fig:bt} we observe that to satisfy various experimental constraints on the PBH relic abundance for $M\sim M_{\odot}$, we should require $\beta(M)\lesssim 10^{-8}-10^{-7}$, corresponding to $x=\delta_c/(\sqrt{2}\sigma)\gtrsim 4$. The density fluctuations deep in the tails of the probability distribution exponentially 
suppress the production of PBHs via the critical density collapse mechanism (see Fig.~\ref{fig:bt} in 
appendix~\ref{app:btf}). The parameters that give the probability of a density contrast within $3\sigma$ are strongly 
constrained by the PBH abundance. To estimate these bounds, we note that the PBH abundance has a sharp spike at the annihilation temperature.
Therefore, we calculate the $f_w$-dependent critical density contrast $\delta_c$ with Eqs.~\eqref{eq:weff} and~\eqref{eq:fit} and determine $\Omega_{\rm PBH}(f_w(T_{\rm ann}),T_{\rm ann})$ with Eq.~\eqref{eq:omgpbh} at $T_{\rm ann}$.
In table~\ref{tab:fwupper}, we provide the upper limits on the domain wall energy fraction $f_w$ at the annihilation 
temperature $T_{\rm ann}=0.1$~GeV from the condition $\Omega_{\rm PBH}\leq 10^{-3}$ for various values of the domain wall number $N$.  In Fig.~\ref{fig:fTN1} and Fig.~\ref{fig:fTN2}, we show the contours of PBH relic abundance with $\Omega_{\rm PBH}=10^{-3}$ (green solid line) and 
$\Omega_{\rm PBH}=0.25=\Omega_{\rm DM}$ (gray dot line) in the $f_w(T_{\rm ann})-T_{\rm ann}$ plane.
We observe from Fig.~\ref{fig:fTN1} that the scenario for PTAs with the domain wall number $N=1$ (corresponding to models with a $Z_2$ symmetry) has been excluded by the PBH abundance from domain wall number density fluctuations. A marginal space out of $2\sigma$ CL remains available for the case with $N=2$. We therefore also exclude such models for explaining the nano-Hz SGWB.
For the cases of $N=3$ and $N=4$, most of the parameter space for NG15 and about half of the space region for IPTA-DR2 are 
excluded by the PBH relic abundance.
Finally, for models with $N=6$ and $N=10$, we show in Fig.~\ref{fig:fTN2} that most of the parameter space for IPTA-DR2 and about half of the space region for NG15 are not constrained by the PBH formation.
Models for $N\sim 10$ domain walls include the clockwork axion, in which several domain walls are produced 
following the spontaneous breakdown of a set of discrete shift symmetries.

\begin{figure}[t!]
    \centering
    \includegraphics[width=0.75\textwidth]{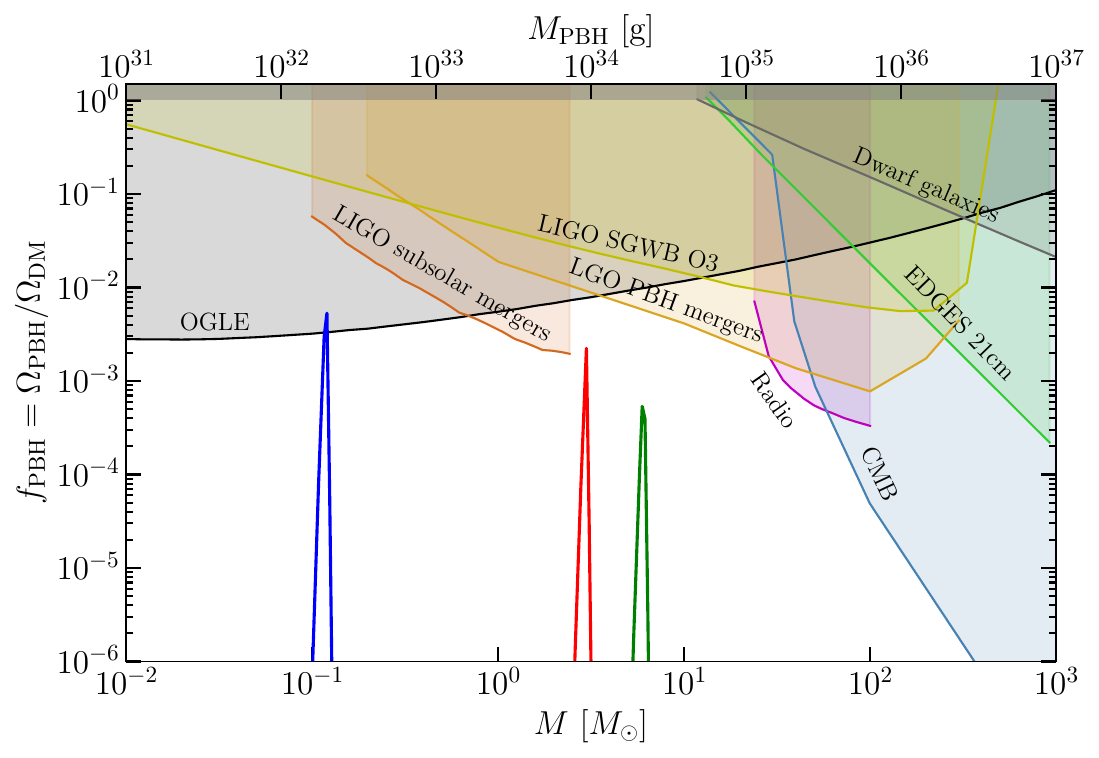}
    \caption{The PBH energy fraction for the monochromatic PBH mass spectrum. The colored regions represent the experimental constraints from OGLE~\cite{Mroz:2024wag}, LIGO~\cite{LIGOScientific:2019kan}, galactic radio observations~\cite{Manshanden:2018tze}, EDGES experiments~\cite{Hektor:2018qqw}, CMB spectrum measurements~\cite{Serpico:2020ehh}, and ultra-faint dwarfs dynamics~\cite{Brandt:2016aco}. The blue, red, and green peaks denote the results for the three benchmarks with $T_{\rm ann}=500$~MeV, 100~MeV, and 70~MeV, respectively. The energy fraction of the domain wall and the domain number are fixed as $f_w(T_{\rm ann})=0.125$ and $N=4$.}
     \label{fig:constraints}
\end{figure}
In Fig.~\ref{fig:constraints}, the blue, red, and green peaks represent the PBH energy fraction for the monochromatic PBH mass spectrum, corresponding to three benchmarks that can account for the nano-Hz stochastic gravitational wave background (SGWB) observations. We utilize the {\tt PBHbounds} package~\cite{Green:2020jor} to illustrate various experimental constraints on the PBH energy fraction, as indicated by the colored regions as follows:
\begin{itemize}
    \item The grey region denotes the most recent OGLE bounds from 20-year observations on PBH-induced microlensing events~\cite{Mroz:2024wag}.
    \item The yellow area represents constraints from LIGO experiments, including observations on SGWB~\cite{Nitz:2022ltl}, direct searches for gravitational waves from subsolar masses~\cite{Nitz:2022ltl}, and direct constraints on PBH-PBH mergers with LIGO~\cite{LIGOScientific:2019kan}.
    \item The purple region indicates constraints from galactic PBH accretion based on radio observations~\cite{Manshanden:2018tze}.
    \item The light-blue region represents constraints from distortions in the CMB spectrum~\cite{Serpico:2020ehh}.
    \item The green region corresponds to constraints from energy injection in the 21-cm signal by EDGES experiments~\cite{Hektor:2018qqw} due to accreting PBHs.
    \item The grey region also includes constraints from ultra-faint dwarf dynamics~\cite{Brandt:2016aco}.
\end{itemize}
We observe that the benchmarks denoted by the green and red curves, which align with NANOGrav and IPTA SGWB observations, are well below the PBH abundance constraints. In contrast, the blue curve is marginally constrained by the OGLE bounds.

\section{Conclusions and discussions}\label{sec:conclusion}

In this work, we propose that PBHs can be formed from the number density fluctuations of the domain walls in a late stage. 
The network, consisting of domain walls that are initially superhorizon-sized, gives rise to large Poisson fluctuations when these walls re-enter the Hubble horizon, which can subsequently lead to the formation of PBHs if the density contrast exceeds the critical value.
We show that the PBH formation outburst occurs when the fraction of domain wall energy density increases to 
$f_w\sim 0.1$. In addition, the domain wall with negative pressure could soften the equation of state and reduce the gas pressure that resists the collapse of density perturbations, increasing the probability of PBH formation.

With a fit to the NANOGrav and IPTA-DR2 datasets, we showed that the gravitational waves from domain wall annihilation could well explain the nano-Hz SGWB if the domain walls account for $\sim 10$\% of the total energy density at the annihilation temperature $T_{\rm ann}\sim 0.1$~GeV.
We also demonstrated that the abundance of PBHs can impose stringent constraints on particle models proposed to explain PTA observations of nano-Hz SGWB. For models with a domain wall number \( N \lesssim 2 \), the parameter space compatible with PTA observations is excluded due to the overproduction of PBHs. The models with \( N \gtrsim 3 \) also face strong constraints from the PBH overabundance, significantly limiting their viability.

We note that our constraints are applicable to models with a domain wall number \( N \lesssim 10 \) (the realizations of such models can be found in appendix~\ref{app:models}). For models with a domain wall number \( N \gg 10 \), the production of PBHs from domain wall number density fluctuations is strongly suppressed by Poisson statistics. Consequently, the constraints on such models become insignificant.

Finally, let us briefly review uncertainties that commonly exist in the critical collapse PBH formation mechanism~\cite{Carr:2016drx}. 
{\it (i) Non-sphericity:} 
We adopt the assumption of a homogeneous and spherically symmetric perturbation. 
Simulations regarding the evolution of domain walls indicate that such walls exhibit a tendency to form isotropic and closed structures in the course of their evolution~\cite{Lazanu:2015fua}. 
The non-sphericality of the overdense regions induced by domain wall fluctuations is expected to be of $\mathcal{O}(1)$ in the radiation-dominated Universe. 
Furthermore, simulations~\cite{Yoo:2020lmg,Escriva:2024aeo} concerning PBH formation demonstrate that in a radiation-dominated Universe, non-spherical effects are negligible with respect to PBH formation, as gas pressure governs the collapse process~\cite{Harada:2013epa}.
{\it (ii) Non-Gaussianity:}
We have assumed a Gaussian probability density distribution for the density contrast. The critical collapse mechanism for PBH 
formation relies on the exponential suppression in the upper tail of the probability distribution. Therefore, a non-Gaussian component in the distribution could exponentially change the abundance of PBH. 
Further investigations on the non-Gaussianity in the domain wall fluctuations are needed. 
Nevertheless, the fluctuations do not affect the CMB-scale observations if the domain walls annihilate long before the onset of BBN.

This paper is a joint work of Ref.~\cite{Lu:2024szr}, in which we seek to explore a common primordial origin for both PTA observations and LIGO-observed black hole merger events, taking into account black hole accretion.

\section{Acknowledgments}

BQL is supported in part by the National Natural Science Foundation of China under Grant No.~12405058 and No.~12575082 and by the Zhejiang Provincial Natural Science Foundation of China under Grant No.~LQ23A050002. CWC is supported in part by the National Science and Technology Council under Grant Nos.~NSTC-111-2112-M-002-018-MY3 and 114-2112-M-002-020-MY3. TL is supported in part by the National Key Research and Development Program of China Grant No. 2020YFC2201504, by the Project No.~12275333 supported by the National Natural Science Foundation of China, by the Scientific Instrument Developing Project of the Chinese Academy of Sciences, Grant No.~YJKYYQ20190049, and by the International Partnership Program of Chinese Academy of Sciences for Grand Challenges, Grant No.~112311KYSB20210012.

\appendix

\section{Domain wall energy tension in toy model}\label{app:toy}

Let's consider the following toy model for the real scalar $\phi$ as an example~\cite{Vilenkin:1984ib,Gelmini:1988sf,Saikawa:2017hiv}:
\begin{equation}\label{eq:lagraphi}
    \mathcal{L}=\frac{1}{2}\left(\partial_\mu \phi\right)^2-V(\phi),
\end{equation}
where
\begin{eqnarray}
    V(\phi)=\frac{1}{4} \lambda\left(\phi^2-v^2\right)^2.
\end{eqnarray}
The $Z_2$ symmetry of the model is spontaneously broken when $\phi$ obtains vacuum expectation values $\pm v$.
The manifold $\mathcal{M}$ of the degenerate vacua contains two points, $\langle\phi\rangle=+v$ and $\langle\phi\rangle=-v$, 
which are separated by the domain wall of the false vacuum with $\langle\phi\rangle=0$.

Using the Lagrangian density~\eqref{eq:lagraphi}, the description of the domain wall can be obtained by solving the 
equation of motion
\begin{equation}
    \square \phi+\lambda \phi\left(\phi^2-v^2\right)=0,
\end{equation}
where the d'Alambertian $\square=\partial_\mu \partial^\mu=\partial_0^2-\nabla^2$.
Suppose the domain wall is infinite and lies in the $yz$ plane at $x=0$. 
The solution that satisfies the boundary conditions $\phi(x=+\infty)=+v$ and $\phi(x=-\infty)=-v$ is found to be
\begin{equation}\label{eq:DWsolution}
    \phi(z)=v \tanh (x/\delta),
\end{equation}
where $\delta=1/(\sqrt{\lambda/2}v)$ denotes the thickness of the wall.
The energy-momentum tensor for the domain wall solution~\eqref{eq:DWsolution} is given by 
\begin{equation}
    T_{\mu \nu}=\partial_\mu \phi \partial_\nu \phi-g_{\mu \nu} \mathcal{L}.
\end{equation}
With the wall configuration~\eqref{eq:DWsolution} and $\partial_\mu\phi(x)=0$ for $\mu\neq x$, we have
\begin{eqnarray}
    T_{00}=-T_{22}=-T_{33}=\frac{1}{2}\left( \partial_x\phi \right)^2+V(\phi)~~{\rm and}~~T_{11}=0.
\end{eqnarray}
Then we obtain
\begin{equation}\label{eq:tensor2}
    T_{\mu\nu}=\frac{\lambda}{2} v^4[{\rm sech}(x / \delta)]^{4} \operatorname{diag}(1,0,-1,-1) .
\end{equation}
We can define the energy-momentum surface density tensor of the walls as
\begin{equation}\label{eq:S1}
    S_{\mu\nu}=\int dx T_{\mu\nu}.
\end{equation}
Using Eq.~\eqref{eq:tensor2}, the surface energy density of the wall is given by 
\begin{equation}\label{eq:S2}
    \sigma_w=S_{00}=-S_{22}=-S_{33}=\frac{4}{3} \sqrt{\frac{\lambda}{2}} v^3,~~{\rm and}~~S_{11}=0.
\end{equation}

\section{Surface energy-momentum tensor}\label{app:surfaceenergy}
In the wall rest frame $K'$, the surface tensor is given by Eqs.~\eqref{eq:S1} and~\eqref{eq:S2}. 
In the observer's rest frame $K$, $K'$ moves along the $\mathbf{x}$ axis with velocity $\overline{v}_w$, 
then the surface density tensor $S_{\mu\nu}^x(\overline{v}_w)$ measured by the observer is obtained by performing the Lorentz 
transformation 
\begin{equation}
    \begin{aligned}
    S_{00}^{x}&=\gamma^2 (S_{00}^{\prime}+2\overline{v}_w S_{01}^{\prime}+{\overline{v}_w}^2 S_{11}^{\prime}),\\
    S_{11}^{x}&=\gamma^2 (S_{11}^{\prime}+2\overline{v}_w S_{01}^{\prime}+{\overline{v}_w}^2 S_{00}^{\prime}),\\
    S_{01}^{x}&=S_{10}^x=\gamma^2 (S_{01}^{\prime}(1+{\overline{v}_w}^2)+\overline{v}_w S_{00}^{\prime}+{\overline{v}_w} S_{11}^{\prime}),\\
    S_{22}^{x}&=S_{22}^{\prime},~~S_{33}^{x}=S_{33}^{\prime},
    \end{aligned}
\end{equation}
where $\gamma\equiv 1/\sqrt{1-\overline{v}_w^2}$ and $'$ denotes the value in the $K'$ frame. The wall plane may also move along the $\mathbf{y}$ or $\mathbf{z}$ axis. The surface energy-momentum tensors measured by the observer are
\begin{equation}
    S_{\mu \nu}^x=\left(\begin{array}{cccc}
    \gamma^2 \sigma_w & \gamma^2 \sigma_w \overline{v}_w & 0 & 0 \\
    \gamma^2 \sigma_w \overline{v}_w & \gamma^2 \sigma_w \overline{v}_w^2 & 0 & 0 \\
    0 & 0 & -\sigma_w & 0 \\
    0 & 0 & 0 & -\sigma_w
    \end{array}\right),
\end{equation} 
\begin{equation}
    S_{\mu \nu}^y=\left(\begin{array}{cccc}
        \gamma^2 \sigma_w & 0 & \gamma^2 \sigma_w \overline{v}_w & 0 \\
        0 & -\sigma_w & 0 & 0 \\
        \gamma^2 \sigma_w \overline{v}_w & 0 & \gamma^2 \sigma_w \overline{v}_w^2 & 0 \\
        0 & 0 & 0 & -\sigma_w
        \end{array}\right),
\end{equation}
\begin{equation}
    S_{\mu \nu}^z=\left(\begin{array}{cccc}
        \gamma^2 \sigma_w & 0 & 0 & \gamma^2 \sigma_w \overline{v}_w \\
        0 & -\sigma_w & 0 & 0 \\
        0 & 0 & -\sigma_w & 0 \\
        \gamma^2 \sigma_w \overline{v}_w & 0 & 0 & \gamma^2 \sigma_w \overline{v}_w^2
        \end{array}\right).
\end{equation}
In addition, in each axis, there are two directions available for the wall plane. The corresponding tensors can be obtained by the replacement $\overline{v}_w\to -\overline{v}_w$ in the above results. 
Averaged over the six directions, the off-diagonal terms cancel out, and the average surface energy tensor
\begin{equation}
    S_{\mu\nu}(\overline{v}_w)=\frac{\sigma_w}{3}{\rm diag}(3\gamma^2,\overline{v}_w^2\gamma^2-2,
    \overline{v}_w^2\gamma^2-2,\overline{v}_w^2\gamma^2-2).
\end{equation}

\section{Wall evolution model}\label{app:VOS}

We follow Ref.~\cite{Martins:2016book} to review the VOS model for describing the evolution of the wall.
Let us start from the world volume (Dirac) action~\cite{Martins:2016book}
\begin{equation}
    S=-\int \mathcal{L} d^3 \sigma=-\sigma_w \int \sqrt{\gamma} d^3 \sigma,
\end{equation}
where $\gamma=\frac{1}{3!} \varepsilon^{a b} \varepsilon^{c d} \gamma_{a c} \gamma_{b d}$, with 
$\gamma_{a b}=g_{\mu \nu} x_{, a}^\mu x_{, b}^\nu$ and $x_{, a}^\mu=\frac{\partial x^{\mu}}{\partial\sigma^a}$,
$x_{\mu}$ and $\sigma_a$ are respectively the spacetime coordinates and worldsheet (two-dimensional surface) coordinates, 
$g_{\mu\nu}$ and $\gamma_{ab}$ are respectively the 4D spacetime and 2D worldsheet metrics.
The equation of motion (EOM) in the flat FRW Universe is given by
\begin{equation}\label{eq:DWEOM}
    \frac{\dot{a}}{a} \delta_{0 \lambda} \sqrt{\gamma} \gamma^{a b} \gamma_{a b}-\partial_c\left(\sqrt{\gamma} \gamma^{a b} 
    g_{\mu \lambda} x_{, a}^\mu \delta_b^c\right)=0 .
\end{equation}
Introduce the averaged energy density and the root-mean-squared (RMS) velocity of the wall
\begin{equation}
    \bar{\rho}_w\equiv \frac{E}{V}=\frac{\sigma_w a^2}{V} \int \varepsilon d^2 \sigma, \quad 
    \bvw^2\simeq v_{\rm rms}^2=\frac{\int \dot{x}^2 \varepsilon d^2 \sigma}{\int \varepsilon d^2 \sigma},
\end{equation}
where $\varepsilon=\sqrt{\gamma}\gamma^{00}/a$ with $a$ being the cosmic scale factor.
By averaging the temporal and spatial components of the EOM~\eqref{eq:DWEOM}, we obtan~\cite{Martins:2016book}
\begin{equation}\label{eq:EOS1}
    \begin{aligned}
    \frac{d \brw}{d t}&=-H \brw\left(1+3 \bvw^2\right)=-3(1+w)H\brw, \\
    \frac{d \bvw}{d t}&=\left(1-\bvw^2\right)\left(\frac{k_w}{\bL}-3 H \bvw\right),
    \end{aligned}
\end{equation}
where $t$ is the physical time, $H=\dot{a}/a$ is the Hubble parameter, and the momentum parameter $k_w=0.66\pm 0.04$ 
from the simulations~\cite{Martins:2016ois}. 
It is assumed in Eq.~\eqref{eq:EOS1} that the curvature radii are identical to the correlation length $\bL$.
Taking the one-scale assumption $\brw=\sgw/\bL$ and adding the energy loss processes (damping effects) by hand to Eq.~\eqref{eq:EOS1}, 
one obtains the VOS model for the description of domain wall as~\cite{Martins:2016book}
\begin{equation}
    \begin{aligned}
    \frac{d \bL}{d t}&=H \bL+\frac{\bL}{\ell_d} \bvw^2+c_w \bvw, \\
    \frac{d \bvw}{d t}&=\left(1-\bvw^2\right)\left(\frac{k_w}{\bL}-\frac{\bvw}{\ell_d}\right),
    \end{aligned}
\end{equation}
where the  chopping parameter $c_w=0.81\pm 0.04$ from the simulations of domain wall evolution in the $Z_2$ model~\cite{Martins:2016ois}.
The damping length is defined as 
\begin{equation}
    \frac{1}{\ell_d}=3 H+\frac{1}{\ell_f} ,
\end{equation}
which includes the damping effects on the network from Hubble drag and particle friction, which is related to the particle pressure 
$\Delta P$ by~\cite{Blasi:2022ayo}
\begin{equation}
    \frac{1}{\ell_f}=\frac{\Delta P}{\bvw \sigma_w}.
\end{equation}

\section{Beta function}\label{app:btf}

In this section, we provide more information on the $\beta(x)$ function since it is critical to determine the PBH abundance.
The $\beta(x)$ function is given by 
\begin{equation}
    \beta(x) = \gamma_p {\rm erfc}(x),
\end{equation}
where the complementary error function is given by 
\begin{equation}
    {\rm erfc}(x)=2\int_{x}^{\infty}\frac{1}{\sqrt{\pi}}\exp(-u^2)du.
\end{equation}
The complementary error function represents the area under the two tails of a zero-mean Gaussian probability distribution 
with variance $\sigma^2=1/2$. The error function is related to the complementary error function via  
\begin{eqnarray}
    {\rm erf}(x)=1-{\rm erfc}(x).
\end{eqnarray}

\begin{figure}[t!]
    \centering
    \includegraphics[width=0.49\textwidth]{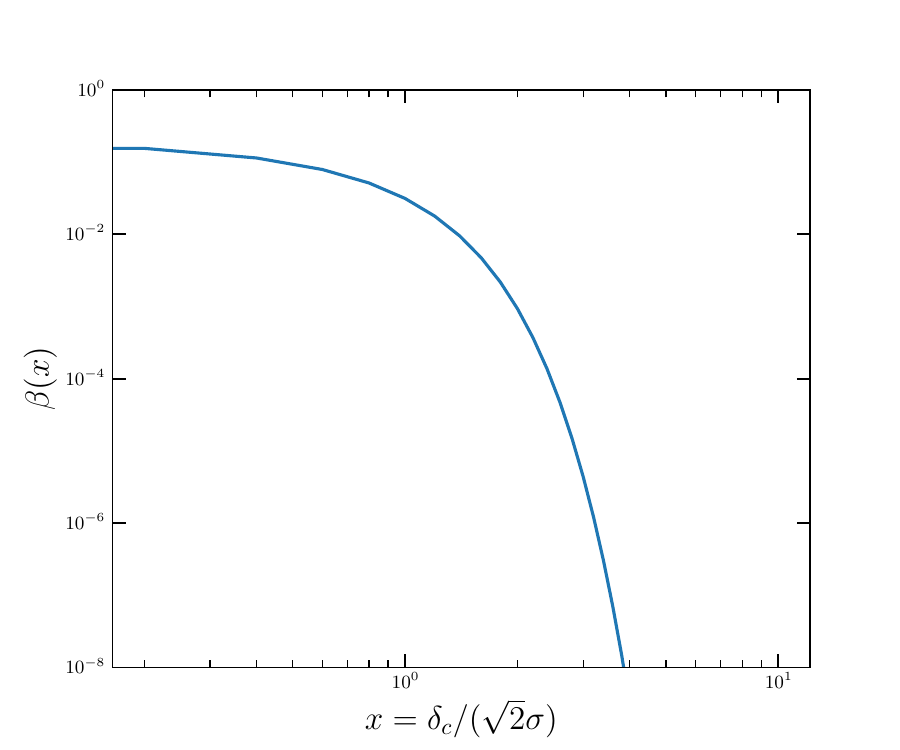}
    \includegraphics[width=0.49\textwidth]{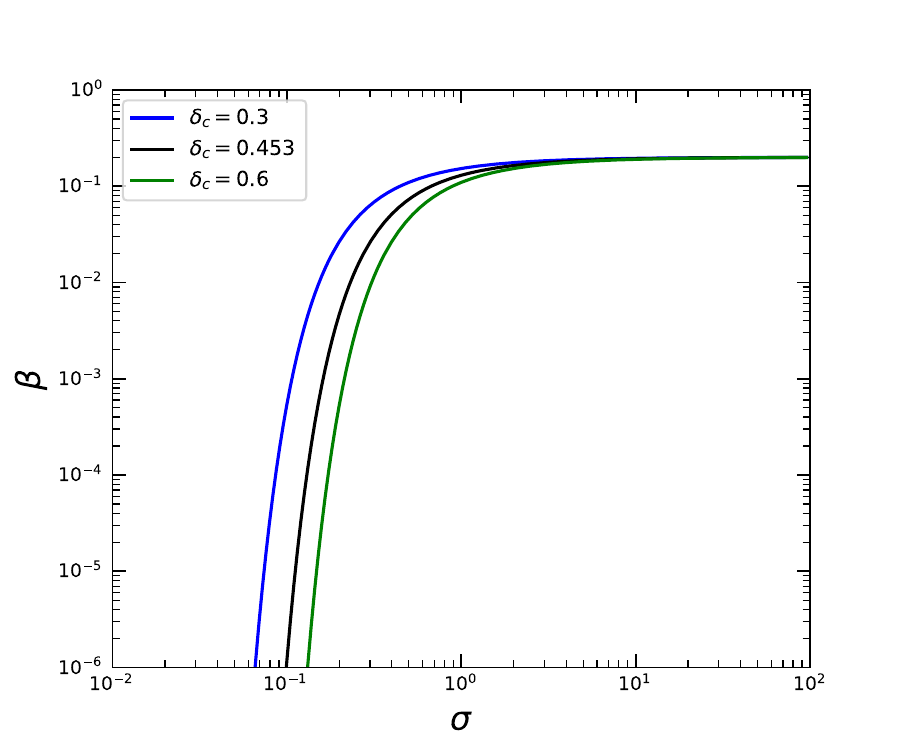}
    \caption{Left: The $\beta(x)$ function as a function of $x$.
    Right: The value of $\beta$ as a function of the variance $\sigma$.}
    \label{fig:bt}
\end{figure}

We plot $\beta(x)$ as a function of $x=\delta_c/(\sqrt{2}\sigma)$ in Fig.~\ref{fig:bt}.
We observe that to obtain an acceptable PBH abundance, $x$ should fall in the range of $3-5$ to exponentially suppress the PBH production.
Therefore, the formation of PBHs via the critical density collapse mechanism arises from the tail of the probability density function.

\section{Strategy for PTA}\label{app:params}

In this section, we provide further details of the PTA data analysis. We follow the analysis procedure encoded in {\tt PTArcade},
which wraps {\tt enterprise} and {\tt Ceffyl} to allow us to perform a standard search strategy for the PTA datasets (see Ref.~\cite{Mitridate:2023oar} for more details).
The searches for PTA stochastic signals adopt the pulsars' timing residuals, which are modeled as 
\begin{equation}\label{eq:dt}
    \overrightarrow{\delta t}=\vec{n}+\boldsymbol{F} \vec{a}+\boldsymbol{M} \vec{\epsilon} ,
\end{equation}
where the terms on the right-hand side of Eq.~\eqref{eq:dt} are the white noise, red noise, and small errors in the fit to
the timing-ephemeris parameter~\cite{NANOGrav:2020tig}, respectively.

Three types of white noise per backend/receiver system are included in the analysis, i.e., EFAC $E_k$, EQUAD $Q_k$[s], and ECORR $J_k$[s].
The pulsar-intrinsic red noise and GW background signals are included in the second term on the right-hand side of Eq.~\eqref{eq:dt}, 
which describes time-correlated stochastic processes. We adopt a power-law red noise with two parameters per pulsar for each dataset:
the amplitude $A_{\rm red}$ and the spectral index $\gamma_{\rm red}$. For the GW signal, we adopt the domain wall annihilation scenario
with four parameters in both datasets: the domain wall energy fraction $f_w(T_{\rm ann})$ at the annihilation temperature $T_{\rm ann}$,
the domain wall annihilation temperature $T_{\rm ann}$ in units of GeV, the spectral index above the peak frequency $c_1$, and the 
parameter that represents the width of the spectral index $c_2$. The spectral $\gamma_c$ is fixed at 3 to respect the causality. 
In table~\ref{tab:priors}, we provide the priors for the parameters of white and red 
noises, as well as the priors for parameters of the GW background signal from the annihilation of the domain walls.

\begin{figure}[t!]
    \centering
    \includegraphics[width=0.8\textwidth]{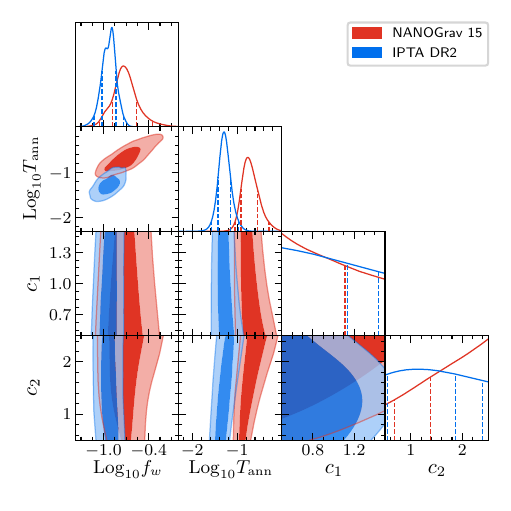}
    \caption{Corner plot of the parameter posterior distributions with $1\sigma$ and $2\sigma$ contours.}
     \label{fig:conN1}
\end{figure}

In table~\ref{tab:posteriors} and Fig.~\ref{fig:conN1}, we report the posterior distributions for the 
parameters of the domain wall annihilations. The posteriors are obtained by {\tt GetDist}. 
In comparison with the fit to the IPTA-DR2 data, the interpretation for NG15 data requires a larger domain wall energy density and a higher annihilation temperature. 
We also observe from Fig.~\ref{fig:conN1} that the posterior distributions of $f_w$ and $T_{\rm ann}$ at 
$1\sigma$ and $2\sigma$ are centralized around the mean values while the posterior of the spectral parameters $c_1$ and $c_2$ have 
broad distributions, which may be due to large uncertainties in the data at high frequencies.

\begin{table}[tbp]
    \centering
    \caption{\label{tab:priors} The noise and signal parameters and their prior ranges.}
    \begin{threeparttable}
    \renewcommand\arraystretch{1.0}
    \fontsize{9}{10}\selectfont
    \begin{tabular}{|llll|}
    \hline
    \hline
    Parameter   & Description   & Prior   &  Comments \\
    \hline
    {\bf White noise } & & & \\
    $E_k$   & EFAC per backend/receiver system   & Uniform $[0,~10]$   & Single-pulsar analysis only \\
    $Q_k$[s]   & EQUAD per backend/receiver system   & Log-uniform $[-8.5,~-5]$   & Single-pulsar analysis only \\
    $J_k$[s]   & ECORR per backend/receiver system   & Log-uniform $[-8.5,~-5]$   & Single-pulsar analysis only \\
    \hline
    {\bf Red noise } & & & \\
    $A_{\rm red}$   & Red noise power-law amplitude   & Log-uniform $[-20,~-11]$   & one parameter per pulsar \\
    $\gamma_{\rm red}$   & Red noise power-law spectral index   & Uniform $[0,~7]$   & one parameter per pulsar \\
    \hline
    {\bf DW } & & & \\
    $f_w(T_{\rm ann})$   & Domain wall energy fraction at $T_{\rm ann}$   & Log-Uniform $[-2, 0]$   & one parameter per PTA \\
    $T_{\rm ann}$   & Domain wall annihilation temperature   & Log-Uniform $[-2.3,~0]$   & one parameter per PTA \\
    $c_1$   & GW spectral index above peak frequency   & Uniform $[0.5,~1.5]$   & one parameter per PTA \\
    $c_2$   & Width of spectral index  & Uniform $[0.5,~2.5]$   & one parameter per PTA \\
    \hline
    \hline
    \end{tabular}
    \end{threeparttable}
  \end{table} 
\begin{table}[tbp]
    \centering
    \caption{\label{tab:posteriors} Bayes estimators for the fit to NG15 and IPTA-DR2 datasets.}
    \begin{threeparttable}
    \renewcommand\arraystretch{1.0}
    \begin{tabular}{|lcc|}
    \hline
    \hline
    & \makecell[c]{{\bf NG15}}    & \makecell[c]{{\bf IPTA-DR2}}      \\
    \centering 
    ${\rm Log}_{10}f_w(T_{\rm ann})$    & \makecell[c]{$-0.71\pm 0.17$} & \makecell[c]{$-0.93\pm 0.10$} \\
    ${\rm Log}_{10}[T_{\rm ann}/\rm GeV]$    & \makecell[c]{$-0.70\pm 0.19$ } & \makecell[c]{$-1.28\pm 0.15$}  \\
    $c_1$     & \makecell[c]{0.95$\pm$0.29} & \makecell[c]{0.97$\pm$0.29}  \\
    $c_2$     & \makecell[c]{1.66$\pm$0.55} & \makecell[c]{1.48$\pm$0.58}  \\
    \hline
    \hline
    \end{tabular} 
    \end{threeparttable}
  \end{table} 

\section{Models for $N$ domain walls}\label{app:models}

To realize a model with $\mathcal{N}$ domains (and the domain wall number $N\sim\mathcal{N}$),
we can consider the model with a $U(1)$ symmetry, which is softly broken by the term~\cite{Chiang:2019oms,Bai:2023cqj}
\begin{equation}
    V(\Phi)\supset \frac{1}{M^{\mathcal{N}-4}}(\Phi^{\mathcal{N}}+\Phi^{\dagger\mathcal{N}})
\end{equation}
to a $Z_{\mathcal{N}}$ symmetry,
where $\Phi$ is a complex scalar $\Phi$, $M$ has a mass dimension. For a renormalization model with $\mathcal{N}\leq 4$, the PBH production can be significant, and therefore, is strongly constrained.

One can consider $\mathcal{N}$ copies of complex scalars $\Phi_i$ (with $i=1,...,\mathcal{N}$). The global $U(1)^{\mathcal{N}}$ symmetry is spontaneously broken when each of $\Phi_i$ obtains a common vacuum expectation value. We can further explicitly break the $\mathcal{N}$ $U(1)$ symmetries to $\mathcal{N}-1$ discrete shift symmetries and one $U(1)$ symmetry by a term $\Phi_i^{\dagger}\Phi_{i+1}^{q}$, which then leads to the formation of $\mathcal{N}-1$ domains. Such symmetries have been naturally embedded in the framework of the clockwork axion models~\cite{Kaplan:2015fuy} and deconstruction of extradimensional grand unified theories (GUTs)~\cite{Arkani-Hamed:2001kyx,Li:2002xd}. 

Another realization of the $Z_{\mathcal{N}}$ symmetry is the axion-like models (see, for example, Ref.~\cite {ZambujalFerreira:2021cte}). 
Consider a hidden non-Abelian $SU(n)$ gauge sector consisting of a complex scalar $\Phi$ and a set of vector-like quarks. 
The potential of $\Phi$ possesses a global $U(1)$ symmetry, which is spontaneously broken by the Higgs mechanism.
If the vector-like quarks are charged under the $U(1)$ transformation, then the quantum anomaly of $U(1)$ with respect to the 
$SU(n)$ gauge group can break the $U(1)$ symmetry to its $Z_{\mathcal{N}}$ subgroup via the instanton effect during the confinement 
of the $SU(n)$ symmetry, which gives rise to an axion-like periodic potential 
\begin{equation}
    V(a)\simeq \Lambda_n^4\left(1-\cos\left(\frac{\mathcal{N}a}{v}\right)\right),
\end{equation}
where $a$ denotes the axion-like particle, $\Lambda_n\gg \Lambda_{\rm QCD}$ is the $SU(n)$ confinement scale, and $v$ is the vacuum expectation value of the scalar.

\bibliographystyle{JHEP}
\bibliography{reference}

@article{Lee:1973iz,
  author  = {Lee, T. D.},
  editor  = {Feinberg, G.},
  title   = {{A Theory of Spontaneous T Violation}},
  doi     = {10.1103/PhysRevD.8.1226},
  journal = {Phys. Rev. D},
  volume  = {8},
  pages   = {1226--1239},
  year    = {1973}
}

@article{Stauffer:1978kr,
  author  = {Stauffer, D.},
  title   = {{Scaling theory of percolation clusters}},
  doi     = {10.1016/0370-1573(79)90060-7},
  journal = {Phys. Rept.},
  volume  = {54},
  pages   = {1--74},
  year    = {1979}
}

@article{Zeldovich:1974uw,
  author       = {Zeldovich, Ya. B. and Kobzarev, I. Yu. and Okun, L. B.},
  title        = {{Cosmological Consequences of the Spontaneous Breakdown of Discrete Symmetry}},
  reportnumber = {SLAC-TRANS-0165, IPM-MOSCOW-15},
  journal      = {Zh. Eksp. Teor. Fiz.},
  volume       = {67},
  pages        = {3--11},
  year         = {1974}
}

@article{Mulki:2022tfw,
  author        = {Mulki, Fargiza A. M. and Wulandari, Hesti and Hidayat, Taufiq},
  title         = {{Noncanonical Domain Wall as a Unified Model of Dark Energy and Dark Matter: I. Cosmic Dynamics}},
  eprint        = {2211.07139},
  archiveprefix = {arXiv},
  primaryclass  = {astro-ph.CO},
  doi           = {10.1134/S0202289324010092},
  journal       = {Grav. Cosmol.},
  volume        = {30},
  number        = {1},
  pages         = {89--106},
  year          = {2024}
}

@article{Friedland:2002qs,
  author        = {Friedland, Alexander and Murayama, Hitoshi and Perelstein, Maxim},
  title         = {{Domain walls as dark energy}},
  eprint        = {astro-ph/0205520},
  archiveprefix = {arXiv},
  reportnumber  = {LBNL-50255},
  doi           = {10.1103/PhysRevD.67.043519},
  journal       = {Phys. Rev. D},
  volume        = {67},
  pages         = {043519},
  year          = {2003}
}

@article{Conversi:2004pi,
  author        = {Conversi, Luca and Melchiorri, Alessandro and Mersini-Houghton, Laura and Silk, Joseph},
  title         = {{Are domain walls ruled out?}},
  eprint        = {astro-ph/0402529},
  archiveprefix = {arXiv},
  reportnumber  = {SU-GP-02-11-1, SU-4252-771},
  doi           = {10.1016/j.astropartphys.2004.02.006},
  journal       = {Astropart. Phys.},
  volume        = {21},
  pages         = {443--449},
  year          = {2004}
}

@article{Vilenkin:1984ib,
  author       = {Vilenkin, Alexander},
  title        = {{Cosmic Strings and Domain Walls}},
  reportnumber = {PRINT-84-0840 (TUFTS)},
  doi          = {10.1016/0370-1573(85)90033-X},
  journal      = {Phys. Rept.},
  volume       = {121},
  pages        = {263--315},
  year         = {1985}
}

@article{Gelmini:1988sf,
  author       = {Gelmini, Graciela B. and Gleiser, Marcelo and Kolb, Edward W.},
  title        = {{Cosmology of Biased Discrete Symmetry Breaking}},
  reportnumber = {NSF-ITP-88-148, FERMILAB-PUB-88-151-A},
  doi          = {10.1103/PhysRevD.39.1558},
  journal      = {Phys. Rev. D},
  volume       = {39},
  pages        = {1558},
  year         = {1989}
}

@article{Ivanov:1994pa,
  author       = {Ivanov, P. and Naselsky, P. and Novikov, I.},
  title        = {{Inflation and primordial black holes as dark matter}},
  reportnumber = {NORDITA-94-12-A},
  doi          = {10.1103/PhysRevD.50.7173},
  journal      = {Phys. Rev. D},
  volume       = {50},
  pages        = {7173--7178},
  year         = {1994}
}

@article{Randall:1995dj,
  author        = {Randall, Lisa and Soljacic, Marin and Guth, Alan H.},
  title         = {{Supernatural inflation: Inflation from supersymmetry with no (very) small parameters}},
  eprint        = {hep-ph/9512439},
  archiveprefix = {arXiv},
  reportnumber  = {MIT-CTP-2501},
  doi           = {10.1016/0550-3213(96)00174-5},
  journal       = {Nucl. Phys. B},
  volume        = {472},
  pages         = {377--408},
  year          = {1996}
}

@article{Garcia-Bellido:1996mdl,
  author        = {Garcia-Bellido, Juan and Linde, Andrei D. and Wands, David},
  title         = {{Density perturbations and black hole formation in hybrid inflation}},
  eprint        = {astro-ph/9605094},
  archiveprefix = {arXiv},
  reportnumber  = {SU-ITP-96-20, SUSSEX-AST-96-5-1},
  doi           = {10.1103/PhysRevD.54.6040},
  journal       = {Phys. Rev. D},
  volume        = {54},
  pages         = {6040--6058},
  year          = {1996}
}

@article{Saikawa:2017hiv,
  author        = {Saikawa, Ken'ichi},
  title         = {{A review of gravitational waves from cosmic domain walls}},
  eprint        = {1703.02576},
  archiveprefix = {arXiv},
  primaryclass  = {hep-ph},
  reportnumber  = {DESY-17-036},
  doi           = {10.3390/universe3020040},
  journal       = {Universe},
  volume        = {3},
  number        = {2},
  pages         = {40},
  year          = {2017}
}

@article{Kibble:1976sj,
  author       = {Kibble, T. W. B.},
  title        = {{Topology of Cosmic Domains and Strings}},
  reportnumber = {ICTP/75/5},
  doi          = {10.1088/0305-4470/9/8/029},
  journal      = {J. Phys. A},
  volume       = {9},
  pages        = {1387--1398},
  year         = {1976}
}

@book{Martins:2016book,
  author         = {C. J. A. P. Martins},
  editor         = {E. Babaev et al.},
  publisher      = {Springer},
  title          = {Defect Evolution in Cosmology and Condensed Matter: Quantitative Analysis with the Velocity-Dependent One-Scale Model},
  year           = {2016}
}

@article{Martins:2016ois,
  author        = {Martins, C. J. A. P. and Rybak, I. Yu. and Avgoustidis, A. and Shellard, E. P. S.},
  title         = {{Extending the velocity-dependent one-scale model for domain walls}},
  eprint        = {1602.01322},
  archiveprefix = {arXiv},
  primaryclass  = {hep-ph},
  doi           = {10.1103/PhysRevD.93.043534},
  journal       = {Phys. Rev. D},
  volume        = {93},
  number        = {4},
  pages         = {043534},
  year          = {2016}
}

@article{Hawking:1971ei,
  author  = {Hawking, Stephen},
  title   = {{Gravitationally collapsed objects of very low mass}},
  doi     = {10.1093/mnras/152.1.75},
  journal = {Mon. Not. Roy. Astron. Soc.},
  volume  = {152},
  pages   = {75},
  year    = {1971}
}

@article{Carr:1975qj,
  author  = {Carr, Bernard J.},
  title   = {{The Primordial black hole mass spectrum}},
  doi     = {10.1086/153853},
  journal = {Astrophys. J.},
  volume  = {201},
  pages   = {1--19},
  year    = {1975}
}

@article{Yokoyama:1995ex,
  author        = {Yokoyama, Junichi},
  title         = {{Formation of MACHO primordial black holes in inflationary cosmology}},
  eprint        = {astro-ph/9509027},
  archiveprefix = {arXiv},
  reportnumber  = {YITP-U-95-26},
  journal       = {Astron. Astrophys.},
  volume        = {318},
  pages         = {673},
  year          = {1997}
}

@article{Harrison:1969fb,
  author  = {Harrison, Edward R.},
  title   = {{Fluctuations at the threshold of classical cosmology}},
  doi     = {10.1103/PhysRevD.1.2726},
  journal = {Phys. Rev. D},
  volume  = {1},
  pages   = {2726--2730},
  year    = {1970}
}

@article{Sasaki:2018dmp,
  author        = {Sasaki, Misao and Suyama, Teruaki and Tanaka, Takahiro and Yokoyama, Shuichiro},
  title         = {{Primordial black holes\textemdash{}perspectives in gravitational wave astronomy}},
  eprint        = {1801.05235},
  archiveprefix = {arXiv},
  primaryclass  = {astro-ph.CO},
  doi           = {10.1088/1361-6382/aaa7b4},
  journal       = {Class. Quant. Grav.},
  volume        = {35},
  number        = {6},
  pages         = {063001},
  year          = {2018}
}

@article{Carr:2020gox,
  author        = {Carr, Bernard and Kohri, Kazunori and Sendouda, Yuuiti and Yokoyama, Jun'ichi},
  title         = {{Constraints on primordial black holes}},
  eprint        = {2002.12778},
  archiveprefix = {arXiv},
  primaryclass  = {astro-ph.CO},
  reportnumber  = {RESCEU-03/20; KEK-Cosmo-249; KEK-TH-2199; IPMU20-0024},
  doi           = {10.1088/1361-6633/ac1e31},
  journal       = {Rept. Prog. Phys.},
  volume        = {84},
  number        = {11},
  pages         = {116902},
  year          = {2021}
}

@article{Choptuik:1992jv,
  author       = {Choptuik, Matthew W.},
  title        = {{Universality and scaling in gravitational collapse of a massless scalar field}},
  reportnumber = {FPRINT-92-33},
  doi          = {10.1103/PhysRevLett.70.9},
  journal      = {Phys. Rev. Lett.},
  volume       = {70},
  pages        = {9--12},
  year         = {1993}
}

@article{Evans:1994pj,
  author        = {Evans, Charles R. and Coleman, Jason S.},
  title         = {{Observation of critical phenomena and selfsimilarity in the gravitational collapse of radiation fluid}},
  eprint        = {gr-qc/9402041},
  archiveprefix = {arXiv},
  reportnumber  = {TAR-039-UNC},
  doi           = {10.1103/PhysRevLett.72.1782},
  journal       = {Phys. Rev. Lett.},
  volume        = {72},
  pages         = {1782--1785},
  year          = {1994}
}

@article{Koike:1995jm,
  author        = {Koike, Tatsuhiko and Hara, Takashi and Adachi, Satoshi},
  title         = {{Critical behavior in gravitational collapse of radiation fluid: A Renormalization group (linear perturbation) analysis}},
  eprint        = {gr-qc/9503007},
  archiveprefix = {arXiv},
  reportnumber  = {TIT-HEP-284, COSMO-53, TIT-HEP-284-COSMO-53},
  doi           = {10.1103/PhysRevLett.74.5170},
  journal       = {Phys. Rev. Lett.},
  volume        = {74},
  pages         = {5170--5173},
  year          = {1995}
}

@article{Byrnes:2018clq,
  author        = {Byrnes, Christian T. and Hindmarsh, Mark and Young, Sam and Hawkins, Michael R. S.},
  title         = {{Primordial black holes with an accurate QCD equation of state}},
  eprint        = {1801.06138},
  archiveprefix = {arXiv},
  primaryclass  = {astro-ph.CO},
  doi           = {10.1088/1475-7516/2018/08/041},
  journal       = {JCAP},
  volume        = {08},
  pages         = {041},
  year          = {2018}
}

@article{Carr:1974nx,
  author  = {Carr, Bernard J. and Hawking, S. W.},
  title   = {{Black holes in the early Universe}},
  doi     = {10.1093/mnras/168.2.399},
  journal = {Mon. Not. Roy. Astron. Soc.},
  volume  = {168},
  pages   = {399--415},
  year    = {1974}
}

@article{Niemeyer:1997mt,
  author        = {Niemeyer, Jens C. and Jedamzik, K.},
  title         = {{Near-critical gravitational collapse and the initial mass function of primordial black holes}},
  eprint        = {astro-ph/9709072},
  archiveprefix = {arXiv},
  doi           = {10.1103/PhysRevLett.80.5481},
  journal       = {Phys. Rev. Lett.},
  volume        = {80},
  pages         = {5481--5484},
  year          = {1998}
}

@article{Kopp:2010sh,
  author        = {Kopp, Michael and Hofmann, Stefan and Weller, Jochen},
  title         = {{Separate Universes Do Not Constrain Primordial Black Hole Formation}},
  eprint        = {1012.4369},
  archiveprefix = {arXiv},
  primaryclass  = {astro-ph.CO},
  doi           = {10.1103/PhysRevD.83.124025},
  journal       = {Phys. Rev. D},
  volume        = {83},
  pages         = {124025},
  year          = {2011}
}

@article{Musco:2012au,
  author        = {Musco, Ilia and Miller, John C.},
  title         = {{Primordial black hole formation in the early universe: critical behaviour and self-similarity}},
  eprint        = {1201.2379},
  archiveprefix = {arXiv},
  primaryclass  = {gr-qc},
  doi           = {10.1088/0264-9381/30/14/145009},
  journal       = {Class. Quant. Grav.},
  volume        = {30},
  pages         = {145009},
  year          = {2013}
}

@article{Green:2004wb,
  author        = {Green, Anne M. and Liddle, Andrew R. and Malik, Karim A. and Sasaki, Misao},
  title         = {{A New calculation of the mass fraction of primordial black holes}},
  eprint        = {astro-ph/0403181},
  archiveprefix = {arXiv},
  doi           = {10.1103/PhysRevD.70.041502},
  journal       = {Phys. Rev. D},
  volume        = {70},
  pages         = {041502},
  year          = {2004}
}

@article{Stewart:1997wg,
  author        = {Stewart, Ewan D.},
  title         = {{Flattening the inflaton's potential with quantum corrections. 2.}},
  eprint        = {hep-ph/9703232},
  archiveprefix = {arXiv},
  reportnumber  = {RESCEU-9-97},
  doi           = {10.1103/PhysRevD.56.2019},
  journal       = {Phys. Rev. D},
  volume        = {56},
  pages         = {2019--2023},
  year          = {1997}
}

@article{Drees:2011hb,
  author        = {Drees, Manuel and Erfani, Encieh},
  title         = {{Running-Mass Inflation Model and Primordial Black Holes}},
  eprint        = {1102.2340},
  archiveprefix = {arXiv},
  primaryclass  = {hep-ph},
  doi           = {10.1088/1475-7516/2011/04/005},
  journal       = {JCAP},
  volume        = {04},
  pages         = {005},
  year          = {2011}
}

@article{Alabidi:2012ex,
  author        = {Alabidi, Laila and Kohri, Kazunori and Sasaki, Misao and Sendouda, Yuuiti},
  title         = {{Observable Spectra of Induced Gravitational Waves from Inflation}},
  eprint        = {1203.4663},
  archiveprefix = {arXiv},
  primaryclass  = {astro-ph.CO},
  doi           = {10.1088/1475-7516/2012/09/017},
  journal       = {JCAP},
  volume        = {09},
  pages         = {017},
  year          = {2012}
}

@article{Silk:1986vc,
  author       = {Silk, Joseph and Turner, Michael S.},
  title        = {{Double Inflation}},
  reportnumber = {FERMILAB-PUB-86-062-A},
  doi          = {10.1103/PhysRevD.35.419},
  journal      = {Phys. Rev. D},
  volume       = {35},
  pages        = {419},
  year         = {1987}
}

@article{GarciaBellido:1996mdl,
  author        = {Garcia-Bellido, Juan and Linde, Andrei D. and Wands, David},
  title         = {Density perturbations and black hole formation in hybrid inflation},
  eprint        = {astro-ph/9605094},
  archiveprefix = {arXiv},
  reportnumber  = {SU-ITP-96-20, SUSSEX-AST-96-5-1},
  doi           = {10.1103/PhysRevD.54.6040},
  journal       = {Phys. Rev. D},
  volume        = {54},
  pages         = {6040--6058},
  year          = {1996}
}

@article{Clesse:2015wea,
  author        = {Clesse, S\'ebastien and Garc\'\i{}a-Bellido, Juan},
  title         = {{Massive Primordial Black Holes from Hybrid Inflation as Dark Matter and the seeds of Galaxies}},
  eprint        = {1501.07565},
  archiveprefix = {arXiv},
  primaryclass  = {astro-ph.CO},
  doi           = {10.1103/PhysRevD.92.023524},
  journal       = {Phys. Rev. D},
  volume        = {92},
  number        = {2},
  pages         = {023524},
  year          = {2015}
}

@article{Lazanu:2015fua,
  author        = {Lazanu, A. and Martins, C. J. A. P. and Shellard, E. P. S.},
  title         = {Contribution of domain wall networks to the CMB power spectrum},
  eprint        = {1505.03673},
  archiveprefix = {arXiv},
  primaryclass  = {astro-ph.CO},
  doi           = {10.1016/j.physletb.2015.06.034},
  journal       = {Phys. Lett. B},
  volume        = {747},
  pages         = {426--432},
  year          = {2015}
}

@article{Carr:2009jm,
  author        = {Carr, B. J. and Kohri, Kazunori and Sendouda, Yuuiti and Yokoyama, Jun'ichi},
  title         = {{New cosmological constraints on primordial black holes}},
  eprint        = {0912.5297},
  archiveprefix = {arXiv},
  primaryclass  = {astro-ph.CO},
  reportnumber  = {RESCEU-31-09, TU-852, YITP-09-112},
  doi           = {10.1103/PhysRevD.81.104019},
  journal       = {Phys. Rev. D},
  volume        = {81},
  pages         = {104019},
  year          = {2010}
}

@article{Jedamzik:1999am,
  author        = {Jedamzik, K. and Niemeyer, Jens C.},
  title         = {{Primordial black hole formation during first order phase transitions}},
  eprint        = {astro-ph/9901293},
  archiveprefix = {arXiv},
  doi           = {10.1103/PhysRevD.59.124014},
  journal       = {Phys. Rev. D},
  volume        = {59},
  pages         = {124014},
  year          = {1999}
}

@article{Jedamzik:2024wtq,
  author        = {Jedamzik, Karsten},
  title         = {{Primordial black hole formation during cosmic phase transitions}},
  eprint        = {2406.11417},
  archiveprefix = {arXiv},
  primaryclass  = {astro-ph.CO},
  month         = {6},
  year          = {2024}
}

@article{Hiramatsu:2013qaa,
  author        = {Hiramatsu, Takashi and Kawasaki, Masahiro and Saikawa, Ken'ichi},
  title         = {{On the estimation of gravitational wave spectrum from cosmic domain walls}},
  eprint        = {1309.5001},
  archiveprefix = {arXiv},
  primaryclass  = {astro-ph.CO},
  reportnumber  = {ICRR-REPORT-659-2013-8, IPMU13-0182, YITP-13-87},
  doi           = {10.1088/1475-7516/2014/02/031},
  journal       = {JCAP},
  volume        = {02},
  pages         = {031},
  year          = {2014}
}

@article{Kawasaki:2014sqa,
  author        = {Kawasaki, Masahiro and Saikawa, Ken'ichi and Sekiguchi, Toyokazu},
  title         = {{Axion dark matter from topological defects}},
  eprint        = {1412.0789},
  archiveprefix = {arXiv},
  primaryclass  = {hep-ph},
  reportnumber  = {ICRR-REPORT-696-2014-22, IPMU14-0348},
  doi           = {10.1103/PhysRevD.91.065014},
  journal       = {Phys. Rev. D},
  volume        = {91},
  number        = {6},
  pages         = {065014},
  year          = {2015}
}

@article{Caprini:2009fx,
  author        = {Caprini, Chiara and Durrer, Ruth and Konstandin, Thomas and Servant, Geraldine},
  title         = {{General Properties of the Gravitational Wave Spectrum from Phase Transitions}},
  eprint        = {0901.1661},
  archiveprefix = {arXiv},
  primaryclass  = {astro-ph.CO},
  doi           = {10.1103/PhysRevD.79.083519},
  journal       = {Phys. Rev. D},
  volume        = {79},
  pages         = {083519},
  year          = {2009}
}

@article{Ferreira:2022zzo,
  author        = {Ferreira, Ricardo Z. and Notari, Alessio and Pujolas, Oriol and Rompineve, Fabrizio},
  title         = {{Gravitational waves from domain walls in Pulsar Timing Array datasets}},
  eprint        = {2204.04228},
  archiveprefix = {arXiv},
  primaryclass  = {astro-ph.CO},
  reportnumber  = {CERN-TH-2022-214},
  doi           = {10.1088/1475-7516/2023/02/001},
  journal       = {JCAP},
  volume        = {02},
  pages         = {001},
  year          = {2023}
}

@article{Hiramatsu:2012sc,
  author        = {Hiramatsu, Takashi and Kawasaki, Masahiro and Saikawa, Ken'ichi and Sekiguchi, Toyokazu},
  title         = {{Axion cosmology with long-lived domain walls}},
  eprint        = {1207.3166},
  archiveprefix = {arXiv},
  primaryclass  = {hep-ph},
  reportnumber  = {ICRR-REPORT-620-2012-9, IPMU12-0140, YITP-12-58},
  doi           = {10.1088/1475-7516/2013/01/001},
  journal       = {JCAP},
  volume        = {01},
  pages         = {001},
  year          = {2013}
}

@misc{IPTA:datalink,
  author       = {S. Ransom and the IPTADR2 team},
  howpublished = {\url{https://gitlab.com/IPTA/DR2/-/tree/master/release/VersionB}}
}

@article{NANOGrav:2023gor,
  author        = {Agazie, Gabriella and others},
  collaboration = {NANOGrav},
  title         = {{The NANOGrav 15 yr Data Set: Evidence for a Gravitational-wave Background}},
  eprint        = {2306.16213},
  archiveprefix = {arXiv},
  primaryclass  = {astro-ph.HE},
  doi           = {10.3847/2041-8213/acdac6},
  journal       = {Astrophys. J. Lett.},
  volume        = {951},
  number        = {1},
  pages         = {L8},
  year          = {2023}
}

@article{Ellis:2020Zenodo,
  author  = {Justin A. Ellis and Michele Vallisneri and Stephen R. Taylor and Paul T. Baker},
  doi     = {10.5281/zenodo.4059815},
  journal = {Zenodo},
  year    = {2014}
}

@article{Lamb:2023jls,
  author        = {Lamb, William G. and Taylor, Stephen R. and van Haasteren, Rutger},
  title         = {{The Need For Speed: Rapid Refitting Techniques for Bayesian Spectral Characterization of the Gravitational Wave Background Using PTAs}},
  eprint        = {2303.15442},
  archiveprefix = {arXiv},
  primaryclass  = {astro-ph.HE},
  month         = {3},
  year          = {2023}
}

@article{Planck:2018vyg,
  author        = {Aghanim, N. and others},
  collaboration = {Planck},
  title         = {{Planck 2018 results. VI. Cosmological parameters}},
  eprint        = {1807.06209},
  archiveprefix = {arXiv},
  primaryclass  = {astro-ph.CO},
  doi           = {10.1051/0004-6361/201833910},
  journal       = {Astron. Astrophys.},
  volume        = {641},
  pages         = {A6},
  year          = {2020},
  note          = {[Erratum: Astron.Astrophys. 652, C4 (2021)]}
}

@article{Mitridate:2023oar,
  author        = {Mitridate, Andrea and Wright, David and von Eckardstein, Richard and Schr\"oder, Tobias and Nay, Jonathan and Olum, Ken and Schmitz, Kai and Trickle, Tanner},
  title         = {{PTArcade}},
  eprint        = {2306.16377},
  archiveprefix = {arXiv},
  primaryclass  = {hep-ph},
  month         = {6},
  year          = {2023}
}

@misc{Ellis:2017,
  author       = {J. Ellis and R. van Haasteren},
  howpublished = {\url{jellis18/ptmcmcsampler: Official release}},
  year         = {2017}
}

@article{Antoniadis:2022pcn,
  author        = {Antoniadis, J. and others},
  title         = {{The International Pulsar Timing Array second data release: Search for an isotropic gravitational wave background}},
  eprint        = {2201.03980},
  archiveprefix = {arXiv},
  primaryclass  = {astro-ph.HE},
  doi           = {10.1093/mnras/stab3418},
  journal       = {Mon. Not. Roy. Astron. Soc.},
  volume        = {510},
  number        = {4},
  pages         = {4873--4887},
  year          = {2022}
}

@article{Lewis:2019xzd,
  author        = {Lewis, Antony},
  title         = {{GetDist: a Python package for analysing Monte Carlo samples}},
  eprint        = {1910.13970},
  archiveprefix = {arXiv},
  primaryclass  = {astro-ph.IM},
  month         = {10},
  year          = {2019}
}

@article{Chiang:2020aui,
  author        = {Chiang, Cheng-Wei and Lu, Bo-Qiang},
  title         = {{Testing clockwork axion with gravitational waves}},
  eprint        = {2012.14071},
  archiveprefix = {arXiv},
  primaryclass  = {hep-ph},
  doi           = {10.1088/1475-7516/2021/05/049},
  journal       = {JCAP},
  volume        = {05},
  pages         = {049},
  year          = {2021}
}

@article{Lu:2023mcz,
  author        = {Lu, Bo-Qiang and Chiang, Cheng-Wei and Li, Tianjun},
  title         = {{Clockwork axion footprint on nanohertz stochastic gravitational wave background}},
  eprint        = {2307.00746},
  archiveprefix = {arXiv},
  primaryclass  = {hep-ph},
  doi           = {10.1103/PhysRevD.109.L101304},
  journal       = {Phys. Rev. D},
  volume        = {109},
  number        = {10},
  pages         = {L101304},
  year          = {2024}
}

@article{Ferrer:2018uiu,
  author        = {Ferrer, Francesc and Masso, Eduard and Panico, Giuliano and Pujolas, Oriol and Rompineve, Fabrizio},
  title         = {{Primordial Black Holes from the QCD axion}},
  eprint        = {1807.01707},
  archiveprefix = {arXiv},
  primaryclass  = {hep-ph},
  doi           = {10.1103/PhysRevLett.122.101301},
  journal       = {Phys. Rev. Lett.},
  volume        = {122},
  number        = {10},
  pages         = {101301},
  year          = {2019}
}

@article{SimonsObservatory:2018koc,
  author        = {Ade, Peter and others},
  collaboration = {Simons Observatory},
  title         = {{The Simons Observatory: Science goals and forecasts}},
  eprint        = {1808.07445},
  archiveprefix = {arXiv},
  primaryclass  = {astro-ph.CO},
  doi           = {10.1088/1475-7516/2019/02/056},
  journal       = {JCAP},
  volume        = {02},
  pages         = {056},
  year          = {2019}
}

@article{CMB-S4:2022ght,
  author        = {Abazajian, Kevork and others},
  collaboration = {CMB-S4},
  title         = {{Snowmass 2021 CMB-S4 White Paper}},
  eprint        = {2203.08024},
  archiveprefix = {arXiv},
  primaryclass  = {astro-ph.CO},
  month         = {3},
  year          = {2022}
}

@article{Caprini:2018mtu,
  author        = {Caprini, Chiara and Figueroa, Daniel G.},
  title         = {{Cosmological Backgrounds of Gravitational Waves}},
  eprint        = {1801.04268},
  archiveprefix = {arXiv},
  primaryclass  = {astro-ph.CO},
  doi           = {10.1088/1361-6382/aac608},
  journal       = {Class. Quant. Grav.},
  volume        = {35},
  number        = {16},
  pages         = {163001},
  year          = {2018}
}

@article{NANOGrav:2023hvm,
  author        = {Afzal, Adeela and others},
  collaboration = {NANOGrav},
  title         = {{The NANOGrav 15 yr Data Set: Search for Signals from New Physics}},
  eprint        = {2306.16219},
  archiveprefix = {arXiv},
  primaryclass  = {astro-ph.HE},
  doi           = {10.3847/2041-8213/acdc91},
  journal       = {Astrophys. J. Lett.},
  volume        = {951},
  number        = {1},
  pages         = {L11},
  year          = {2023}
}

@article{Xu:2023wog,
  author        = {Xu, Heng and others},
  title         = {{Searching for the Nano-Hertz Stochastic Gravitational Wave Background with the Chinese Pulsar Timing Array Data Release I}},
  eprint        = {2306.16216},
  archiveprefix = {arXiv},
  primaryclass  = {astro-ph.HE},
  doi           = {10.1088/1674-4527/acdfa5},
  journal       = {Res. Astron. Astrophys.},
  volume        = {23},
  number        = {7},
  pages         = {075024},
  year          = {2023}
}

@article{EPTA:2023fyk,
  author        = {Antoniadis, J. and others},
  collaboration = {EPTA, InPTA:},
  title         = {{The second data release from the European Pulsar Timing Array - III. Search for gravitational wave signals}},
  eprint        = {2306.16214},
  archiveprefix = {arXiv},
  primaryclass  = {astro-ph.HE},
  doi           = {10.1051/0004-6361/202346844},
  journal       = {Astron. Astrophys.},
  volume        = {678},
  pages         = {A50},
  year          = {2023}
}

@article{Reardon:2023gzh,
  author        = {Reardon, Daniel J. and others},
  title         = {{Search for an Isotropic Gravitational-wave Background with the Parkes Pulsar Timing Array}},
  eprint        = {2306.16215},
  archiveprefix = {arXiv},
  primaryclass  = {astro-ph.HE},
  doi           = {10.3847/2041-8213/acdd02},
  journal       = {Astrophys. J. Lett.},
  volume        = {951},
  number        = {1},
  pages         = {L6},
  year          = {2023}
}

@article{Ellis:2023oxs,
  author        = {Ellis, John and Fairbairn, Malcolm and Franciolini, Gabriele and H\"utsi, Gert and Iovino, Antonio and Lewicki, Marek and Raidal, Martti and Urrutia, Juan and Vaskonen, Ville and Veerm\"ae, Hardi},
  title         = {{What is the source of the PTA GW signal?}},
  eprint        = {2308.08546},
  archiveprefix = {arXiv},
  primaryclass  = {astro-ph.CO},
  reportnumber  = {KCL-PH-TH/2023-43, CERN-TH-2023-153, AION-REPORT/2023-08},
  doi           = {10.1103/PhysRevD.109.023522},
  journal       = {Phys. Rev. D},
  volume        = {109},
  number        = {2},
  pages         = {023522},
  year          = {2024}
}

@article{Ipser:1983db,
  author       = {Ipser, J. and Sikivie, P.},
  title        = {{The Gravitationally Repulsive Domain Wall}},
  reportnumber = {UFTP-83-18},
  doi          = {10.1103/PhysRevD.30.712},
  journal      = {Phys. Rev. D},
  volume       = {30},
  pages        = {712},
  year         = {1984}
}

@article{Li:2002xd,
  author        = {Li, Tian-jun and Liu, Tao},
  title         = {{Deconstruction of gauge symmetry breaking by discrete symmetry and G**N unification}},
  eprint        = {hep-th/0204128},
  archiveprefix = {arXiv},
  reportnumber  = {UPR-988-T},
  doi           = {10.1140/epjc/s2003-01211-8},
  journal       = {Eur. Phys. J. C},
  volume        = {28},
  pages         = {545--555},
  year          = {2003}
}

@article{Kaplan:2015fuy,
  author        = {Kaplan, David E. and Rattazzi, Riccardo},
  title         = {{Large field excursions and approximate discrete symmetries from a clockwork axion}},
  eprint        = {1511.01827},
  archiveprefix = {arXiv},
  primaryclass  = {hep-ph},
  doi           = {10.1103/PhysRevD.93.085007},
  journal       = {Phys. Rev. D},
  volume        = {93},
  number        = {8},
  pages         = {085007},
  year          = {2016}
}

@article{ZambujalFerreira:2021cte,
  author        = {Zambujal Ferreira, Ricardo and Notari, Alessio and Pujol\`as, Oriol and Rompineve, Fabrizio},
  title         = {{High Quality QCD Axion at Gravitational Wave Observatories}},
  eprint        = {2107.07542},
  archiveprefix = {arXiv},
  primaryclass  = {hep-ph},
  doi           = {10.1103/PhysRevLett.128.141101},
  journal       = {Phys. Rev. Lett.},
  volume        = {128},
  number        = {14},
  pages         = {141101},
  year          = {2022}
}

@article{Bai:2023cqj,
  author        = {Bai, Yang and Chen, Ting-Kuo and Korwar, Mrunal},
  title         = {{QCD-collapsed domain walls: QCD phase transition and gravitational wave spectroscopy}},
  eprint        = {2306.17160},
  archiveprefix = {arXiv},
  primaryclass  = {hep-ph},
  doi           = {10.1007/JHEP12(2023)194},
  journal       = {JHEP},
  volume        = {12},
  pages         = {194},
  year          = {2023}
}

@article{Carr:2016drx,
  author        = {Carr, Bernard and Kuhnel, Florian and Sandstad, Marit},
  title         = {{Primordial Black Holes as Dark Matter}},
  eprint        = {1607.06077},
  archiveprefix = {arXiv},
  primaryclass  = {astro-ph.CO},
  reportnumber  = {NORDITA-2016-83},
  doi           = {10.1103/PhysRevD.94.083504},
  journal       = {Phys. Rev. D},
  volume        = {94},
  number        = {8},
  pages         = {083504},
  year          = {2016}
}

@article{Zeldovich:1967lct,
  author  = {Zel'dovich, Ya. B. and Novikov, I. D.},
  title   = {{The Hypothesis of Cores Retarded during Expansion and the Hot Cosmological Model}},
  journal = {Sov. Astron.},
  volume  = {10},
  pages   = {602},
  year    = {1967}
}

@article{Liu:2019lul,
  author        = {Liu, Jing and Guo, Zong-Kuan and Cai, Rong-Gen},
  title         = {{Primordial Black Holes from Cosmic Domain Walls}},
  eprint        = {1908.02662},
  archiveprefix = {arXiv},
  primaryclass  = {astro-ph.CO},
  doi           = {10.1103/PhysRevD.101.023513},
  journal       = {Phys. Rev. D},
  volume        = {101},
  number        = {2},
  pages         = {023513},
  year          = {2020}
}

@article{Gouttenoire:2023gbn,
  author        = {Gouttenoire, Yann and Vitagliano, Edoardo},
  title         = {{Primordial black holes and wormholes from domain wall networks}},
  eprint        = {2311.07670},
  archiveprefix = {arXiv},
  primaryclass  = {hep-ph},
  doi           = {10.1103/PhysRevD.109.123507},
  journal       = {Phys. Rev. D},
  volume        = {109},
  number        = {12},
  pages         = {123507},
  year          = {2024}
}

@article{Ge:2023rrq,
  author        = {Ge, Shuailiang and Guo, Jinhui and Liu, Jia},
  title         = {{New mechanism for primordial black hole formation from the QCD axion}},
  eprint        = {2309.01739},
  archiveprefix = {arXiv},
  primaryclass  = {hep-ph},
  doi           = {10.1103/PhysRevD.109.123030},
  journal       = {Phys. Rev. D},
  volume        = {109},
  number        = {12},
  pages         = {123030},
  year          = {2024}
}

@article{Preskill:1992ck,
  author        = {Preskill, John and Vilenkin, Alexander},
  title         = {{Decay of metastable topological defects}},
  eprint        = {hep-ph/9209210},
  archiveprefix = {arXiv},
  reportnumber  = {HUTP-92-A018, CALT-68-1786},
  doi           = {10.1103/PhysRevD.47.2324},
  journal       = {Phys. Rev. D},
  volume        = {47},
  pages         = {2324--2342},
  year          = {1993}
}

@article{Dunsky:2021tih,
  author        = {Dunsky, David I. and Ghoshal, Anish and Murayama, Hitoshi and Sakakihara, Yuki and White, Graham},
  title         = {{GUTs, hybrid topological defects, and gravitational waves}},
  eprint        = {2111.08750},
  archiveprefix = {arXiv},
  primaryclass  = {hep-ph},
  doi           = {10.1103/PhysRevD.106.075030},
  journal       = {Phys. Rev. D},
  volume        = {106},
  number        = {7},
  pages         = {075030},
  year          = {2022}
}

@article{Preskill:1991kd,
  author       = {Preskill, John and Trivedi, Sandip P. and Wilczek, Frank and Wise, Mark B.},
  title        = {{Cosmology and broken discrete symmetry}},
  reportnumber = {IASSNS-HEP-91-11, CALT-68-1718},
  doi          = {10.1016/0550-3213(91)90241-O},
  journal      = {Nucl. Phys. B},
  volume       = {363},
  pages        = {207--220},
  year         = {1991}
}

@article{Gunion:2002zf,
  author        = {Gunion, John F. and Haber, Howard E.},
  title         = {{The CP conserving two Higgs doublet model: The Approach to the decoupling limit}},
  eprint        = {hep-ph/0207010},
  archiveprefix = {arXiv},
  reportnumber  = {SCIPP-02-10},
  doi           = {10.1103/PhysRevD.67.075019},
  journal       = {Phys. Rev. D},
  volume        = {67},
  pages         = {075019},
  year          = {2003}
}

@article{Sikivie:1982qv,
  author       = {Sikivie, P.},
  title        = {{Of Axions, Domain Walls and the Early Universe}},
  reportnumber = {UFTP-82-3},
  doi          = {10.1103/PhysRevLett.48.1156},
  journal      = {Phys. Rev. Lett.},
  volume       = {48},
  pages        = {1156--1159},
  year         = {1982}
}

@article{King:2023ayw,
  author        = {King, Stephen F. and Roshan, Rishav and Wang, Xin and White, Graham and Yamazaki, Masahito},
  title         = {{Quantum Gravity Effects on Dark Matter and Gravitational Waves}},
  eprint        = {2308.03724},
  archiveprefix = {arXiv},
  primaryclass  = {hep-ph},
  month         = {8},
  year          = {2023}
}

@article{Chiang:2019oms,
  author        = {Chiang, Cheng-Wei and Lu, Bo-Qiang},
  title         = {{First-order electroweak phase transition in a complex singlet model with $\mathbb{Z}_3$ symmetry}},
  eprint        = {1912.12634},
  archiveprefix = {arXiv},
  primaryclass  = {hep-ph},
  doi           = {10.1007/JHEP07(2020)082},
  journal       = {JHEP},
  volume        = {07},
  pages         = {082},
  year          = {2020}
}

@article{Dvali:1994wv,
  author        = {Dvali, G. R. and Tavartkiladze, Z. and Nanobashvili, J.},
  title         = {{Biased discrete symmetry and domain wall problem}},
  eprint        = {hep-ph/9411387},
  archiveprefix = {arXiv},
  reportnumber  = {IFUP-TH-60-94},
  doi           = {10.1016/0370-2693(95)00511-I},
  journal       = {Phys. Lett. B},
  volume        = {352},
  pages         = {214--219},
  year          = {1995}
}

@article{Abel:1995wk,
  author        = {Abel, S. A. and Sarkar, Subir and White, P. L.},
  title         = {{On the cosmological domain wall problem for the minimally extended supersymmetric standard model}},
  eprint        = {hep-ph/9506359},
  archiveprefix = {arXiv},
  reportnumber  = {OUTP-95-22-P, RAL-TR-95-019},
  doi           = {10.1016/0550-3213(95)00483-9},
  journal       = {Nucl. Phys. B},
  volume        = {454},
  pages         = {663--684},
  year          = {1995}
}

@article{Larsson:1996sp,
  author        = {Larsson, Sebastian E. and Sarkar, Subir and White, Peter L.},
  title         = {{Evading the cosmological domain wall problem}},
  eprint        = {hep-ph/9608319},
  archiveprefix = {arXiv},
  reportnumber  = {OUTP-96-11-P},
  doi           = {10.1103/PhysRevD.55.5129},
  journal       = {Phys. Rev. D},
  volume        = {55},
  pages         = {5129--5135},
  year          = {1997}
}

@article{Dine:1993yw,
  author        = {Dine, Michael and Nelson, Ann E.},
  title         = {{Dynamical supersymmetry breaking at low-energies}},
  eprint        = {hep-ph/9303230},
  archiveprefix = {arXiv},
  reportnumber  = {SCIPP-93-03, UCSD-PTH-93-05},
  doi           = {10.1103/PhysRevD.48.1277},
  journal       = {Phys. Rev. D},
  volume        = {48},
  pages         = {1277--1287},
  year          = {1993}
}

@article{Chiang:2020yym,
  author        = {Chiang, Cheng-Wei and Huang, Da and Lu, Bo-Qiang},
  title         = {{Electroweak phase transition confronted with dark matter detection constraints}},
  eprint        = {2009.08635},
  archiveprefix = {arXiv},
  primaryclass  = {hep-ph},
  doi           = {10.1088/1475-7516/2021/01/035},
  journal       = {JCAP},
  volume        = {01},
  pages         = {035},
  year          = {2021}
}

@article{Grzadkowski:2018nbc,
  author        = {Grzadkowski, Bohdan and Huang, Da},
  title         = {{Spontaneous $CP$-Violating Electroweak Baryogenesis and Dark Matter from a Complex Singlet Scalar}},
  eprint        = {1807.06987},
  archiveprefix = {arXiv},
  primaryclass  = {hep-ph},
  doi           = {10.1007/JHEP08(2018)135},
  journal       = {JHEP},
  volume        = {08},
  pages         = {135},
  year          = {2018}
}

@article{Chao:2017vrq,
  author        = {Chao, Wei and Guo, Huai-Ke and Shu, Jing},
  title         = {{Gravitational Wave Signals of Electroweak Phase Transition Triggered by Dark Matter}},
  eprint        = {1702.02698},
  archiveprefix = {arXiv},
  primaryclass  = {hep-ph},
  doi           = {10.1088/1475-7516/2017/09/009},
  journal       = {JCAP},
  volume        = {09},
  pages         = {009},
  year          = {2017}
}

@article{Huang:2017kzu,
  author        = {Huang, Fa Peng and Li, Chong Sheng},
  title         = {{Probing the baryogenesis and dark matter relaxed in phase transition by gravitational waves and colliders}},
  eprint        = {1709.09691},
  archiveprefix = {arXiv},
  primaryclass  = {hep-ph},
  reportnumber  = {CTPU-17-34},
  doi           = {10.1103/PhysRevD.96.095028},
  journal       = {Phys. Rev. D},
  volume        = {96},
  number        = {9},
  pages         = {095028},
  year          = {2017}
}

@article{Madge:2023dxc,
  author        = {Madge, Eric and Morgante, Enrico and Puchades-Ib\'a\~nez, Cristina and Ramberg, Nicklas and Ratzinger, Wolfram and Schenk, Sebastian and Schwaller, Pedro},
  title         = {{Primordial gravitational waves in the nano-Hertz regime and PTA data \textemdash{} towards solving the GW inverse problem}},
  eprint        = {2306.14856},
  archiveprefix = {arXiv},
  primaryclass  = {hep-ph},
  reportnumber  = {MITP-23-029},
  doi           = {10.1007/JHEP10(2023)171},
  journal       = {JHEP},
  volume        = {10},
  pages         = {171},
  year          = {2023}
}

@article{Bian:2023dnv,
  author        = {Bian, Ligong and Ge, Shuailiang and Shu, Jing and Wang, Bo and Yang, Xing-Yu and Zong, Junchao},
  title         = {{Gravitational wave sources for pulsar timing arrays}},
  eprint        = {2307.02376},
  archiveprefix = {arXiv},
  primaryclass  = {astro-ph.HE},
  doi           = {10.1103/PhysRevD.109.L101301},
  journal       = {Phys. Rev. D},
  volume        = {109},
  number        = {10},
  pages         = {L101301},
  year          = {2024}
}

@article{Wu:2023hsa,
  author        = {Wu, Yu-Mei and Chen, Zu-Cheng and Huang, Qing-Guo},
  title         = {{Cosmological interpretation for the stochastic signal in pulsar timing arrays}},
  eprint        = {2307.03141},
  archiveprefix = {arXiv},
  primaryclass  = {astro-ph.CO},
  doi           = {10.1007/s11433-023-2298-7},
  journal       = {Sci. China Phys. Mech. Astron.},
  volume        = {67},
  number        = {4},
  pages         = {240412},
  year          = {2024}
}

@article{Planck:2015fie,
  author        = {Ade, P. A. R. and others},
  collaboration = {Planck},
  title         = {{Planck 2015 results. XIII. Cosmological parameters}},
  eprint        = {1502.01589},
  archiveprefix = {arXiv},
  primaryclass  = {astro-ph.CO},
  doi           = {10.1051/0004-6361/201525830},
  journal       = {Astron. Astrophys.},
  volume        = {594},
  pages         = {A13},
  year          = {2016}
}

@article{King:2014nza,
  author        = {King, Stephen F. and Merle, Alexander and Morisi, Stefano and Shimizu, Yusuke and Tanimoto, Morimitsu},
  title         = {{Neutrino Mass and Mixing: from Theory to Experiment}},
  eprint        = {1402.4271},
  archiveprefix = {arXiv},
  primaryclass  = {hep-ph},
  doi           = {10.1088/1367-2630/16/4/045018},
  journal       = {New J. Phys.},
  volume        = {16},
  pages         = {045018},
  year          = {2014}
}

@article{Bertone:2016nfn,
  author        = {Bertone, Gianfranco and Hooper, Dan},
  title         = {{History of dark matter}},
  eprint        = {1605.04909},
  archiveprefix = {arXiv},
  primaryclass  = {astro-ph.CO},
  reportnumber  = {FERMILAB-PUB-16-157-A},
  doi           = {10.1103/RevModPhys.90.045002},
  journal       = {Rev. Mod. Phys.},
  volume        = {90},
  number        = {4},
  pages         = {045002},
  year          = {2018}
}

@article{Marsh:2015xka,
  author        = {Marsh, David J. E.},
  title         = {{Axion Cosmology}},
  eprint        = {1510.07633},
  archiveprefix = {arXiv},
  primaryclass  = {astro-ph.CO},
  reportnumber  = {KCL-PH-TH-2015-50},
  doi           = {10.1016/j.physrep.2016.06.005},
  journal       = {Phys. Rept.},
  volume        = {643},
  pages         = {1--79},
  year          = {2016}
}

@article{Dine:2003ax,
  author        = {Dine, Michael and Kusenko, Alexander},
  title         = {{The Origin of the matter - antimatter asymmetry}},
  eprint        = {hep-ph/0303065},
  archiveprefix = {arXiv},
  reportnumber  = {SCIPP-2003-08, UCLA-03-TEP-08},
  doi           = {10.1103/RevModPhys.76.1},
  journal       = {Rev. Mod. Phys.},
  volume        = {76},
  pages         = {1},
  year          = {2003}
}

@article{NANOGrav:2020tig,
  author        = {Vallisneri, M. and others},
  collaboration = {NANOGrav},
  title         = {{Modeling the uncertainties of solar-system ephemerides for robust gravitational-wave searches with pulsar timing arrays}},
  eprint        = {2001.00595},
  archiveprefix = {arXiv},
  primaryclass  = {astro-ph.HE},
  doi           = {10.3847/1538-4357/ab7b67},
  month         = {1},
  year          = {2020}
}

@article{Pham:2024vso,
  author        = {Pham, Hieu The and Senaha, Eibun},
  title         = {{Gravitational waves from domain wall collapses and dark matter in the SM with a complex scalar field}},
  eprint        = {2403.16568},
  archiveprefix = {arXiv},
  primaryclass  = {hep-ph},
  doi           = {10.1103/PhysRevD.109.095048},
  journal       = {Phys. Rev. D},
  volume        = {109},
  number        = {9},
  pages         = {095048},
  year          = {2024}
}

@article{Arkani-Hamed:2001kyx,
  author        = {Arkani-Hamed, Nima and Cohen, Andrew G. and Georgi, Howard},
  title         = {{(De)constructing dimensions}},
  eprint        = {hep-th/0104005},
  archiveprefix = {arXiv},
  reportnumber  = {HUTP-01-A015, BUHEP-01-05, LBNL-47676, UCB-PTH-01-11},
  doi           = {10.1103/PhysRevLett.86.4757},
  journal       = {Phys. Rev. Lett.},
  volume        = {86},
  pages         = {4757--4761},
  year          = {2001}
}

@article{Press:1989yh,
    author = "Press, William H. and Ryden, Barbara S. and Spergel, David N.",
    title = "{Dynamical Evolution of Domain Walls in an Expanding Universe}",
    reportNumber = "NSF-ITP-89-51, CFA-1870",
    doi = "10.1086/168151",
    journal = "Astrophys. J.",
    volume = "347",
    pages = "590--604",
    year = "1989"
}

@article{Rubin:2000dq,
    author = "Rubin, S. G. and Khlopov, M. Yu. and Sakharov, A. S.",
    editor = "Khlopov, M. Yu. and Prokhorov, M. E. and Starobinsky, A. A.",
    title = "{Primordial black holes from nonequilibrium second order phase transition}",
    eprint = "hep-ph/0005271",
    archivePrefix = "arXiv",
    journal = "Grav. Cosmol.",
    volume = "6",
    pages = "51--58",
    year = "2000"
}

@article{Blasi:2022ayo,
    author = {Blasi, Simone and Mariotti, Alberto and Rase, A\"aron and Sevrin, Alexander and Turbang, Kevin},
    title = "{Friction on ALP domain walls and gravitational waves}",
    eprint = "2210.14246",
    archivePrefix = "arXiv",
    primaryClass = "hep-ph",
    doi = "10.1088/1475-7516/2023/04/008",
    journal = "JCAP",
    volume = "04",
    pages = "008",
    year = "2023"
}

@article{Lu:2024szr,
    author = "Lu, Bo-Qiang and Chiang, Cheng-Wei and Li, Tianjun",
    title = "{A Common Origin for Nano-Hz Gravitational Wave Background and Black Hole Merger Events}",
    eprint = "2409.10251",
    archivePrefix = "arXiv",
    primaryClass = "astro-ph.CO",
    month = "9",
    year = "2024"
}

@article{Deng:2020mds,
    author = "Deng, Heling",
    title = "{Primordial black hole formation by vacuum bubbles. Part II}",
    eprint = "2006.11907",
    archivePrefix = "arXiv",
    primaryClass = "astro-ph.CO",
    doi = "10.1088/1475-7516/2020/09/023",
    journal = "JCAP",
    volume = "09",
    pages = "023",
    year = "2020"
}

@article{Takahashi:2020tqv,
  author        = {Takahashi, Fuminobu and Yin, Wen},
  title         = {{Kilobyte Cosmic Birefringence from ALP Domain Walls}},
  eprint        = {2012.11576},
  archiveprefix = {arXiv},
  primaryclass  = {hep-ph},
  reportnumber  = {TU-1115, IPMU20-0130},
  doi           = {10.1088/1475-7516/2021/04/007},
  journal       = {JCAP},
  volume        = {04},
  pages         = {007},
  year          = {2021}
}

@article{Kitajima:2022jzz,
  author        = {Kitajima, Naoya and Kozai, Fumiaki and Takahashi, Fuminobu and Yin, Wen},
  title         = {{Power spectrum of domain-wall network, and its implications for isotropic and anisotropic cosmic birefringence}},
  eprint        = {2205.05083},
  archiveprefix = {arXiv},
  primaryclass  = {astro-ph.CO},
  reportnumber  = {TU-1155},
  doi           = {10.1088/1475-7516/2022/10/043},
  journal       = {JCAP},
  volume        = {10},
  pages         = {043},
  year          = {2022}
}

@article{Papanikolaou:2020qtd,
    author = "Papanikolaou, Theodoros and Vennin, Vincent and Langlois, David",
    title = "{Gravitational waves from a universe filled with primordial black holes}",
    eprint = "2010.11573",
    archivePrefix = "arXiv",
    primaryClass = "astro-ph.CO",
    doi = "10.1088/1475-7516/2021/03/053",
    journal = "JCAP",
    volume = "03",
    pages = "053",
    year = "2021"
}

@article{Kodama:1986ud,
  author       = {Kodama, Hideo and Sasaki, Misao},
  title        = {{Evolution of Isocurvature Perturbations. 2. Radiation Dust Universe}},
  reportnumber = {UTAP-41},
  doi          = {10.1142/S0217751X8700020X},
  journal      = {Int. J. Mod. Phys. A},
  volume       = {2},
  pages        = {491},
  year         = {1987}
}

@unpublished{Suonio:2014cp,
    author = {H. Kurki-Suonio},
    title = {{Cosmological Perturbation Theory I}},
    note = {Lecture on cosmological perturbation theory at the University of Helsinki}
}

@article{Mroz:2024wag,
    author = "Mr\'oz, Przemek and others",
    title = "{Microlensing Optical Depth and Event Rate toward the Large Magellanic Cloud Based on 20 yr of OGLE Observations}",
    eprint = "2403.02398",
    archivePrefix = "arXiv",
    primaryClass = "astro-ph.GA",
    doi = "10.3847/1538-4365/ad452e",
    journal = "Astrophys. J. Suppl.",
    volume = "273",
    number = "1",
    pages = "4",
    year = "2024"
}

@article{LIGOScientific:2019kan,
    author = "Abbott, B. P. and others",
    collaboration = "LIGO Scientific, Virgo",
    title = "{Search for Subsolar Mass Ultracompact Binaries in Advanced LIGO\textquoteright{}s Second Observing Run}",
    eprint = "1904.08976",
    archivePrefix = "arXiv",
    primaryClass = "astro-ph.CO",
    reportNumber = "LIGO-P1900037",
    doi = "10.1103/PhysRevLett.123.161102",
    journal = "Phys. Rev. Lett.",
    volume = "123",
    number = "16",
    pages = "161102",
    year = "2019"
}

@article{Nitz:2022ltl,
    author = "Nitz, Alexander H. and Wang, Yi-Fan",
    title = "{Broad search for gravitational waves from subsolar-mass binaries through LIGO and Virgo\textquoteright{}s third observing run}",
    eprint = "2202.11024",
    archivePrefix = "arXiv",
    primaryClass = "astro-ph.HE",
    doi = "10.1103/PhysRevD.106.023024",
    journal = "Phys. Rev. D",
    volume = "106",
    number = "2",
    pages = "023024",
    year = "2022"
}

@article{Manshanden:2018tze,
    author = "Manshanden, Julien and Gaggero, Daniele and Bertone, Gianfranco and Connors, Riley M. T. and Ricotti, Massimo",
    title = "{Multi-wavelength astronomical searches for primordial black holes}",
    eprint = "1812.07967",
    archivePrefix = "arXiv",
    primaryClass = "astro-ph.HE",
    doi = "10.1088/1475-7516/2019/06/026",
    journal = "JCAP",
    volume = "06",
    pages = "026",
    year = "2019"
}

@article{Serpico:2020ehh,
    author = "Serpico, Pasquale D. and Poulin, Vivian and Inman, Derek and Kohri, Kazunori",
    title = "{Cosmic microwave background bounds on primordial black holes including dark matter halo accretion}",
    eprint = "2002.10771",
    archivePrefix = "arXiv",
    primaryClass = "astro-ph.CO",
    reportNumber = "LAPTH-005/20, KEK-Cosmo-248, KEK-TH-2198, IPMU20-0021",
    doi = "10.1103/PhysRevResearch.2.023204",
    journal = "Phys. Rev. Res.",
    volume = "2",
    number = "2",
    pages = "023204",
    year = "2020"
}

@article{Hektor:2018qqw,
    author = {Hektor, Andi and H\"utsi, Gert and Marzola, Luca and Raidal, Martti and Vaskonen, Ville and Veerm\"ae, Hardi},
    title = "{Constraining Primordial Black Holes with the EDGES 21-cm Absorption Signal}",
    eprint = "1803.09697",
    archivePrefix = "arXiv",
    primaryClass = "astro-ph.CO",
    reportNumber = "CERN-TH-2018-073",
    doi = "10.1103/PhysRevD.98.023503",
    journal = "Phys. Rev. D",
    volume = "98",
    number = "2",
    pages = "023503",
    year = "2018"
}

@article{Brandt:2016aco,
    author = "Brandt, Timothy D.",
    title = "{Constraints on MACHO Dark Matter from Compact Stellar Systems in Ultra-Faint Dwarf Galaxies}",
    eprint = "1605.03665",
    archivePrefix = "arXiv",
    primaryClass = "astro-ph.GA",
    doi = "10.3847/2041-8205/824/2/L31",
    journal = "Astrophys. J. Lett.",
    volume = "824",
    number = "2",
    pages = "L31",
    year = "2016"
}

@article{Green:2020jor,
    author = "Green, Anne M. and Kavanagh, Bradley J.",
    title = "{Primordial Black Holes as a dark matter candidate}",
    eprint = "2007.10722",
    archivePrefix = "arXiv",
    primaryClass = "astro-ph.CO",
    doi = "10.1088/1361-6471/abc534",
    journal = "J. Phys. G",
    volume = "48",
    number = "4",
    pages = "043001",
    year = "2021"
}

@article{Yoo:2020lmg,
  author        = {Yoo, Chul-Moon and Harada, Tomohiro and Okawa, Hirotada},
  title         = {{Threshold of Primordial Black Hole Formation in Nonspherical Collapse}},
  eprint        = {2004.01042},
  archiveprefix = {arXiv},
  primaryclass  = {gr-qc},
  reportnumber  = {RUP-20-12},
  doi           = {10.1103/PhysRevD.102.043526},
  journal       = {Phys. Rev. D},
  volume        = {102},
  number        = {4},
  pages         = {043526},
  year          = {2020},
  note          = {[Erratum: Phys.Rev.D 107, 049901 (2023)]}
}

@article{Escriva:2024aeo,
  author        = {Escriv\`a, Albert and Yoo, Chul-Moon},
  title         = {{Non-spherical effects on the mass function of Primordial Black Holes}},
  eprint        = {2410.03451},
  archiveprefix = {arXiv},
  primaryclass  = {gr-qc},
  month         = {10},
  year          = {2024}
}

@article{Harada:2013epa,
  author        = {Harada, Tomohiro and Yoo, Chul-Moon and Kohri, Kazunori},
  title         = {{Threshold of primordial black hole formation}},
  eprint        = {1309.4201},
  archiveprefix = {arXiv},
  primaryclass  = {astro-ph.CO},
  reportnumber  = {RUP-13-9, KEK-COSMO-129, KEK-TH-1668},
  doi           = {10.1103/PhysRevD.88.084051},
  journal       = {Phys. Rev. D},
  volume        = {88},
  number        = {8},
  pages         = {084051},
  year          = {2013},
  note          = {[Erratum: Phys.Rev.D 89, 029903 (2014)]}
}

@book{Vilenkin:2000jqa,
    author = "Vilenkin, A. and Shellard, E. P. S.",
    title = "{Cosmic Strings and Other Topological Defects}",
    isbn = "978-0-521-65476-0",
    publisher = "Cambridge University Press",
    month = "7",
    year = "2000"
}

@article{Dankovsky:2024zvs,
    author = "Dankovsky, I. and Babichev, E. and Gorbunov, D. and Ramazanov, S. and Vikman, A.",
    title = "{Revisiting evolution of domain walls and their gravitational radiation with CosmoLattice}",
    eprint = "2406.17053",
    archivePrefix = "arXiv",
    primaryClass = "astro-ph.CO",
    doi = "10.1088/1475-7516/2024/09/047",
    journal = "JCAP",
    volume = "09",
    pages = "047",
    year = "2024"
}

@article{Ferreira:2024eru,
    author = "Ferreira, Ricardo Z. and Notari, Alessio and Pujol{\`a}s, Oriol and Rompineve, Fabrizio",
    title = "{Collapsing domain wall networks: impact on pulsar timing arrays and primordial black holes}",
    eprint = "2401.14331",
    archivePrefix = "arXiv",
    primaryClass = "astro-ph.CO",
    reportNumber = "CERN-TH-2024-020",
    doi = "10.1088/1475-7516/2024/06/020",
    journal = "JCAP",
    volume = "06",
    pages = "020",
    year = "2024"
}

@article{Notari:2025kqq,
    author = "Notari, Alessio and Rompineve, Fabrizio and Torrenti, Francisco",
    title = "{The spectrum of gravitational waves from annihilating domain walls}",
    eprint = "2504.03636",
    archivePrefix = "arXiv",
    primaryClass = "astro-ph.CO",
    doi = "10.1088/1475-7516/2025/07/049",
    journal = "JCAP",
    volume = "07",
    pages = "049",
    year = "2025"
}
\end{document}